\documentclass[11pt,preprint,natbib209]{aastex}

\usepackage{epsfig}
\usepackage{graphicx}
\usepackage{natbib}
\usepackage{longtable,lscape}

\newcommand{\flux}{erg\,s$^{-1}$\,cm$^{-2}$}
\newcommand{\nh}{$N_{\mathrm H}$}

\begin{document}

\title{The Suzaku View of the Swift/BAT AGNs (II): Time Variability and Spectra of Five ``Hidden'' AGNs}
\author{Lisa M. Winter\altaffilmark{1}, Richard F. Mushotzky\altaffilmark{2}, Yuichi Terashima\altaffilmark{3}, \& Yoshihiro Ueda\altaffilmark{4}}

\email{lisa.winter@colorado.edu}

\altaffiltext{1}{Center for Astrophysics and Space Astronomy, University of Colorado, Boulder, CO 80309}
\altaffiltext{2}{NASA Goddard Space Flight Center, Greenbelt, MD 20771}
\altaffiltext{3}{Department of Physics, Ehime University, Matsuyama, Ehime 790-8577, Japan}
\altaffiltext{4}{Department of Astronomy, University of Kyoto, Kyoto 606-8502, Japan}

\begin{abstract}
The fraction of Compton thick sources is one of the main uncertainties left in understanding the AGN population.  The Swift Burst Alert Telescope (BAT) all-sky survey, for the first time gives us an unbiased sample of AGN for all but the most heavily absorbed sources \nh $ > 10^{25}$\,cm$^{-2}$).  Still, the BAT spectra (14 -- 195\,keV) are time-averaged over months of observations and therefore hard to compare with softer spectra from the Swift XRT or other missions.  This makes it difficult to distinguish between Compton-thin and Compton-thick models.  With Suzaku, we have obtained simultaneous hard ($> 15$\,keV) and soft (0.3 -- 10\,keV) X-ray spectra for 5 Compton-thick candidate sources.  We report on the spectra and a comparison with the BAT and earlier XMM observations.  Based on both flux variability and spectral shape, we conclude that these hidden sources are not Compton-thick.  We also report on a possible correlation between excess variance and Swift BAT luminosity from the 16 d binned light curves, which holds true for a sample of both absorbed (4 sources), unabsorbed (8 sources), and Compton thick (Circinus) AGN, but is weak in the 64\,day binned BAT light curves.
\end{abstract}

\section{Introduction}
The Burst Alert Telescope (BAT) is a very hard X-ray (14 -- 195 keV) telescope surveying the entire sky in search of Gamma Ray Bursts.  Between bursts, it conducts an all-sky survey.  
Since the BAT is sensitive above 14\,keV, it is unbiased towards the absorption which greatly affects surveys in other wavebands \citep{2004ASSL..308...53M}.  Analysis of the properties of an unbiased active galactic nuclei (AGN) sample will allow us to discover heavily absorbed sources with little or no optical indication of AGN emission.  This is important because it will give us an indication of the true ratio of absorbed/unabsorbed AGN in the local universe.  Since the fraction of Compton thick sources is one of the main uncertainties left in understanding the AGN population \citep{2004ASSL..308..245C}, the fraction of Compton thick BAT sources will indicate, at the very least, limits on the flux and spectral shape for local Compton thick populations.  Also, since AGN are likely the major contributors to the cosmic X-ray background \citep{1995AA...296....1C,1999AA...347..424G}, the distribution of properties of the unbiased BAT AGN sources will allow for more accurate models of the X-ray background. 

This work concentrates on five AGN sources detected in the 9-month BAT AGN survey, which were associated with 4 sources in the BAT catalog (one pair of sources, NGC 6921 and MCG +04-48-002, is associated with SWIFT J2028.5+2543), with a detection significance of 4.8-$\sigma$ or above.  There are 153 AGN sources in this 9-month catalog \citep{2008ApJ...681..113T}, all with BAT fluxes (14--195\,keV) in excess of $10^{-11}$\,erg\,s$^{-1}$\,cm$^{-2}$ and an average redshift of 0.03. In this paper, we present Suzaku follow-ups for interesting sources from a previous {\it XMM-Newton} follow-up study.  

In \citet{2008ApJ...674..686W}, we presented the X-ray properties from $\approx 10$\,ks {\it XMM-Newton} observations of 22 previously unobserved (in the X-rays) AGNs from the BAT 9-month catalog.  Of these sources, half were absorbed sources with \nh$ > 10^{23}$\,cm$^{-2}$.  We found that 5 sources had spectra consistent with Compton-thick sources (\nh$ > 1.5 \times 10^{24}$\,cm$^{-2}$) and 4 more were ``hidden'' AGN, sources with large scattering fractions and a low ratio of $F_{0.5-2 keV}/F_{2-10 keV}$ as discussed in \citet{2007ApJ...664L..79U} and \citet{2009ApJ...690.1322W}.  However, with the $\approx 10$\,ks observations these heavily absorbed sources did not have enough counts to distinguish whether a reflection model or a power law model better fit the data.  Further, adding the BAT data to better constrain the fit is potentially problematic since the BAT spectra are time averaged over months of observations.  Only with Suzaku observations could we obtain a simultaneous hard ($> 15$\,keV) and soft ($< 10$\,keV) spectrum, necessary for further investigation of the complex absorbed sources. 

This paper is the second in a series of papers presenting Suzaku follow-ups of Swift BAT-detected AGNs (following paper one: \citet{eguchi2008}). 
In this paper, we present results of our analysis of the Suzaku XIS and HXD PIN spectra for 5 heavily obscured sources (NGC 1142,  Mrk 417,  ESO 506-G027, NGC 6921, and MCG +04-48-002).  In \S~\ref{data}, we present the data analysis.  The spectral analysis, focusing on variability (\S~\ref{var}) and interesting features (\S~\ref{spectra}), follows.  In \S~\ref{summary}, we summarize our results.  Throughout our analysis, we use the {\tt wilm} ISM abundances \citep{2000ApJ...542..914W}.

\section{Data Analysis}\label{data}
Based on successful proposals during Suzaku AO-1 and AO-2, we obtained observations of NGC 1142,  Mrk 417,  ESO 506-G027, NGC 6921, and MCG +04-48-002 in the HXD nominal pointing mode.  All of these sources were absorbed (n$_H > 10^{23}$\,cm$^{-2}$), with complex spectra in our previous XMM-Newton study \citep{2008ApJ...674..686W}.  In  \citet{2008ApJ...674..686W}, we noted that MCG +04-48-002 and NGC 6921 combined are likely the BAT source, since both of these AGN are within the 6\arcmin error circle of BAT.  In our {\it XMM-Newton} analysis, NGC 6921 was the brighter source with a pn count rate of 0.255\,cts\,s$^{-1}$ compared to 0.096\,cts\,s$^{-1}$ for MCG +04-48-002.  However, in the Suzaku observation, NGC 6921 is extremely dim with an XIS1 count rate of 0.003\,cts\,s$^{-1}$ compared to 0.015\,cts\,s$^{-1}$ for  MCG +04-48-002.  With a field of view of 1\degr, the HXD PIN spectrum of this observation is also likely a combination of both AGN.  However, given the faint state of NGC 6921 in the observation, this spectrum is probably dominated by emission from MCG +04-48-002.  Details of the Suzaku observations for NGC 1142,  Mrk 417,  ESO 506-G027, NGC 6921, and MCG +04-48-002 are shown in Table~\ref{tbl-1}.  An analysis of the AO-1 observation of NGC 1142 is also included in \citet{eguchi2008} (paper 1 in this series).  For details of the analysis of the {\it XMM-Newton} and Swift BAT spectra, see \citet{2008ApJ...674..686W}.  

To extract each of the Suzaku spectra, we used the cleaned version 2.0 processed event files supplied by the Suzaku team.  For processing of the XIS data, we combined the 3x3 and 5x5 edit modes for the front-illuminated (FI), XIS0 and XIS3, and back-illuminated (BI), XIS1, CCDs in {\tt XSELECT}.  The source spectra were extracted from a circular 20$\arcsec$ region, centered on the source.   Background spectra were extracted from 40$\arcsec$ regions located nearby regions free of emission.  Response matrices (rmf) and ancillary response matrices (arf) were then generated using the newest versions of {\tt xissimrmfgen} and {\tt xissimarfgen}.  Following this, we combined the two XIS-FI spectra with {\tt MATHPHA} and the response files with {\tt addrmf} and {\tt addarf}.  The spectra and rmf files were rebinned from 4096 to 1024 channels, which is still larger than the resolution of the CCDs, in order to slow down the time the standard spectral fitting software, {\tt XSPEC} \citep{ADASS_96_A}, takes to load and fit the spectra.  The combined FI spectra and XIS1 spectra were each grouped with the response files with 20\,cts/bin, using {\tt grppha}.

For the HXD data, we only used data from the PIN instrument.  The spectra were extracted using {\tt XSELECT} for both the PIN event file and the corresponding instrumental tuned background file, supplied by the Suzaku team.   Both spectra were generated using the same good time intervals.  The source spectrum was further corrected for instrument dead time.  In addition to the instrumental background spectrum, we generated a cosmic X-ray background in {\tt XSPEC} using the model suggested by the Suzaku team (a cutoff power law of the form $CXB(E) = 9.412 \times 10^{-3} \times (E/1 keV)^{-1.29} \times \exp(-E/40 keV)$ photons\,cm$^{-2}$\,s$^{-1}$\,FOV$^{-1}$\,keV$^{-1}$).  We then combined the instrumental and cosmic X-ray backgrounds with {\tt MATHPHA}.  Finally, the spectra were binned, using {\tt grppha}, with the background and standard response files from the Suzaku {\tt CALDB} to a signal-to-noise ratio of 3\,$\sigma$.


\section{Variability}\label{var}
Compton-thick sources are AGN where our line-of-sight to the source is blocked by obscuring matter that has an optical depth of $\tau > 1$ towards Compton scattering (\nh$ > 1.5 \times 10^{24}$\,cm$^{-2}$).  At these optical depths, much of the emission is reflected and not direct emission.  For Compton-thick sources, the column densities are so high that little to no direct emission escapes below 10\,keV. As such, one possible marker of a Compton-thick source, whose emission is dominated by reflection, is a lack of variability.  For instance, analysis of the recent 100\,ks Suzaku observation of the mildly Compton thick source NGC 4945 (\nh $\sim 5 \times 10^{24}$\,cm$^{-2}$) found no variability below 10\,keV, consistent with a reflection dominated spectrum, 
while the high energy spectrum ($> 10$\,keV), where the direct emission is not obscured,  showed a factor of two variability
\citep{2008PASJ...60S.251I}.

Further, one of the closest Compton-thick AGN, located in the nucleus of Circinus, is observed as consistently not variable.  Circinus has been observed extensively with ASCA \citep{1996MNRAS.281L..69M}, {\it XMM-Newton} \citep{2003MNRAS.343L...1M}, Chandra \citep{2001ApJ...546L..13S,2006AA...455..153M}, and Beppo-Sax \citep{1999MNRAS.310...10G}.  Between these observations, which span approximately 9 years, there is no significant change in the shape of the spectrum or the flux of the source.  Similarly, the recent Suzaku observation of Circinus also shows a reflection dominated spectrum with a flux consistent with previous observations \citep{2008arXiv0809.4656Y}.  However, \citet{2008arXiv0809.4656Y} note possible low level (2--15\%) variability in the spectra both above and below 10\,keV, whose uncertainty is high due to uncertainties in the pin background (the source is only 7\% above the background) and possible contamination from ULX sources.  If our target sources are Compton-thick, like the AGN in Circinus, we would expect little variability in spectral shape and flux.  However, if our sources resemble ``changing-look'' sources, sources such as NGC 1365 \citep{2005ApJ...623L..93R} which change for Compton-thick to Compton-thin stages \citep{2003MNRAS.342..422M}, we would expect to find variability in the absorbing column. 

There are two types of variability that we can probe with our Suzaku observations.  These types include variability on the time scale of the observation (short term variability) and variability between observations for an individual source.  In this section, we first characterize short term variability in the Suzaku observations with XIS.  With fewer counts in the HXD spectra, we do not perform a similar analysis at higher energies.  To probe variability between observations for individual sources, we compare variability in spectral parameters, including spectral index, flux, column density, and the fluorescent Fe-K$\alpha$ line, between our previous 10\,ks {\it XMM-Newton} observations and the Suzaku XIS observations.  Finally, we compare variability between the HXD PIN and BAT for NGC 1142, the only source for which we have two HXD PIN observations.

\subsection{Variability during the Observations}\label{var-during}
\subsubsection{Light Curve Analysis}
To test for variability during the individual Suzaku observations, we first constructed 128\,s light curves (0.1 -- 12\,keV) for both the source and background regions used to extract the spectra using {\tt XSELECT}.  Light curves were computed for XIS0, XIS1, and XIS3.  Computing average count rates for
each light curve, we find that the XIS0 and XIS3 source light curves are 4 -- 6 times higher than the background rates for NGC 1142, Mrk 417, and ESO 506-G027.  The XIS1 rates are only 1.5 -- 1.6 times higher than the background.  For MCG +04-48-002 and NGC 6921, the source count rates are lower, compared to the background rates.  MCG +04-48-002 has average XIS0 and XIS3 rates of about 2 times the background with an XIS1 rate of only 1.25 times the background.  NGC 6921 is dimmer, with XIS0 and XIS3 rates 1.4 -- 1.5 times the background and XIS1 average rates only 1.2 times the background.

For each of the XIS light curves, we subtracted the background rates to create a net light curve.  The XIS0, XIS1, and XIS3 net light curves were then combined to create an average net light curve.
For each of these average net light curves, we calculated the normalized excess variance, a measurement of the variability amplitude in the light curve, and $\chi^2$ values.  Excess variability, as defined by \citet{1997ApJ...476...70N}, corresponds to: \begin{equation} \sigma^2_{rms} = \frac{1}{N\mu^2}\displaystyle\sum_{i=1}^N [(X_i - \mu)^2 - \sigma_i^2] .\end{equation}  Here, $\mu$ corresponds to the unweighted mean, $N$ is the number of points in the light curve, $X_i$ is the count rate at $i$, and $\sigma_i$ is the corresponding error in the count rate.  We computed errors on $\sigma^2_{rms}$ using equation 11 from \citet{2003MNRAS.345.1271V}.  Since the excess variability is dependent on observation length \citep{1993ApJ...414L..85L}, we divided the long NGC 1142 (observation 1) observation into two evenly spaced observations, computing $\sigma^2_{rms}$ and $\chi^2$ individually for each half of the observation.  These variability measurements are shown in Table~\ref{tbl-2}.  We also constructed light curves binned by the orbital time scale ($\approx 5760$\,s), with the results of this analysis also shown in Table~\ref{tbl-2}.

\begin{figure}
\begin{center}
\includegraphics[width=7cm]{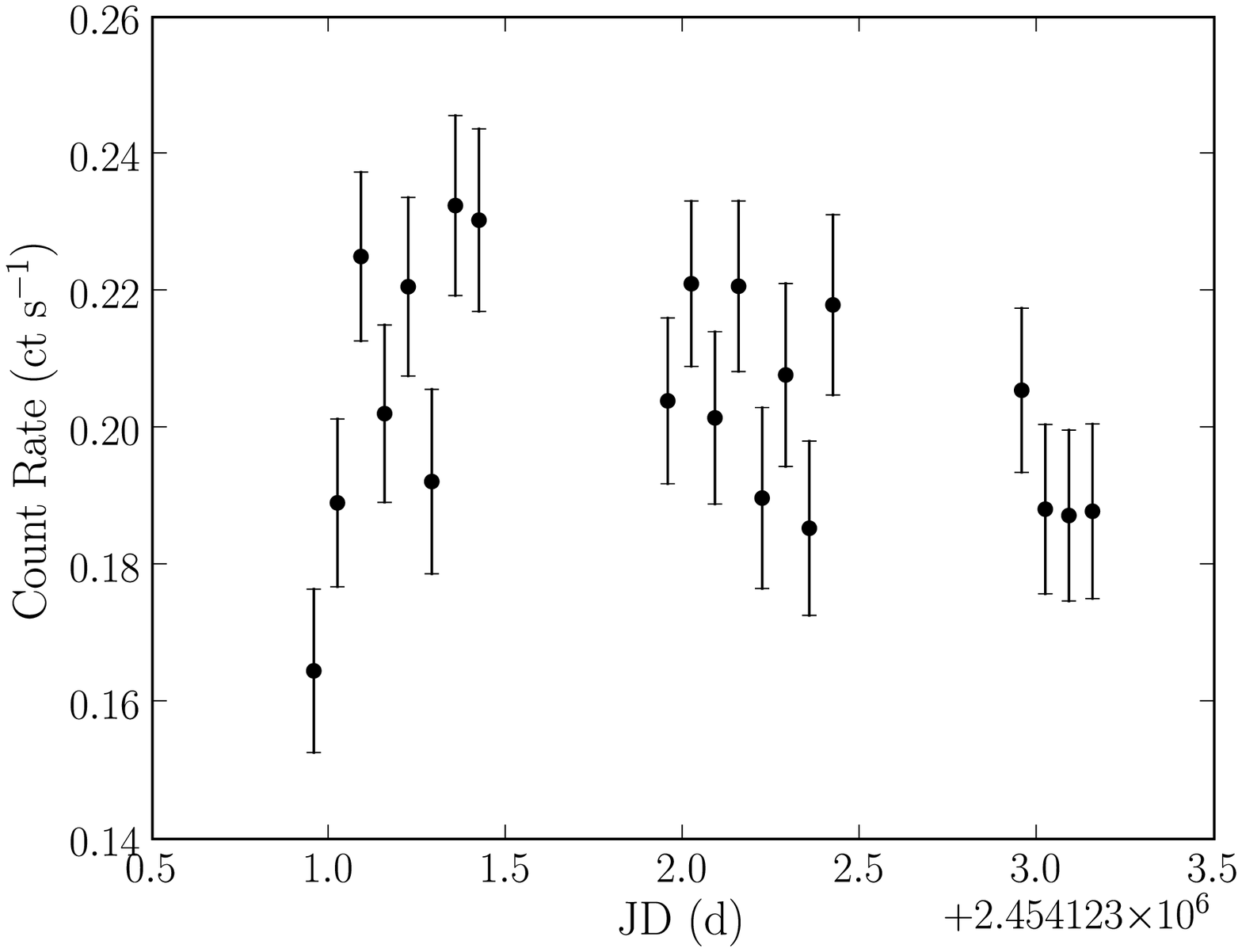}
\hspace{0.25cm}
\includegraphics[width=7cm]{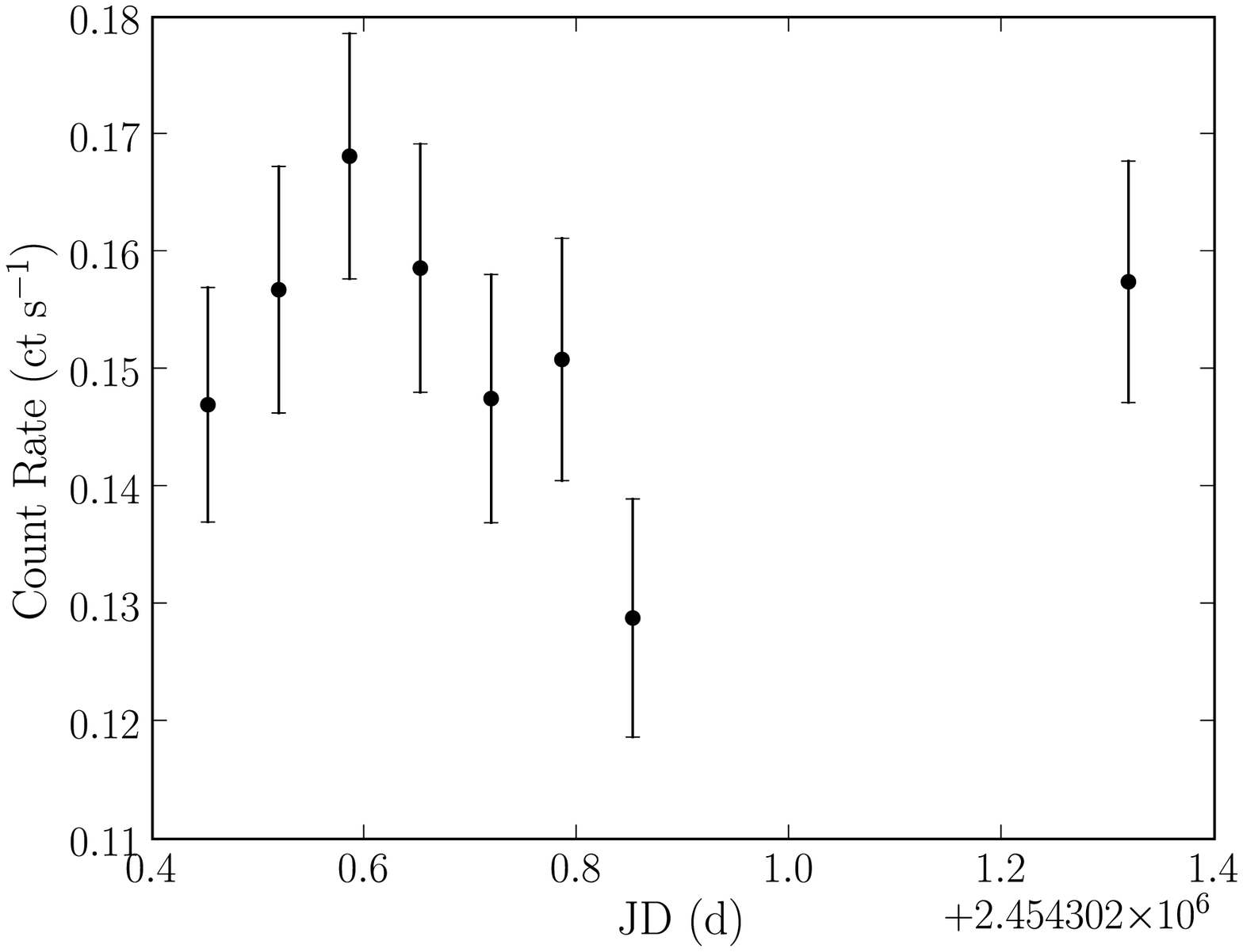}
\vspace{0.5cm}
\includegraphics[width=7cm]{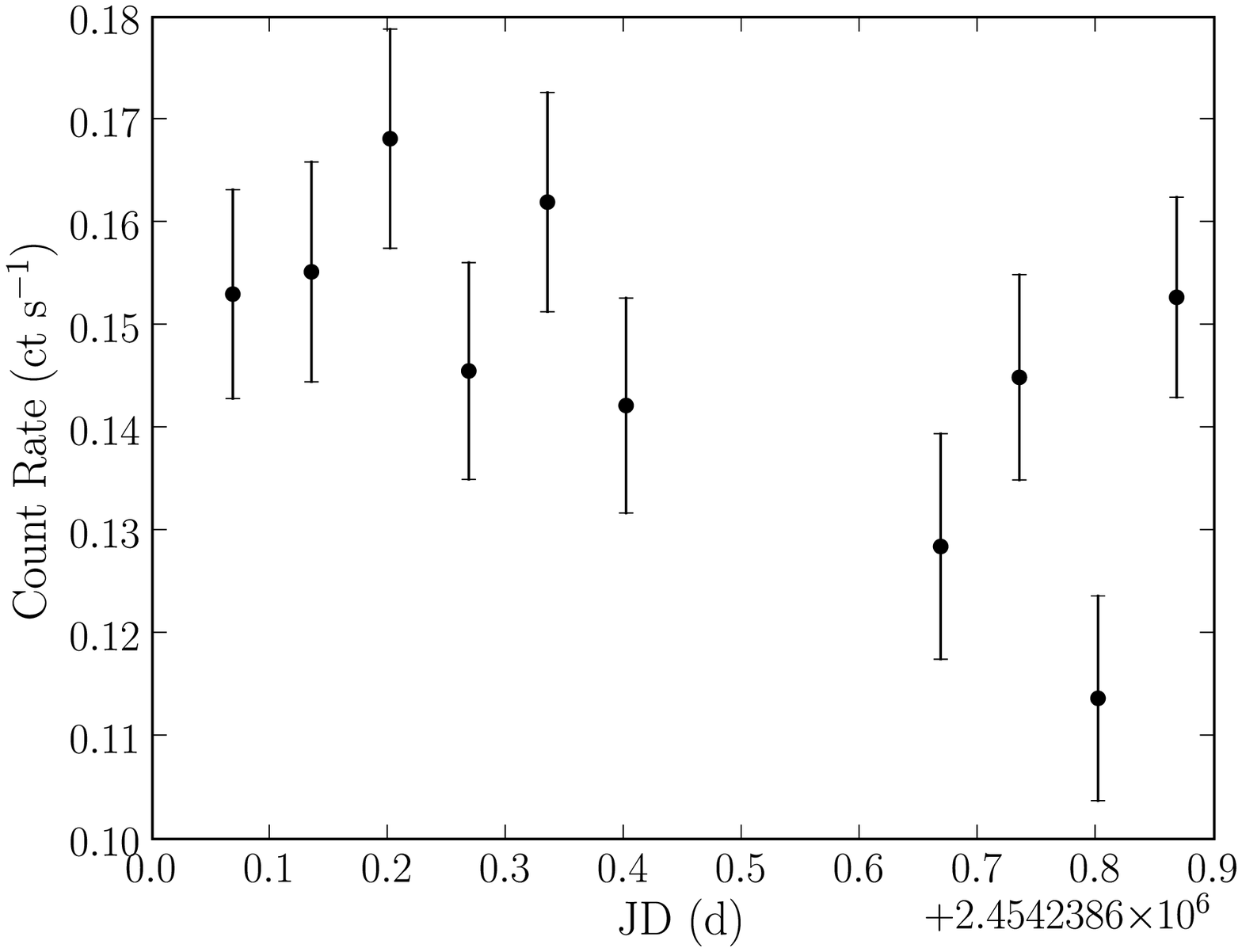}
\hspace{0.25cm}
\includegraphics[width=7cm]{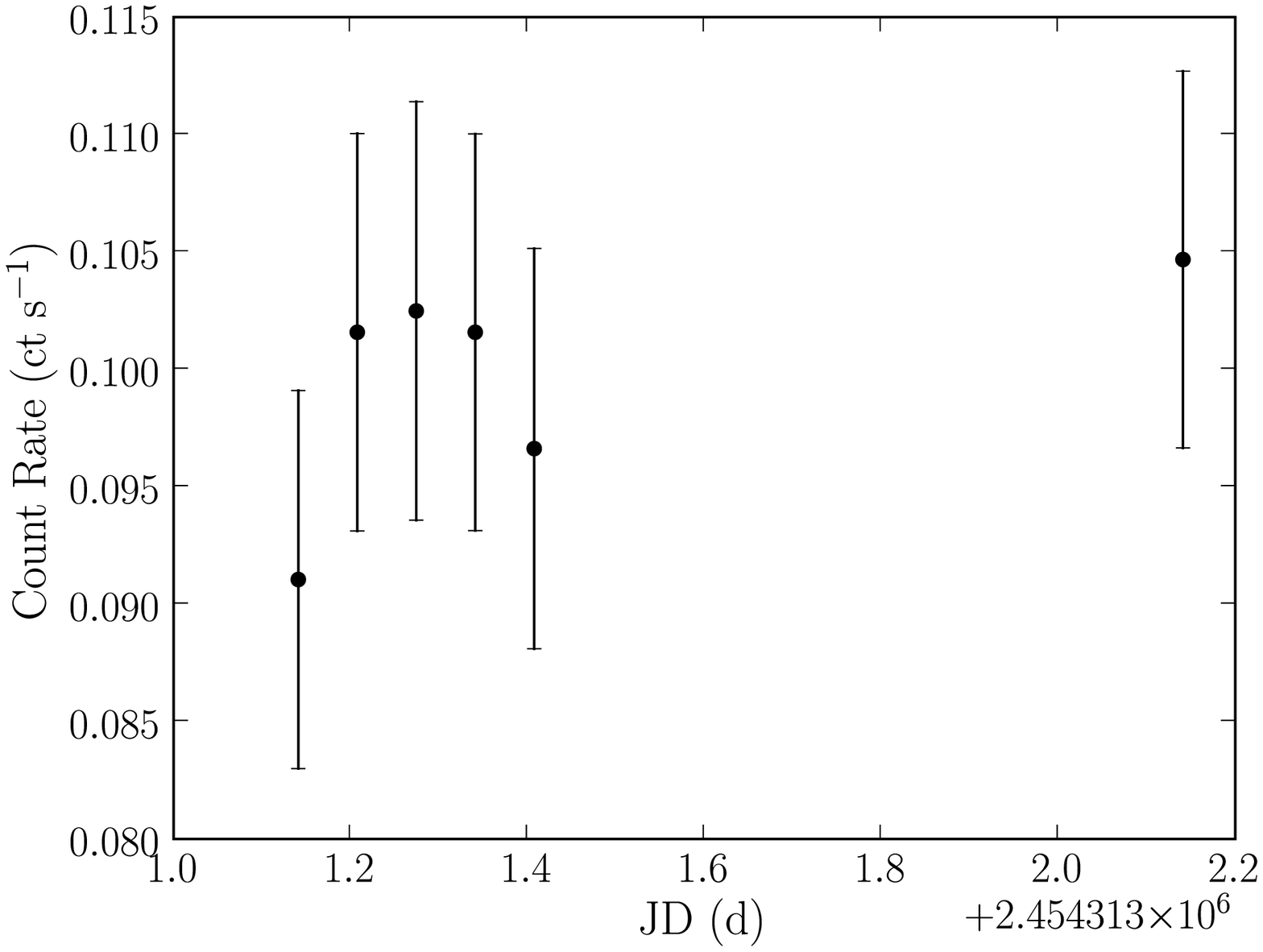}
\vspace{0.5cm}
\includegraphics[width=7cm]{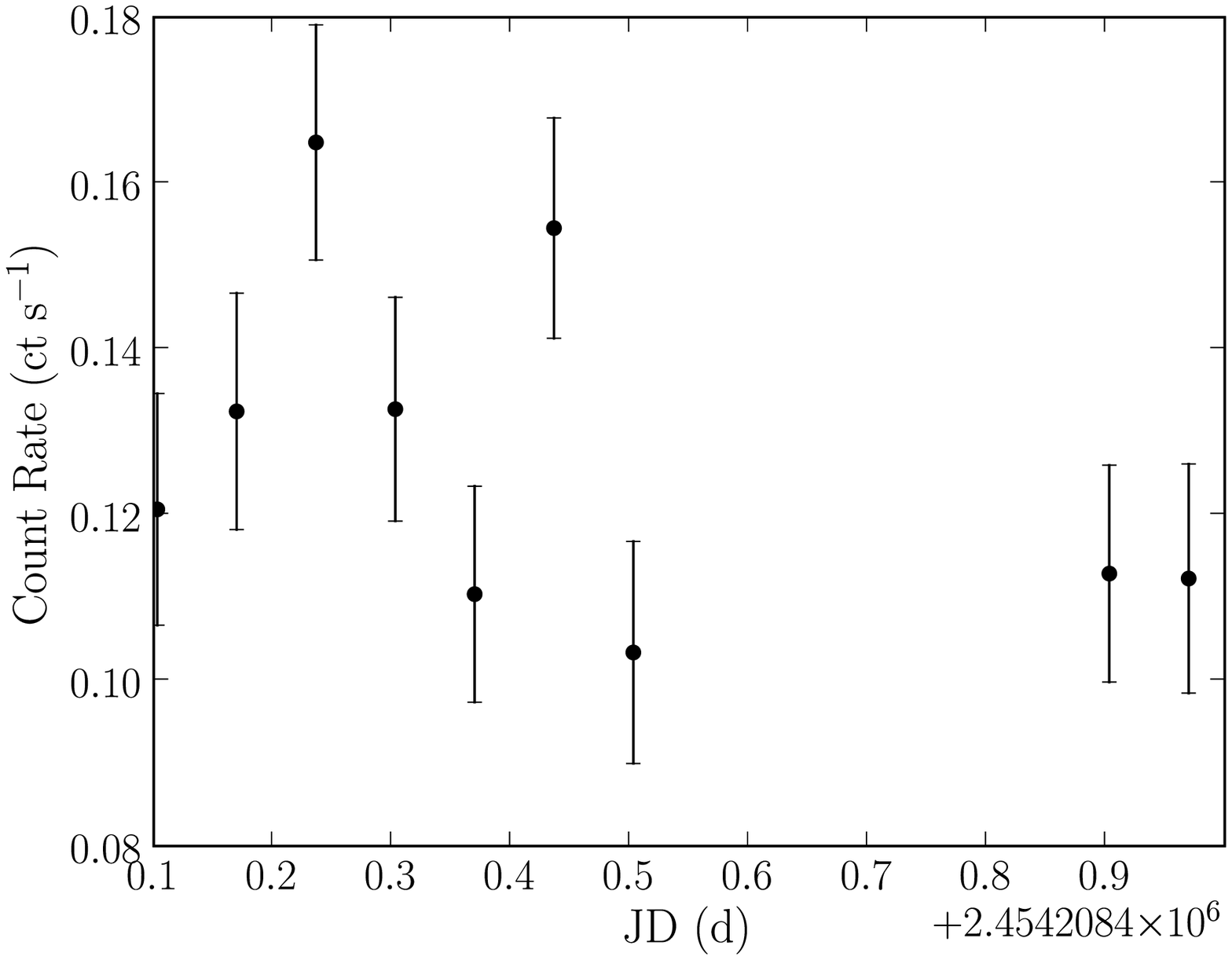}
\hspace{0.25cm}
\includegraphics[width=7cm]{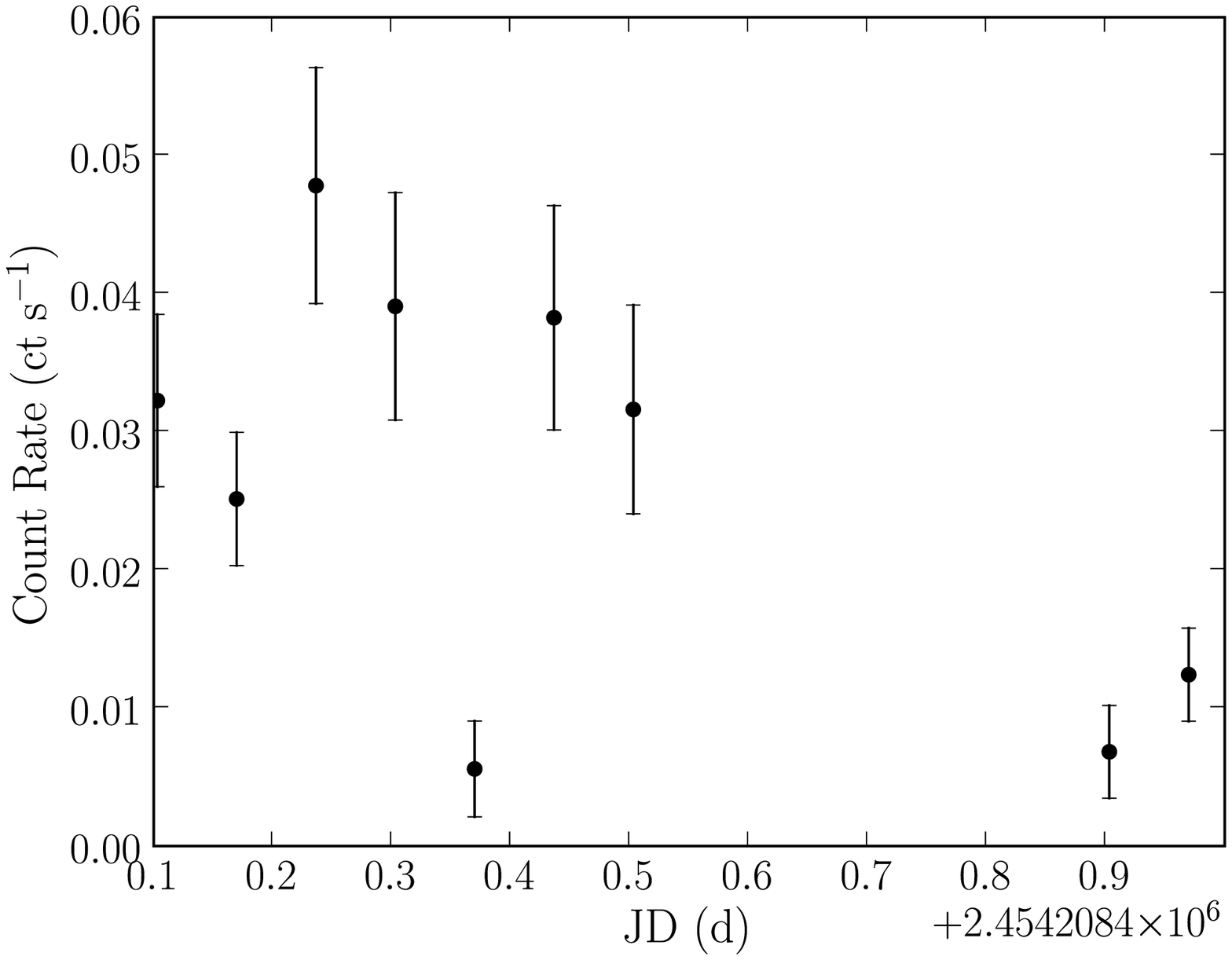}
\end{center}
\caption{Light curves binned by 5760s (orbital period) for NGC 1142 observation 1 (top left), NGC 1142 observation 2 (top right), Mrk 417 (middle left), ESO 506-G027 (middle right), MCG +04-48-02 (bottom left) and NGC 6921 (bottom right).  The light curves are background subtracted and added XIS0 + XIS1 + XIS3 curves.  
\label{fig-128lc}}
\end{figure}

Light curves binned by 5760\,s (the orbital time scale) are shown in Figure~\ref{fig-128lc}.  While little variability exists on rapid time scales sampled by the 128\,s light curves, the 5760\,s light curves do show variability for most of the sources.  The amplitude of this variability, however, is not very high, with changes in count rates spanning $< 0.1$\,cts\,s$^{-1}$ in the combined XIS0 + XIS1 + XIS3 light curves, corresponding to changes on the 60\% or lower scale from the lowest to highest count rates in individual light curves.  The reduced $\chi^2$ values range from 0.31 -- 2.77, with the exception of NGC 6921, whose high value (11.18) may be the result of poorer statistics from the higher ratio of background count rates to source count rates. Clearly, there is no strong variability (factors of 2 or higher) present on these timescales.  While low luminosity AGN, like our obscured target sources, are thought to be more variable than high luminosity AGN \citep{1986Natur.320..421B}, AGN are more variable on long time scales than short time scales  \citep{1986Natur.320..421B,1997ApJ...476...70N} and so the lack of short term variability is not surprising.

In addition to the full 0.1--12\,keV light curves, we constructed 5760\,s binned net light curves for three energy bands: 0.1--3\,keV (L), 3--7\,keV (M), and 7--12\,keV (H).  The average count rate and $\chi^2$ values for each observation are recorded in Table~\ref{tbl-varenergies}.  We chose these energy bands to separate the soft emission, the region including the Fe K-$\alpha$ emission features, and the hard band emission.  From our analysis, we find that variability in these regions is not the same for all of these ``hidden'' sources.  In paticular we find that the observations of NGC 1142 show more variability at the highest energies, while Mrk 417, MCG +04-48-002, and NGC 6921 show more variability in the soft emission.  Further, the region including the Fe K-$\alpha$ emission, which contains the most counts for all observation except that of NGC 6921, shows little to no variability on these timescales.

\subsubsection{$\sigma^2_{rms}$--$L$ Relation}
In order to test whether the anti-correlation between luminosity and $\sigma^2_{rms}$ \citep{1997ApJ...476...70N} is also seen in our sources, we needed to obtain the X-ray fluxes from spectral fits to the energy spectra.  To model the XIS spectra, we simultaneously fit the XIS front-illuminated and back-illuminated spectra with the same standard model used in 
\citet{2008ApJ...674..686W} (0.3--10\,keV spectra).  This model consists of a partially covered power law spectrum with an Fe K -$\alpha$ feature at 6.4\,keV (implemented in {\tt XSPEC} as {\tt pcfabs}*{\tt tbabs}({\tt pow}$+${\tt zgauss})\,{\tt const}).  The partial covering model is a multiplicative model in which the direct AGN emission is partially blocked by material in the line of sight, defined as: \begin{equation} M(E) = f \times e^{-N_{\mathrm H} \sigma(E)} + (1 - f). \end{equation} Here \nh \,is the absorbing column in the line of sight (in units of atoms\,cm$^{-2}$), $\sigma(E)$ is the photo-electric cross section, and $f$ is the covering fraction (ranging from 0 to 1).  A large covering fraction ($f \approx 1$) indicates that either much of the direct emission is blocked or that a very small fraction of the emission is scattered into our line of sight.  Galactic absorption from the Milky Way is accounted for with a second neutral absorber ({\tt tbabs}), fixed to the \citet{1990ARAA..28..215D} value (listed in Table 1 of \citet{2008ApJ...674..686W} along with $z$, Seyfert type, and host galaxy type).  The Galactic \nh~\ values were obtained using the web version of the \nh~\ {\tt ftool}.  Results of these fits are presented in Table~\ref{tbl-pcfabs}.  Due to the low signal-to-noise in the NGC 6921 spectrum, the error bars are not constrained and encompass the full range of values.  Therefore, we fixed the power law component to 1.75 (the average value of $\Gamma$ obtained for the BAT AGNs in \citet{2009ApJ...690.1322W}) and record the upper limit on column density and scattering fraction.

As Table~\ref{tbl-pcfabs} shows, the $\chi^2$ values from this model are not optimal.  The residuals to the model show features not accounted for by this simplified model, for instance there is a soft component evident in the spectra of NGC 1142.  To investigate these features, we present our detailed spectral analyses in \S~\ref{spectra}.  For our current study of variability, this simpler partial covering model is sufficient to determine luminosities and is further used to provide a direct comparison to the $\approx 10$\,ks {\it XMM-Newton} spectra (which do not have simultaneous spectra $> 10$\,keV as in the Suzaku spectra analyzed in \S~\ref{spectra}), investigated in the following subsection (\S~\ref{var-between}).

\begin{figure}
\includegraphics[width=8.25cm]{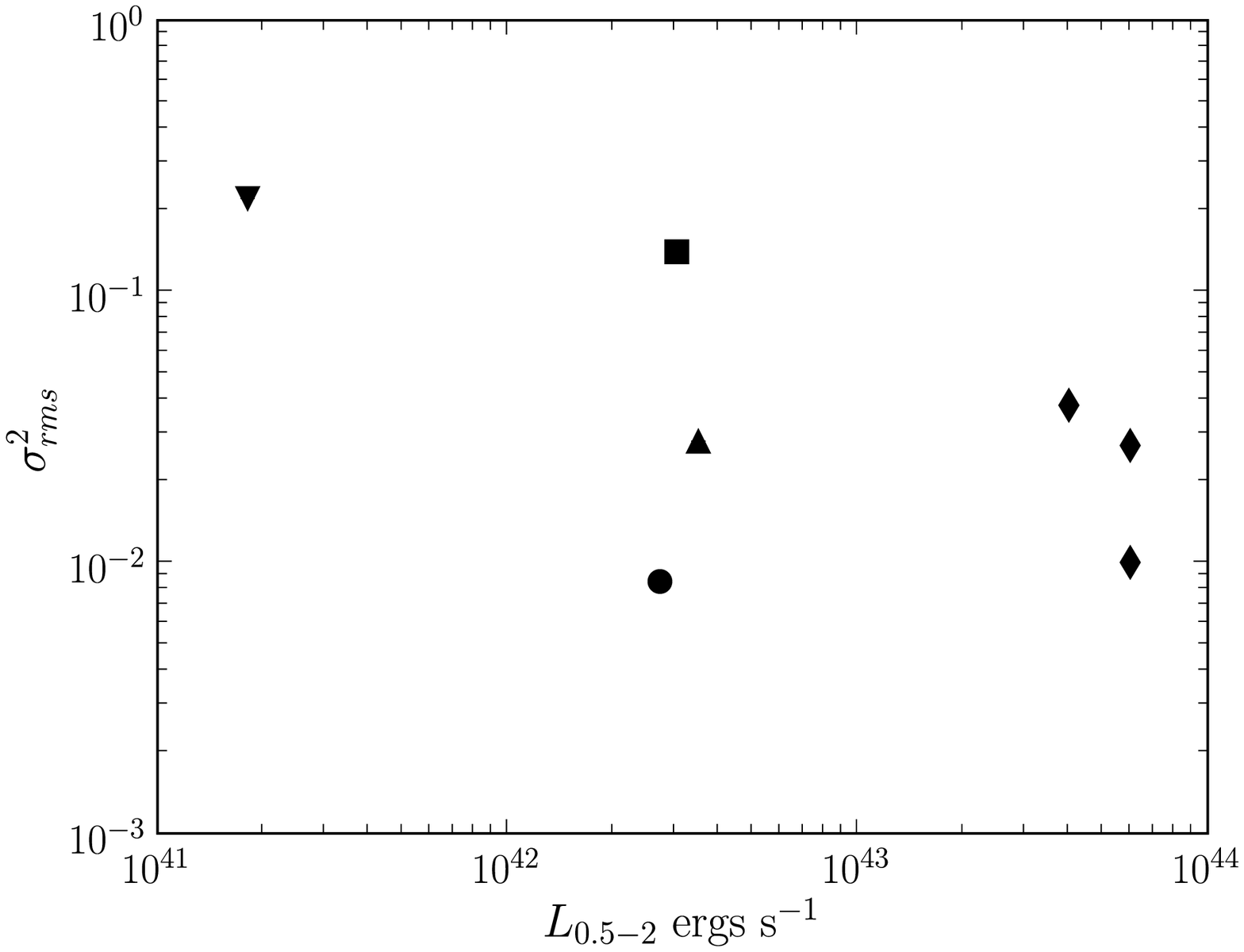}
\hspace{0.25cm}
\includegraphics[width=8.25cm]{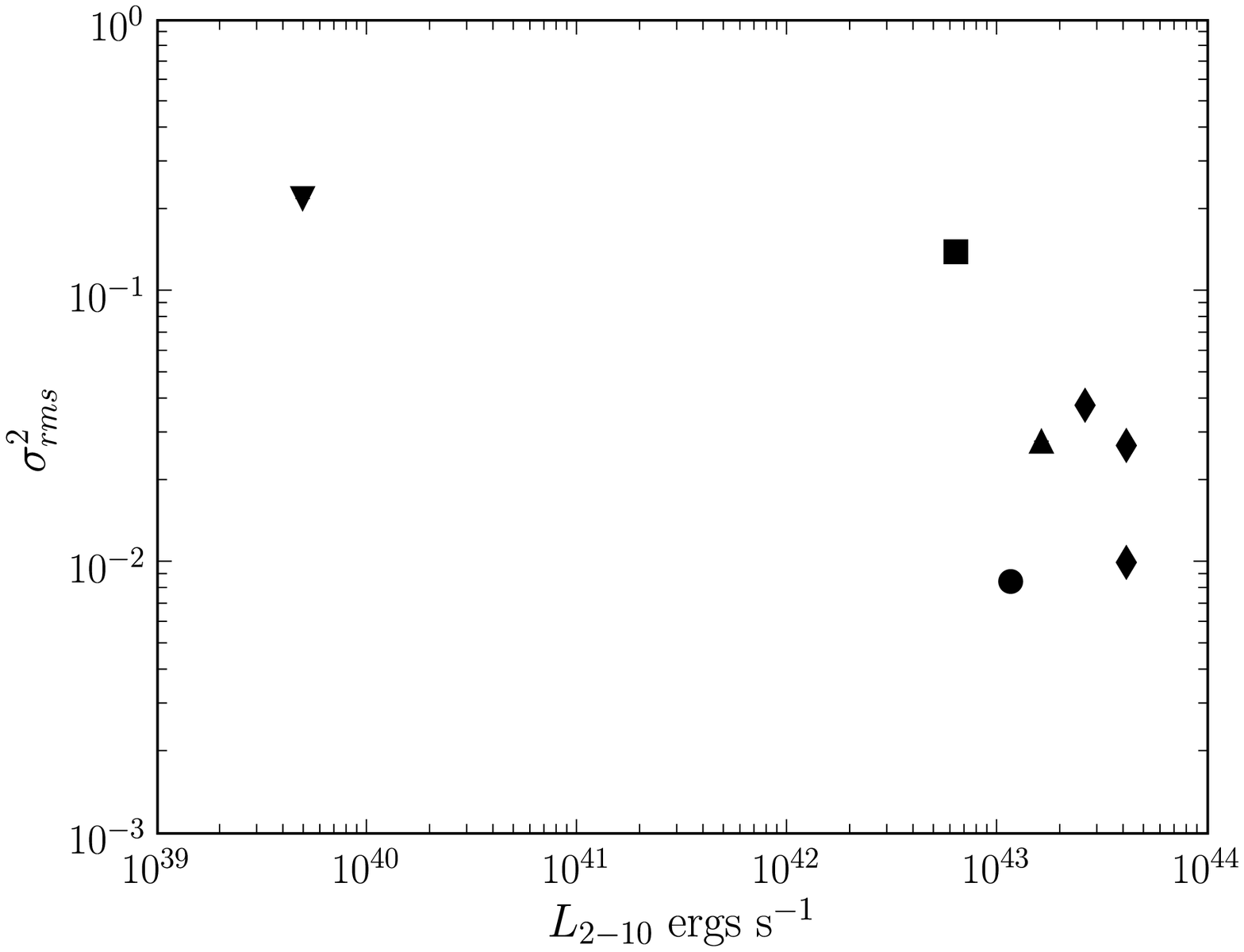}
\caption{Normalized excess variance, as a measure of variability, versus unabsorbed 0.5--2\,keV (left) and unabsorbed 2--10\,keV (right) luminosity.  The individual sources are symbolized as NGC 6921 (upside down triangle), MCG +04-48-002 (square), ESO 506-G027 (circle), Mrk 417 (triangle), and NGC 1142 (diamond).  There is no statistically significant relationship between $\sigma^2_{rms}$ and luminosity, however, the least luminous sources have the highest normalized excess variance measurements.
\label{fig-excess128}
}
\end{figure}

Computing 0.5--2\,keV and 2--10\,keV unabsorbed luminosities, using the redshift values recorded in Table~\ref{tbl-1}, we compare the luminosities with the excess variance measurements in Figure~\ref{fig-excess128}.  These plots show no correlation between normalized excess variance and luminosity, in either band.  However, the least luminous observations (NGC 6921 and MCG +04-48-002) do have the highest $\sigma^2_{rms}$ values, consistent with the anti-correlation found by \citet{1997ApJ...476...70N} and others.  


\subsection{Variability Between the {\it XMM-Newton} and Suzaku Observations}\label{var-between}
In \citet{2008ApJ...674..686W}, we found that 13/16 of the sources with SWIFT XRT observations in addition to the {\it XMM-Newton} follow-ups varied (in flux, power law index, or column density).  In that study, we had very few counts in the XRT spectra of heavily absorbed sources ($< 60$\,counts for sources with \nh$ > 10^{23}$\,cm$^{-2}$), making it difficult to compare spectra for the most absorbed sources.  However, we are now able to determine whether our five AGN target sources are variable between the {\it XMM-Newton} and Suzaku observations.  

\begin{figure}
\begin{center}
\includegraphics[width=7cm]{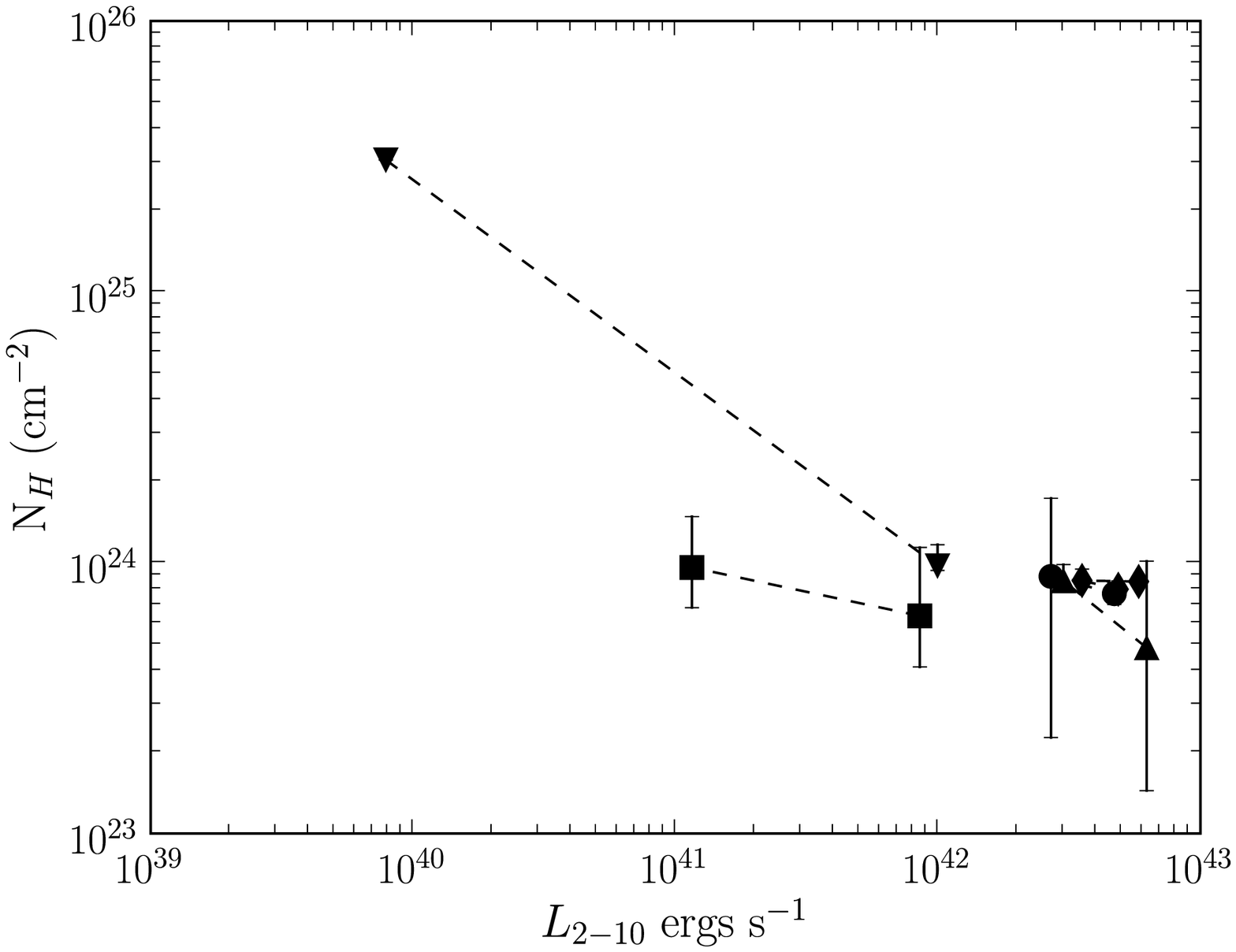}
\includegraphics[width=7cm]{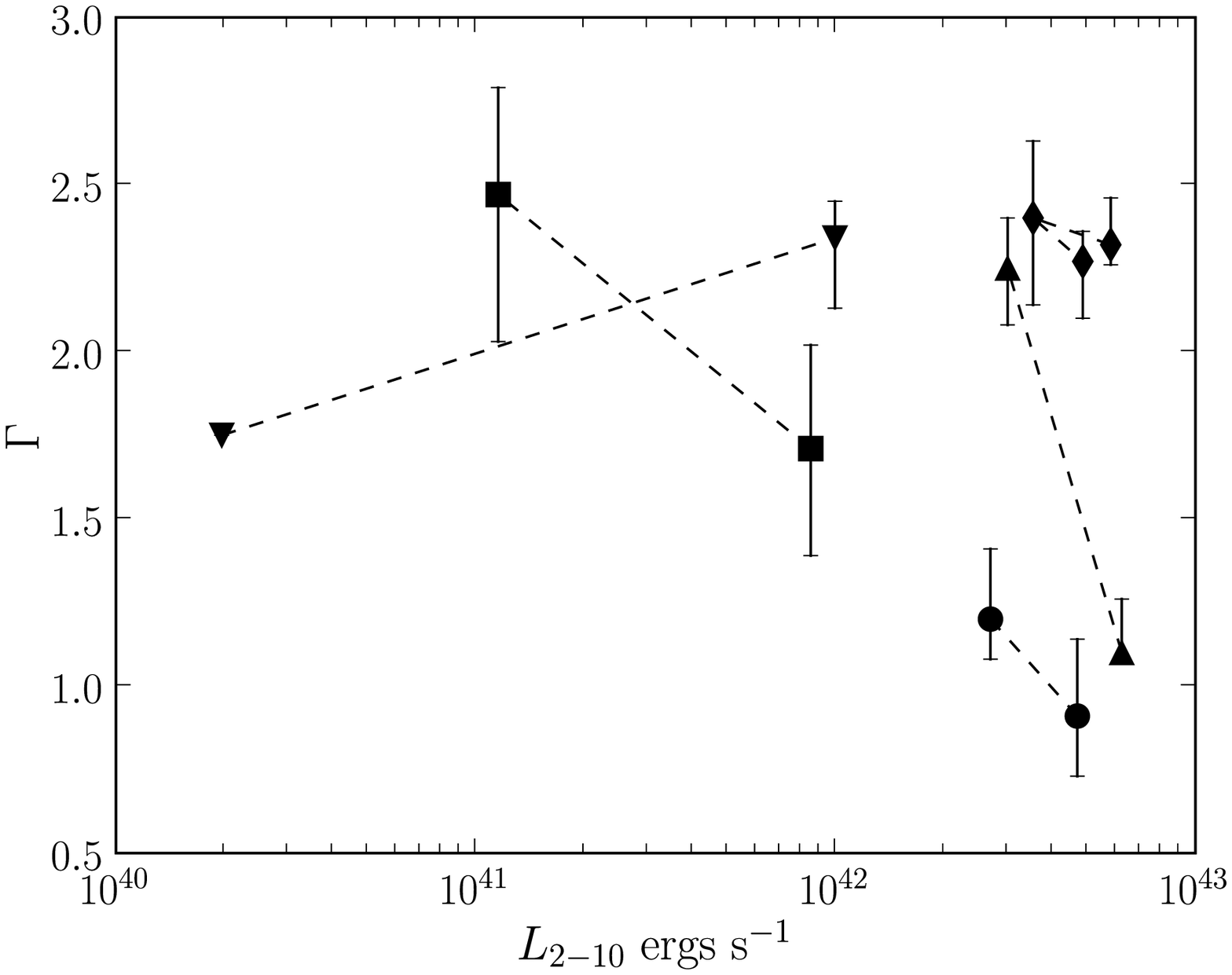} 
\includegraphics[width=7cm]{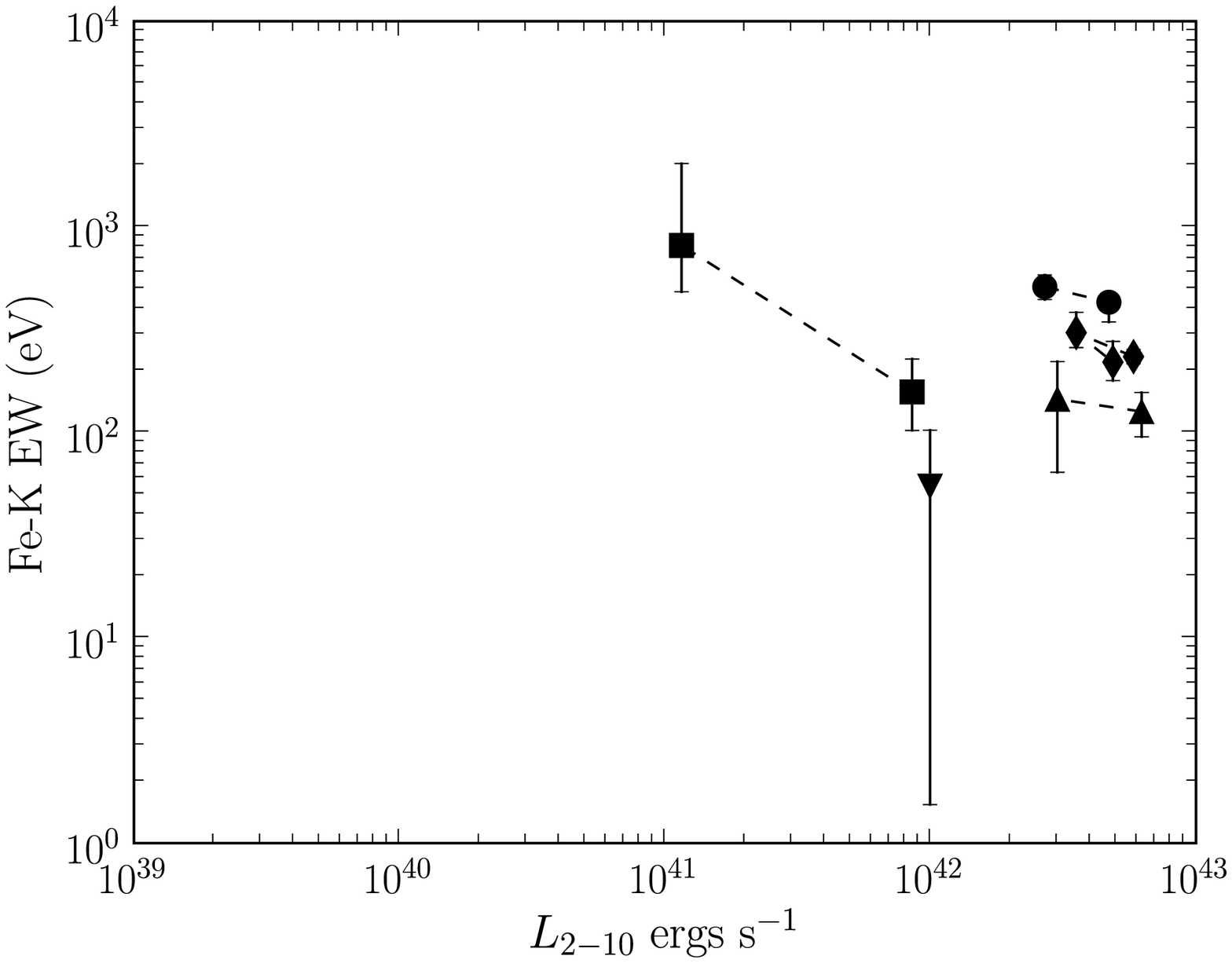}
\end{center}
\caption{Comparison of the {\it XMM-Newton} \citep{2008ApJ...674..686W} and Suzaku spectral fits (\S~\ref{var-during}) in the 0.3--10\,keV band.  We plot \nh, $\Gamma$, and the Fe K-$\alpha$ EW at 6.4\,keV versus $L_{2-10}$.  The individual sources are symbolized as NGC 6921 (upside down triangle), MCG +04-48-002 (square), ESO 506-G027 (circle), Mrk 417 (triangle), and NGC 1142 (diamond).  Dashed lines are used to clearly distinguish observations of the same source.  From the plots, it appears that \nh, $\Gamma$, and Fe K EW are higher at lower luminosities for individual sources.  Note that the exception to this in the $\Gamma$--$L_{2-10}$ plot is NGC 6921, where $\Gamma$ could not be constrained and was fixed to 1.75 in the lower luminosity observation.  For this source, $N_{\rm H}$ is an upper limit and the Fe K$\alpha$ EW could not be determined for the Suzaku observation. 
\label{fig-comparespec}}
\end{figure}

To quantify differences in the spectra, we fit the 0.3--10\,keV XIS spectra with the same partial covering model used for the {\it XMM-Newton} observations in \citet{2008ApJ...674..686W} (\S~\ref{var-during}).  The power law index for NGC 6921 was difficult to constrain and so we again fixed this value to the average AGN photon index of 1.75 \citep{2009ApJ...690.1322W}.
In Figure~\ref{fig-comparespec}, we plot the column density derived from the partial covering model, spectral index ($\Gamma$), and 6.4\,keV equivalent width versus observed 2--10\,keV luminosity for all of the observations.  These plots reveal a number of results.  First, we find that there is a great change in 2--10\,keV luminosity for both NGC 6921 and MCG +04-48-002.  The luminosity of MCG +04-48-002 increases by one magnitude between the {\it XMM-Newton} and Suzaku observations ($\approx 1$ year apart), while NGC 6921's luminosity drops by two magnitudes.  None of the other three sources vary to such a degree, however, all show signs of variability in luminosity between observations.  Therefore, the heavily obscured sources are varying on time scales of $\approx 0.5$--$1.5$ years (the time between the Suzaku and {\it XMM-Newton} observations).

Second, it is clear that \nh, $\Gamma$, and Fe K-$\alpha$ EW are higher during the lower luminosity observation for individual sources.  An anti-correlation between Fe K EW and luminosity is known as the X-ray Baldwin/ ``Iwasawa-Taniguchi'' effect \citep{1993ApJ...413L..15I} and has been seen in a number of AGN samples, for instance in the radio quiet samples of \citet{2006ApJ...644..725J} and \citet{2007AA...467L..19B}.  Thus, we would expect to see this anti-correlation, even amid our heavily obscured sources.  A correlation has also been noted between $\Gamma$ and luminosity \citep{2004AA...422...85P, 2005AA...432...15P, 2006ApJ...646L..29S}.  In \citet{2009ApJ...690.1322W}, we did not find a correlation among the 9-month BAT AGN sample.  However, we noted that this correlation was seen for individual sources in \citet{2008ApJ...674..686W} and therefore suggests that sources do have higher spectral indices at higher luminosities.  It is unclear why we see the opposite effect in our obscured sources (i.e. an anti-correlation -- note that NGC 6921 is the exception, however, $\Gamma$ was fixed to the average value of 1.75 in the lower luminosity observation).  It is possible that this is the result, in part, of unconstrained error bars.  In particular, this could be the case for MCG +04-48-002, whose spectral parameters were difficult to constrain in the lowest luminosity observations.  This may also be the cause of the anti-correlation seen between \nh and luminosity, which is  dominated by NGC 6921 and MCG +04-48-002.  For NGC 6921, however, the column density is poorly constrained.  Further, a value of \nh$ = 3 \times 10^{25}$\,cm$^{-2}$ is beyond the limitations of the {\tt tbabs} model, since it does not treat multiple scatterings, important for \nh$ >> 1 \times 10^{24}$\,cm$^{-2}$.  Another possible cause of the observed anti-correlation between $\Gamma$ and luminosity could be the lack of our inclusion of a reflection model.  It is possible that reflection plays an important role for some of these sources, because of this, we will discuss models including reflection in our detailed fits in \S~\ref{spectra}.

\begin{figure}
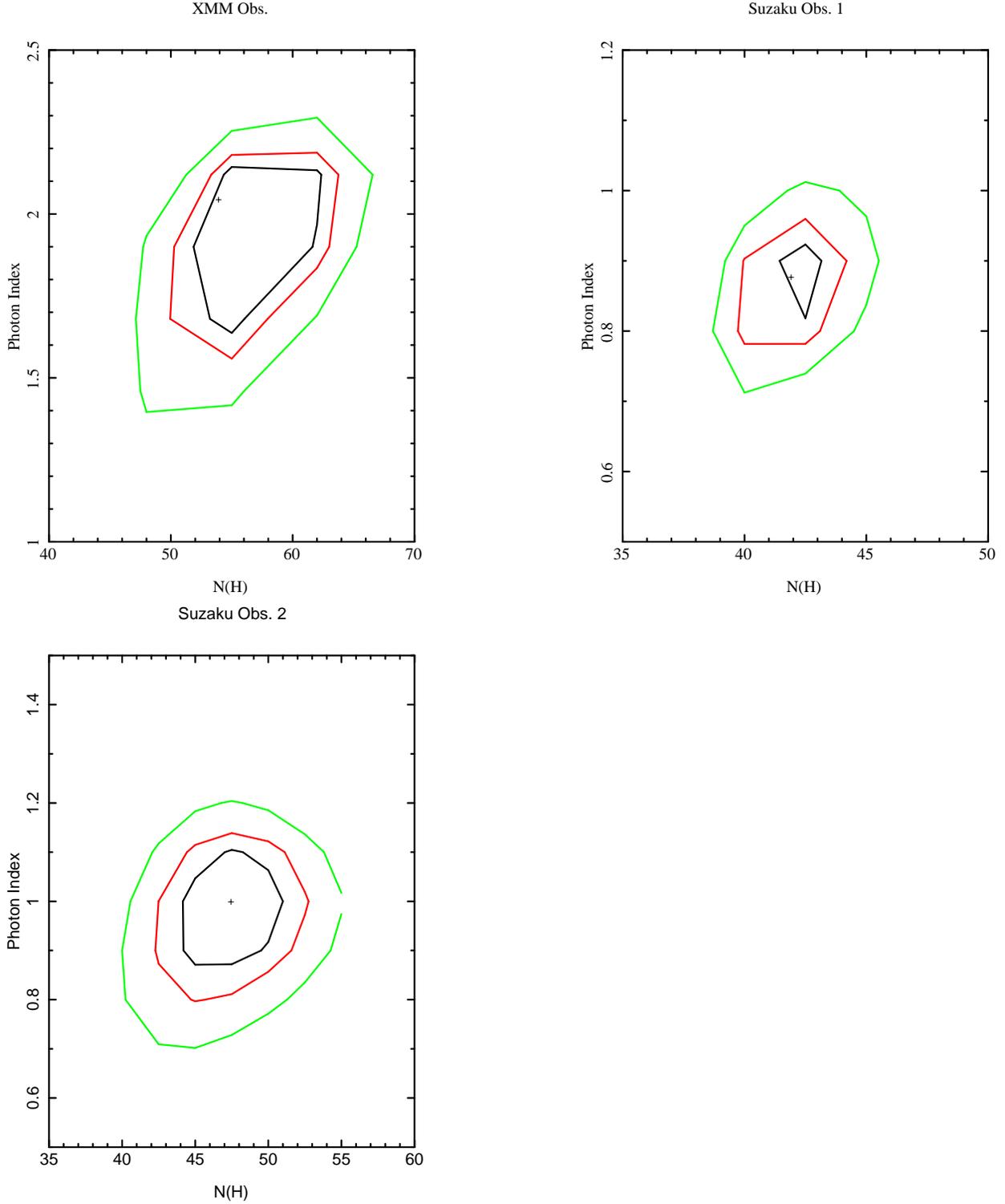

\includegraphics[width=7cm]{f4a.ps}
\includegraphics[width=7cm]{f4b.ps}
\includegraphics[width=7cm]{f4c.ps}
\caption{Shown are the error contours for each of the NGC 1142 spectra simultaneously fit, including the two Suzaku observations and the {\it XMM-Newton} observation.  The errors were computed for the photon index ($\Gamma$) and column density (\nh in units of $10^{22}$\,cm$^{-2}$).  The contour levels shown represent 99\%, 90\%, and 68\% confidence levels. 
\label{fig-contours}}
\end{figure}

To further test variability between the {\it XMM-Newton} and Suzaku observations, we simultaneously fit the {\it XMM-Newton} pn and Suzaku front-illuminated spectra.  Initially, we fixed all the parameters to the same values.  Next, we allowed the model parameters (\nh\, and covering fraction from {\tt pcfabs}, $\Gamma$ and its normalization, and a constant value to account for flux differences ({\tt const} in {\tt xspec})) to vary, recording the change in $\chi^2$.  For all of the observations, we used the partial covered power law model presented in \S~\ref{var-during}.  However, for NGC 1142 we also added a model for collisionally-ionized diffuse gas ({\tt apec} with $kT = 0.63$\,keV and a normalization of 0.0021) and an Fe edge at $E = 7.07$\,keV with $\tau = 0.40$.  Each of these additional models significantly improved the fit ($\Delta\chi^2 = 178$ for {\tt apec} and $129$ for {\tt zedge}) to the NGC 1142 spectra.  In Figure~\ref{fig-contours}, we plot the error contours for the 99\%, 90\%, and 68\% confidence levels between $\Gamma$ and \nh \,for the simultaneous Suzaku (2) and {\it XMM-Newton} fits to the NGC 1142 spectra.  As the figure shows, both \nh\, and $\Gamma$ are constrained to a fairly narrow parameter space.  Also, as we noted in \citet{2008ApJ...674..686W}, the simple model of a partially covered power law yields flat power law indices ($\Gamma \approx 1$) which are not necessarily physical.  Spectral fits with more complex reflection models were not well-constrained with the {\it XMM-Newton} spectra in the 0.3--10\,keV and thus the motivation for obtaining these Suzaku observations was to extend the energy range for spectral fits so that we could test more complex and possibly more physical models (i.e. reflection).  We present the results of these fits, which constrain the power law indices to steeper, more physical values, in \S~\ref{spectra}.

\begin{figure}
\begin{center}
\includegraphics[width=8cm]{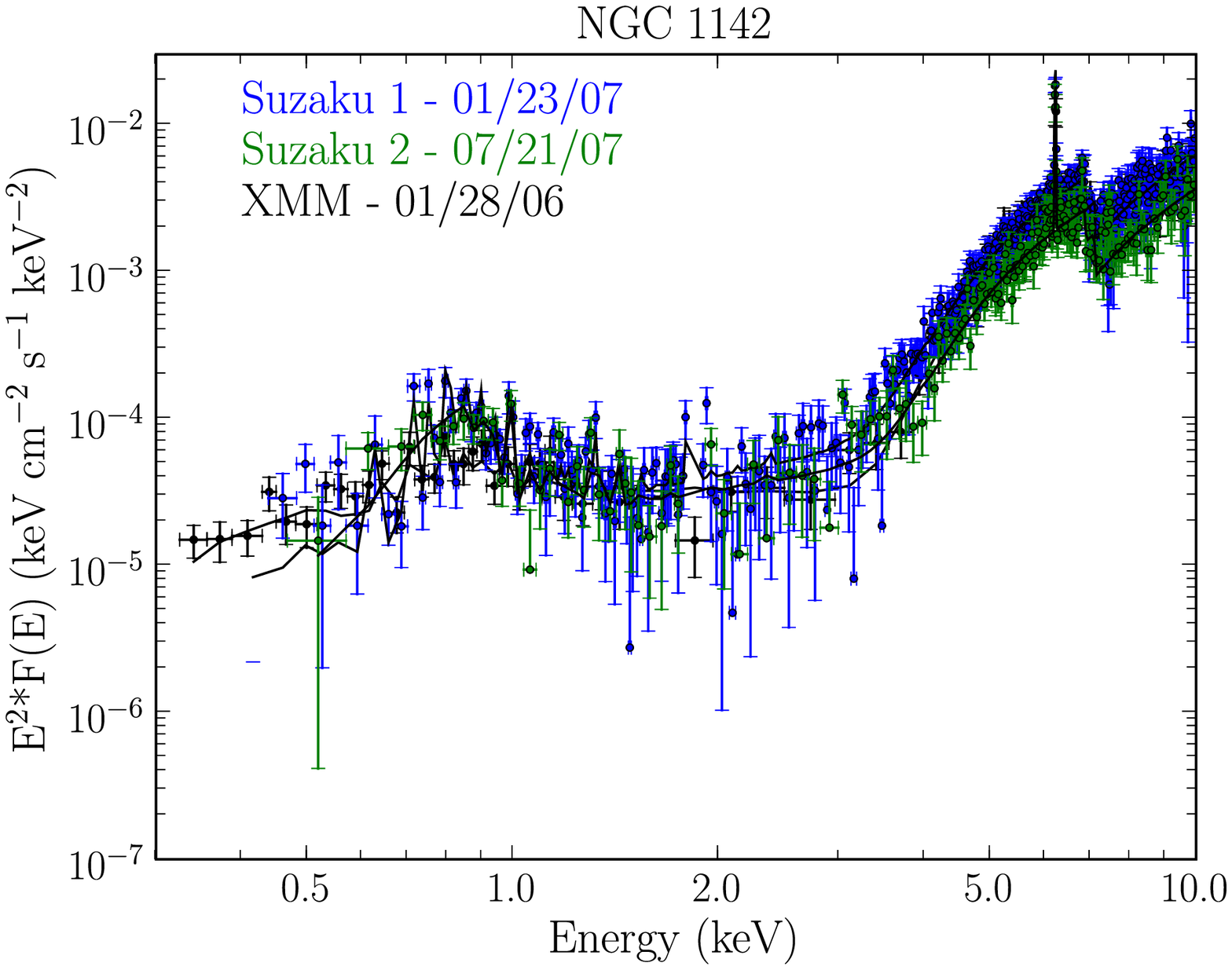}
\hspace{0.25cm}
\includegraphics[width=8cm]{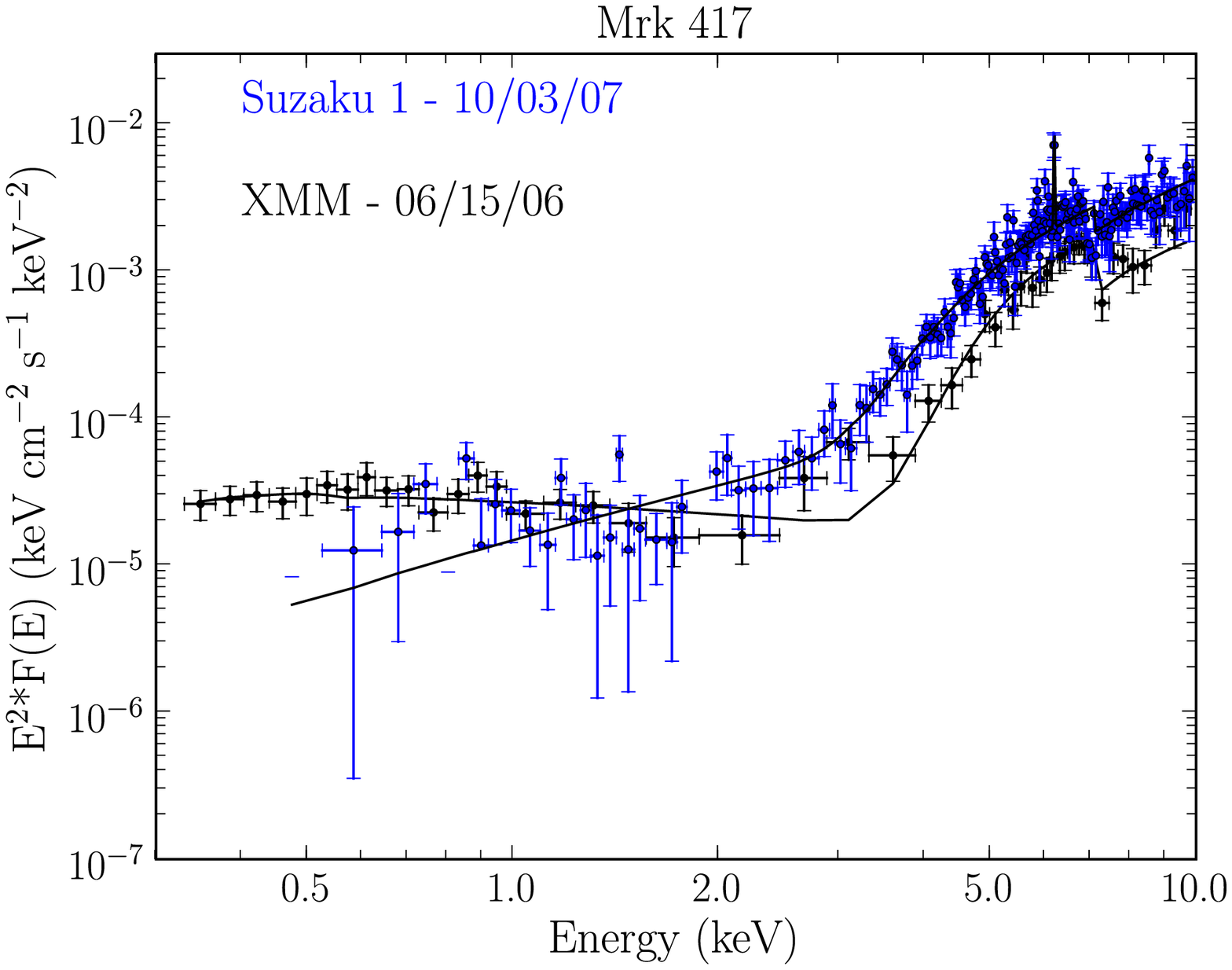}
\vspace{0.5cm}
\includegraphics[width=8cm]{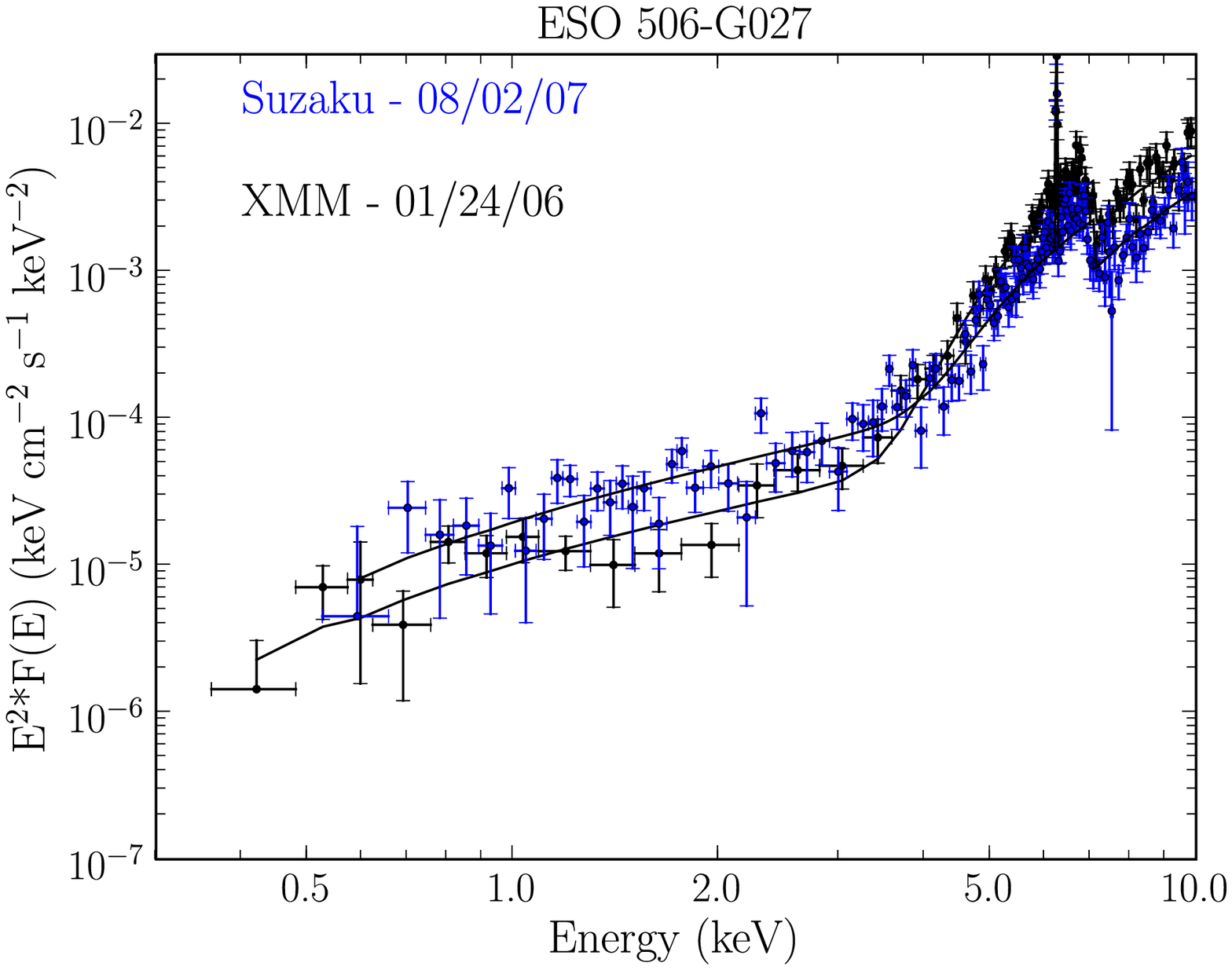}
\hspace{0.25cm}
\includegraphics[width=8cm]{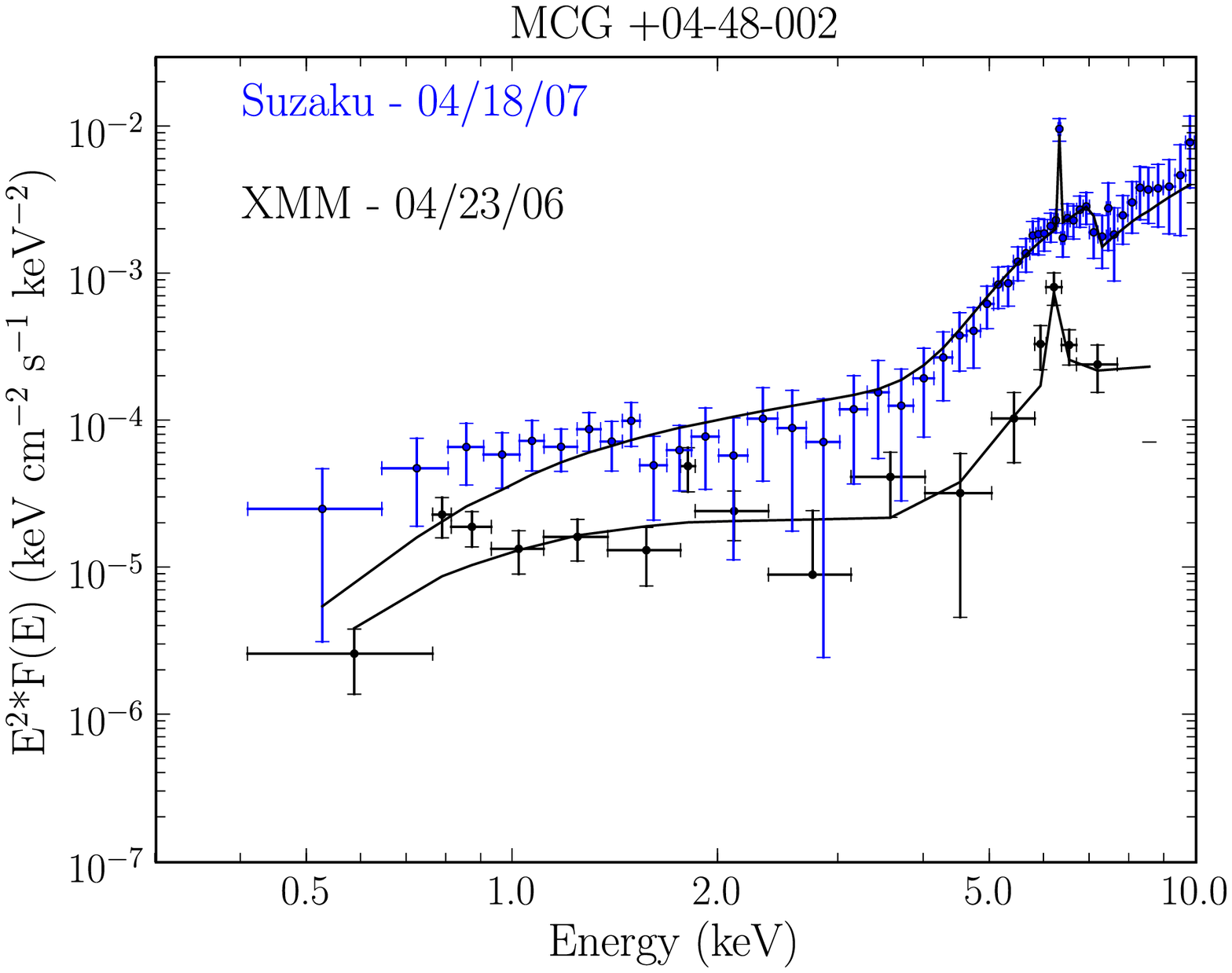}
\vspace{0.5cm}
\includegraphics[width=8cm]{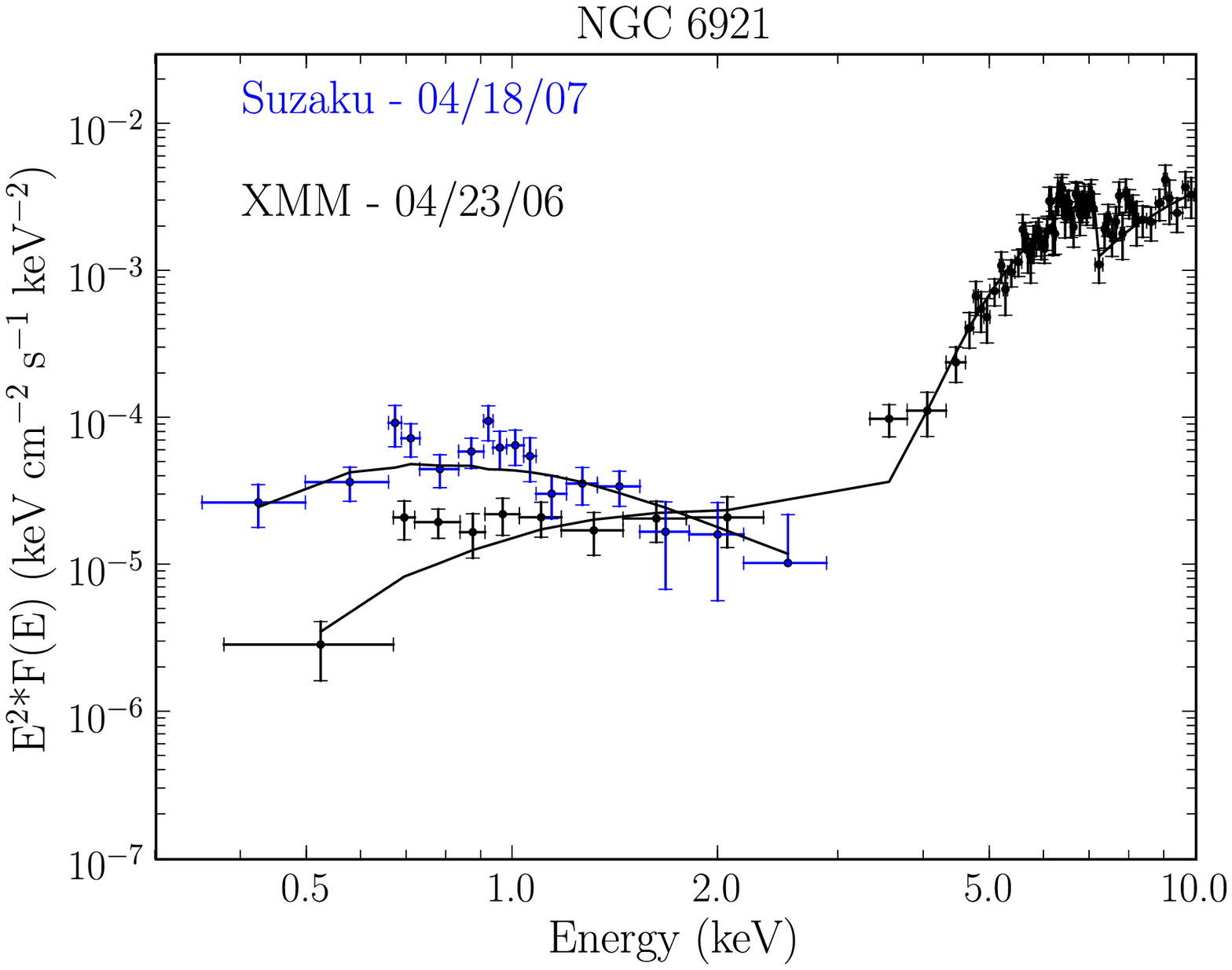} 
\end{center}
\caption{Unfolded $E^2$ spectra showing the {\it XMM-Newton} and Suzaku XIS 0-3 simultaneous fits for our target sources, rebinned by signal-to-noise for illustrative purposes.  NGC 6921 underwent the most change between observations (1 year).  The source is 2 orders of magnitude less luminous in the Suzaku observation and more highly absorbed.  Where the error bars are very large (i.e. extending beyond the plotted range), a horizontal line is used to represent the location of the Suzaku data point.  Similarly, error bars extending below the plotted range in the y-axis are excluded for clarity.\label{fig-simultaneous}}
\end{figure}

Results of the simultaneous fits are presented in Table~\ref{tbl-simultaneous}.  Accompanying plots of the unfolded energy spectra are presented in Figure~\ref{fig-simultaneous}.  In the table, in addition to including the $\Delta\chi^2$ values for allowing the model parameters to vary for the different observations, we include a measurement of the observed flux variability.  The statistic used is $(F_{max} - F_{min})/F_{avg}$, which was computed for both the soft (0.5--2\,keV) and hard (2--10\,keV) band.  These measurements show that the most variation is seen in MCG +04-48-002 and NGC 6921, in both hard and soft bands, and in the hard band for Mrk 417.  The variability in MCG +04-48-002 is purely in the observed flux -- the shape does not change significantly, as indicated by small $\Delta\chi^2$ for changes in absorption, the apparent $\Gamma$ from the partial covering model, and Fe K EW.  However, both the shape and flux of NGC 6921 change drastically, with a very significant $\Delta\chi^2$ in \nh (99) and apparent $\Gamma$ (47).  Finally, the hard flux and apparent $\Gamma$ ($\Delta\chi^2 = 249$) change appreciably in Mrk 417.  Here, the flatter $\Gamma$ ($\approx 1.0$) corresponds to the higher flux observation (the Suzaku observation).  From Figure~\ref{fig-simultaneous}, it appears that the difference in shape is most notable below 2\,keV.

\begin{figure}
\centering
\plotone{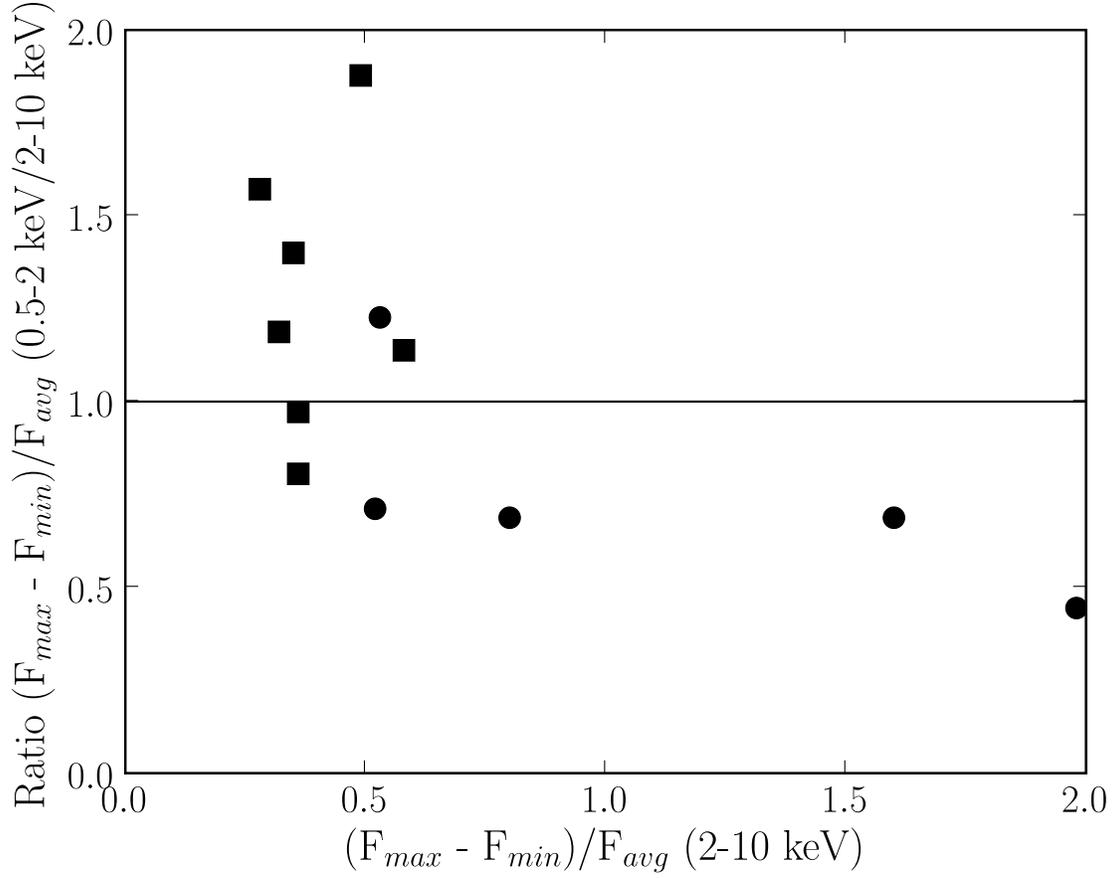}
\caption{The ratio of $(F_{max} - F_{min})/F_{avg} $ in the 0.5-2\,keV band/2-10\,keV band versus the value in the 2-10\,keV band for our obscured targets (circle) and the unobscured sources from \citet{2008ApJ...674..686W} (square).  The line represents values where the soft flux variability and hard flux variability measurements are the same.  This figure shows that there is more hard band variability in the obscured sources.
\label{fig-fmax}
}
\end{figure}

In Figure~\ref{fig-fmax}, we plot the ratio of our statistic $(F_{max} - F_{min})/F_{avg}$ for the 0.5-2\,keV/2-10\,keV bands versus the measurement in the hard band (2-10\,keV).  The values computed for our obscured target sources (circles) are compared to the unobscured sources (square) from \citet{2008ApJ...674..686W}.  From this plot, we find that the ratio of soft to hard variability tends to be lower for obscured sources.  Among our sources, only ESO 506-G027 shows more variability below 2\,keV than above.  One explanation for the difference in the ratio of soft band to hard band variability between obscured and unobscured sources is that the soft band light for the obscured sources is not primarily direct AGN emission, in agreement with \citet{2006AA...448..499B}.  Additional sources of soft X-ray light (e.g. binaries, ionized gas) can dilute the AGN signatures at low energies and thus reduce the amplitude of variability.

In \citet{2009ApJ...690.1322W}, we discussed the fact that emission from X-ray binaries, star formation, or hot ionized gas could cause the soft emission measured for the complex AGN.  The 0.5--2\,keV luminosity of all of our sources lies below $2 \times 10^{41}$\,ergs\,s$^{-1}$, within the range of emission expected from star formation  \citep{2003AA...399...39R}.  To test this further, we added an {\tt apec} model to the simultaneous spectral fits discussed.  The {\tt apec} model is a model for collisionally ionized emission which can also mimic the emission from photoionized gas \citep{2007MNRAS.374.1290G}.  Therefore, the {\tt apec} model can be used to model soft emission that is non-AGN (starburst), AGN (photo-ionized gas), or both.
We found that this model can account for the soft emission in all of the sources except ESO 506-G027, for which the reduced $\chi^2$ value becomes unacceptable at 2.3.  Since the {\tt apec} model well describes the spectra of the remaining sources and the soft emission is not strongly variable, this suggests that for all but ESO 506-G027 the soft emission is dominated by non-AGN emission (i.e. star formation or hot gas), AGN emission (photo-ionized gas), or both.  Photo-ionized emission is extended (e.g., \citet{2001AA...365L.168S,2006AA...448..499B}) and expected to show no variability on time scales of years or less.  Since the observations of NGC 1142 showed no variability in the soft emission (from the 5760\,s binned light curves), it is plausible that the soft emission in this source is from photo-ionized gas.

Another result we find is that the amount of hard variability is lower in the unobscured sources ($(F_{max} - F_{min})/F_{avg} \la 0.5$) than the obscured target sources (see Figure~\ref{fig-fmax}).  While our results on Sy 1s \citep{2008ApJ...674..686W} agreed with those of \citet{1997ApJ...476...70N}, finding more soft band ($< 2$\,keV) variability, the obscured sources are different.  As we have shown, the source of this variability is both from change in flux (for instance in MCG +04-48-002) and shape (i.e., changing \nh\, and/or apparent $\Gamma$).  Since the obscured sources in the BAT sample tend to have lower luminosities than the unobscured sources \citep{2009ApJ...690.1322W}, it is likely that the greater 2--10\,keV variability is the result of the sources having a lower luminosity.  Therefore, this result is consistent with \citet{1986Natur.320..421B} who first showed an anti-correlation between variability amplitude and X-ray luminosity.  This result is also consistent  with the results of \citet{2007arXiv0709.2230B} who found the 14--195\,keV BAT light curves of bright obscured AGN to vary more than unobscured bright AGN.  

\subsection{Variability Above 10\,keV} 
\subsubsection{Swift BAT light curves}
\begin{figure}
\begin{center}
\includegraphics[width=7cm]{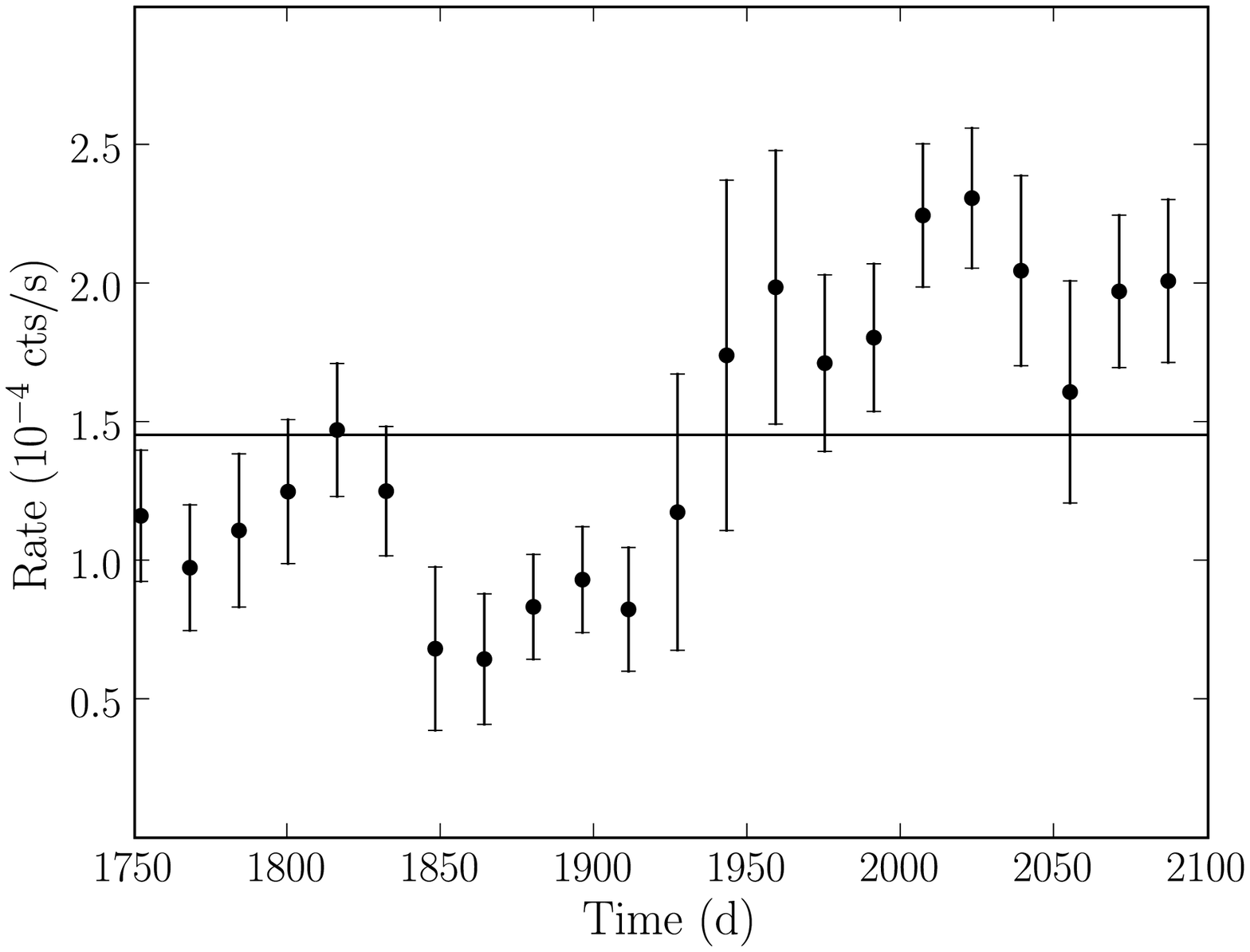} 
\hspace{0.25cm}
\includegraphics[width=7cm]{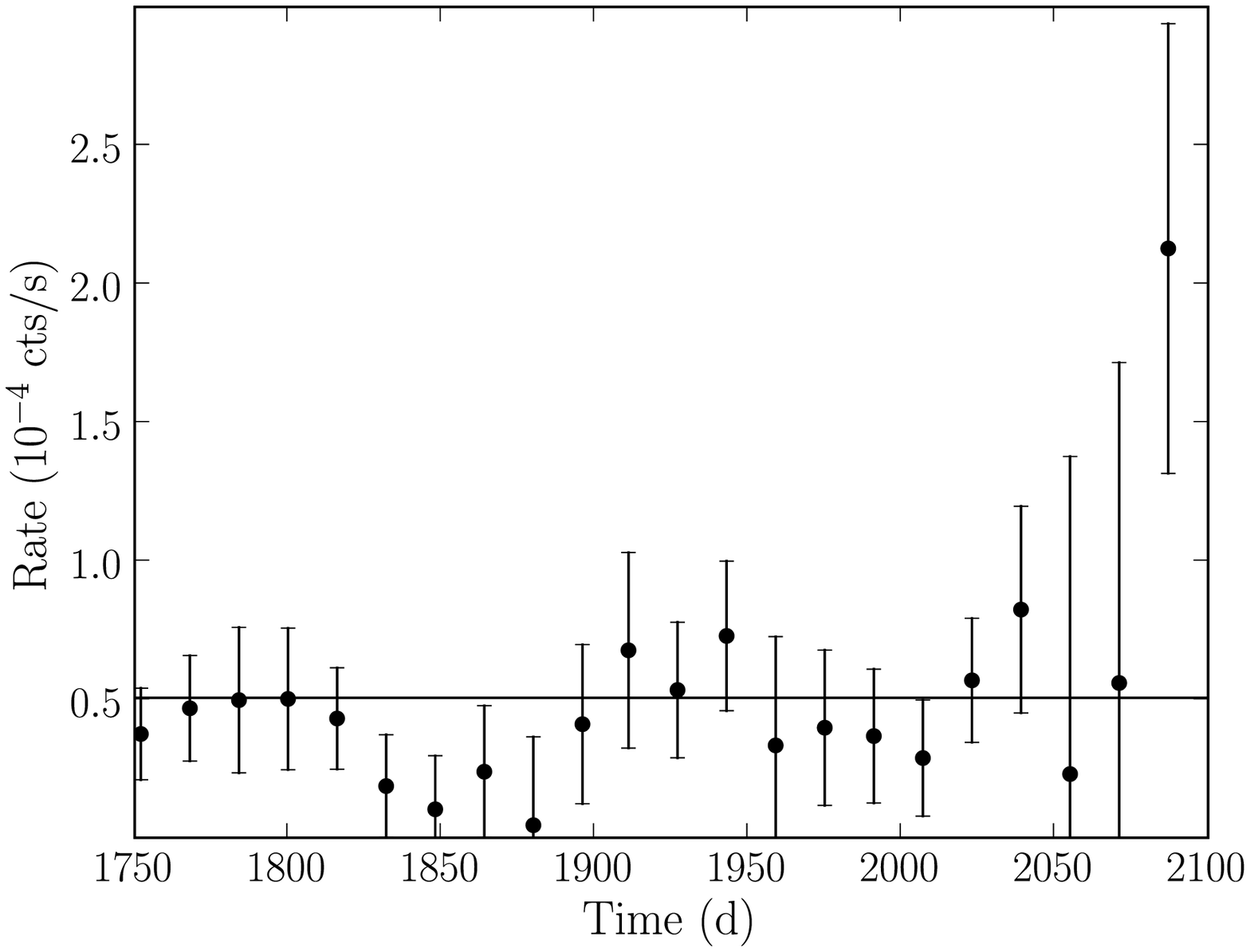} \\

\vspace{1.0cm}
 \includegraphics[width=7cm]{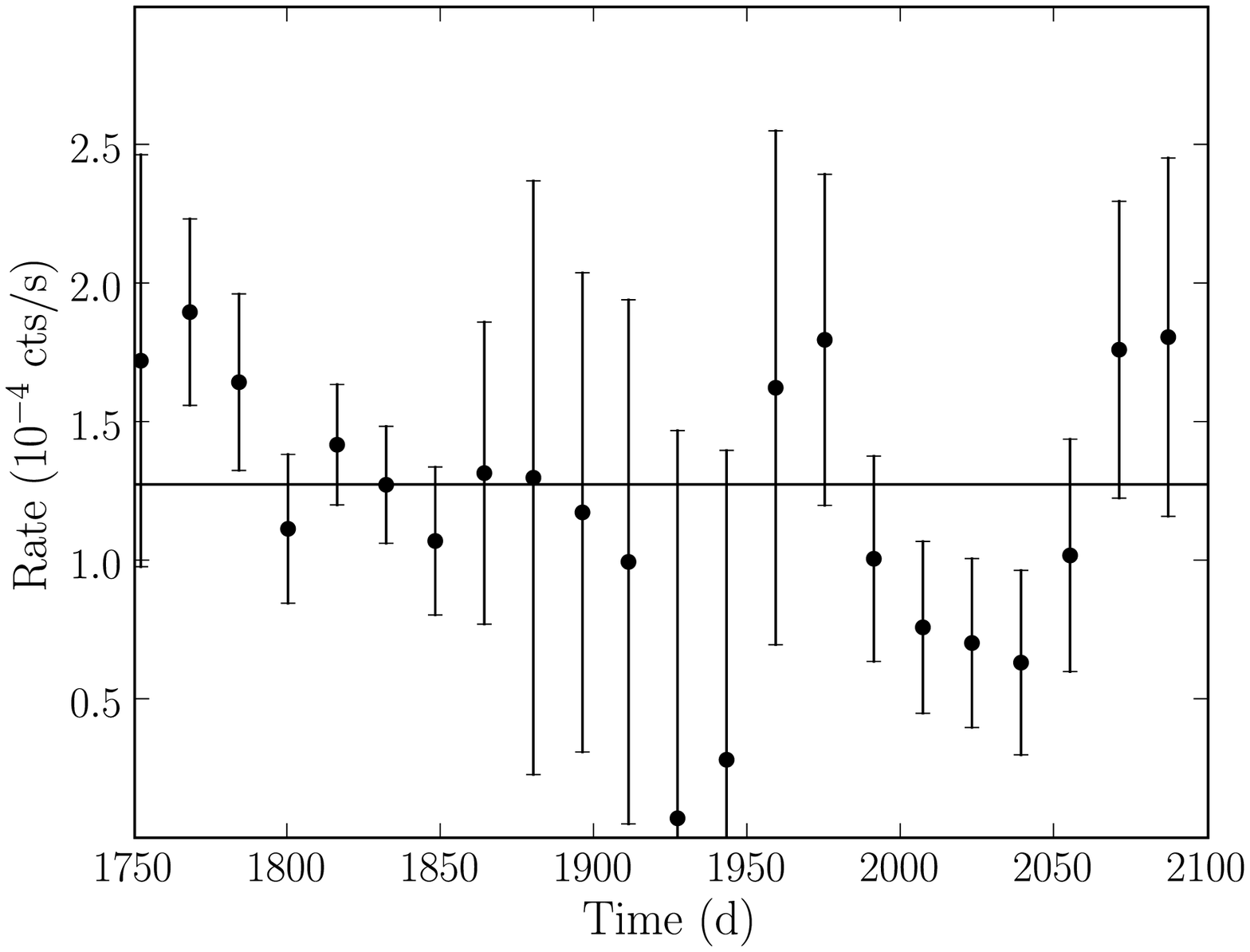}
 \hspace{0.25cm}
 \includegraphics[width=7cm]{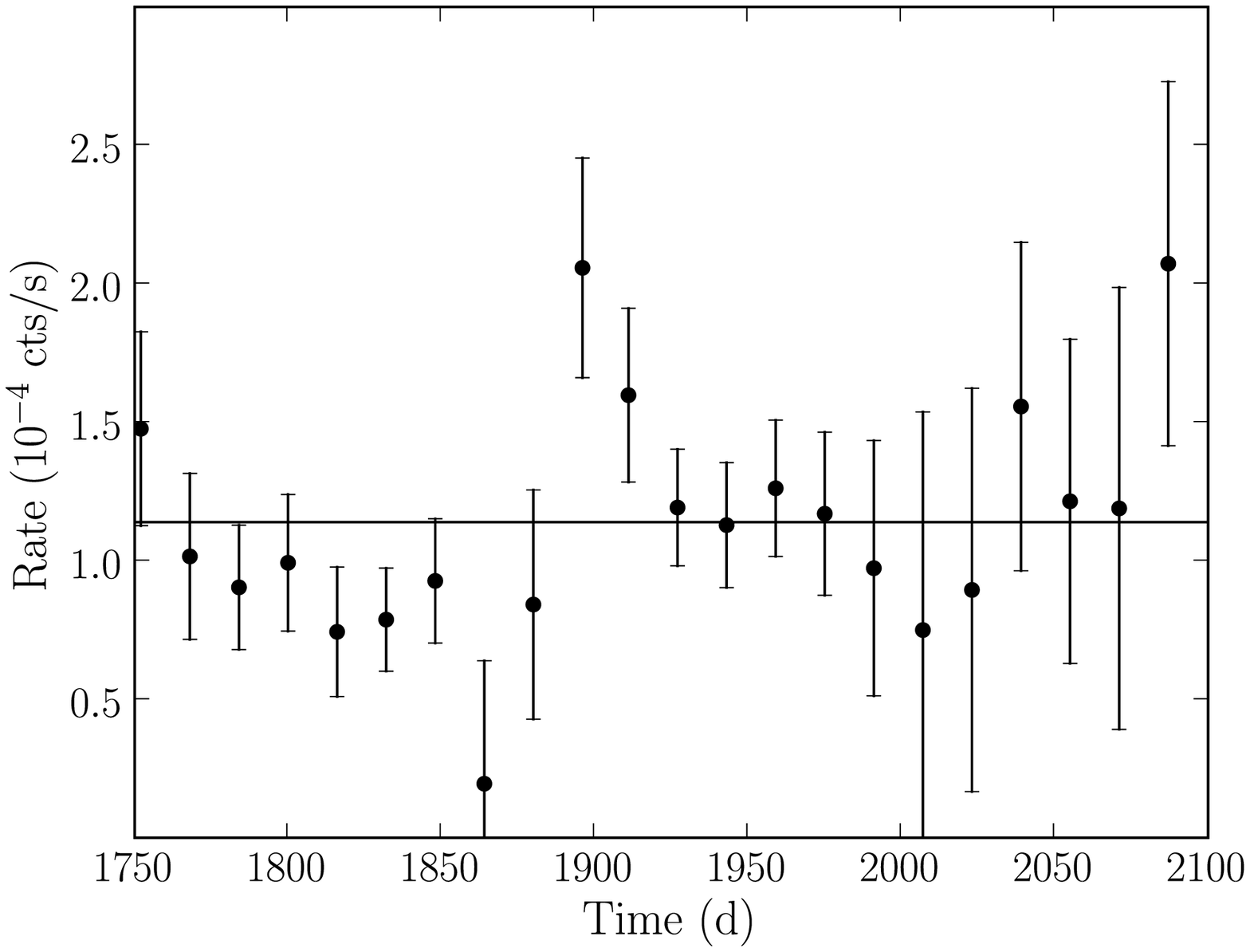} \\
\vspace{1.0cm}
\end{center}
\caption{The Swift BAT light curves for our targets spanning 400\,days and binned by 16\,days.
The light curves represent NGC 1142 (top left), Mrk 417 (top right), ESO 506-G027 (bottom left), and SWIFT J2028.5+2543 (which includes both NGC 6921 and MCG +04-48-002) (bottom right).  All of these sources show variability during the BAT observations.  The line in each plot represents the average count rate for that source.
\label{fig-batlc}}

\end{figure}

To determine whether our sources vary above 10\,keV, we first examined the publicly available 400\,day Swift BAT light curves \citep{2008ATel.1429....1B}.  We chose to bin the light curves by 16 days, roughly half a month, to investigate variability on this timescale, which corresponds to the shortest time scale for which all of the sources are well detected.  Binning the light curves by 16\,days, we computed the average count rate, excess variance, and $\chi^2$ value for our target sources in addition to the unobscured sources from \citet{2008ApJ...674..686W}, for comparison.  We also include the 14--195\,keV luminosity for each source, from \citet{2008ApJ...681..113T}.  Figure~\ref{fig-batlc} shows the 16\,d binned light curves for the 4 Swift BAT AGN sources.  Results of our analysis are recorded in Table~\ref{tbl-batlc}.   While the reduced $\chi^2$ values of 0.99, 1.12, and 0.84 indicate little variability in the 14--195\,keV band for ESO 506-G027, MCG +04-48-002/NGC 6921, and Mrk 417 (NGC 1142 is much higher at 3.87, indicating significant variability), Figure~\ref{fig-batlc} clearly does indicate some variability in all of our target sources.  Similarly, the unobscured comparison sources show the same variability as the obscured sources (indicated by similar $\sigma^2_{rms}$ and reduced $\chi^2$ measurements, as well as the light curves which are not shown).  

\begin{figure}
\plottwo{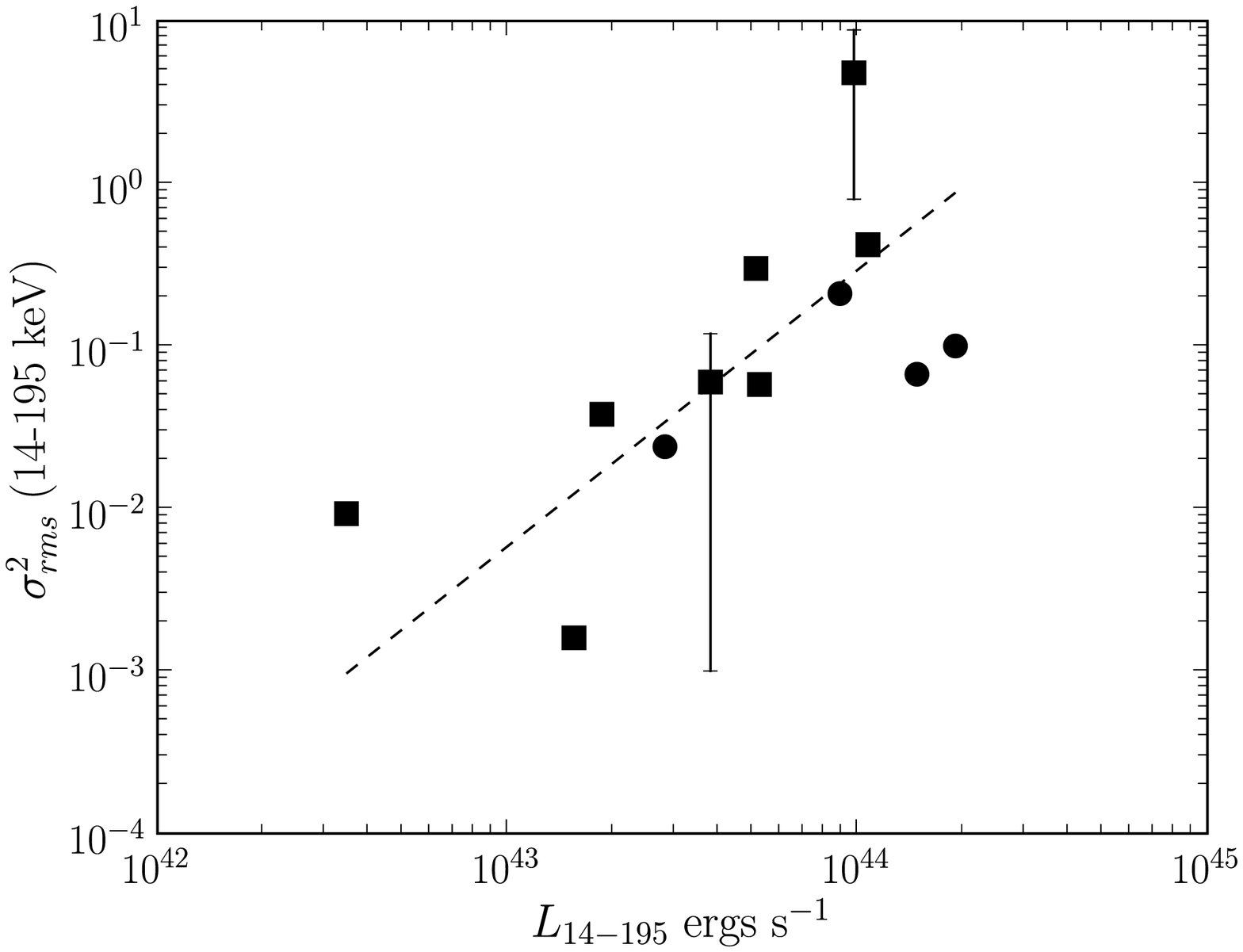}{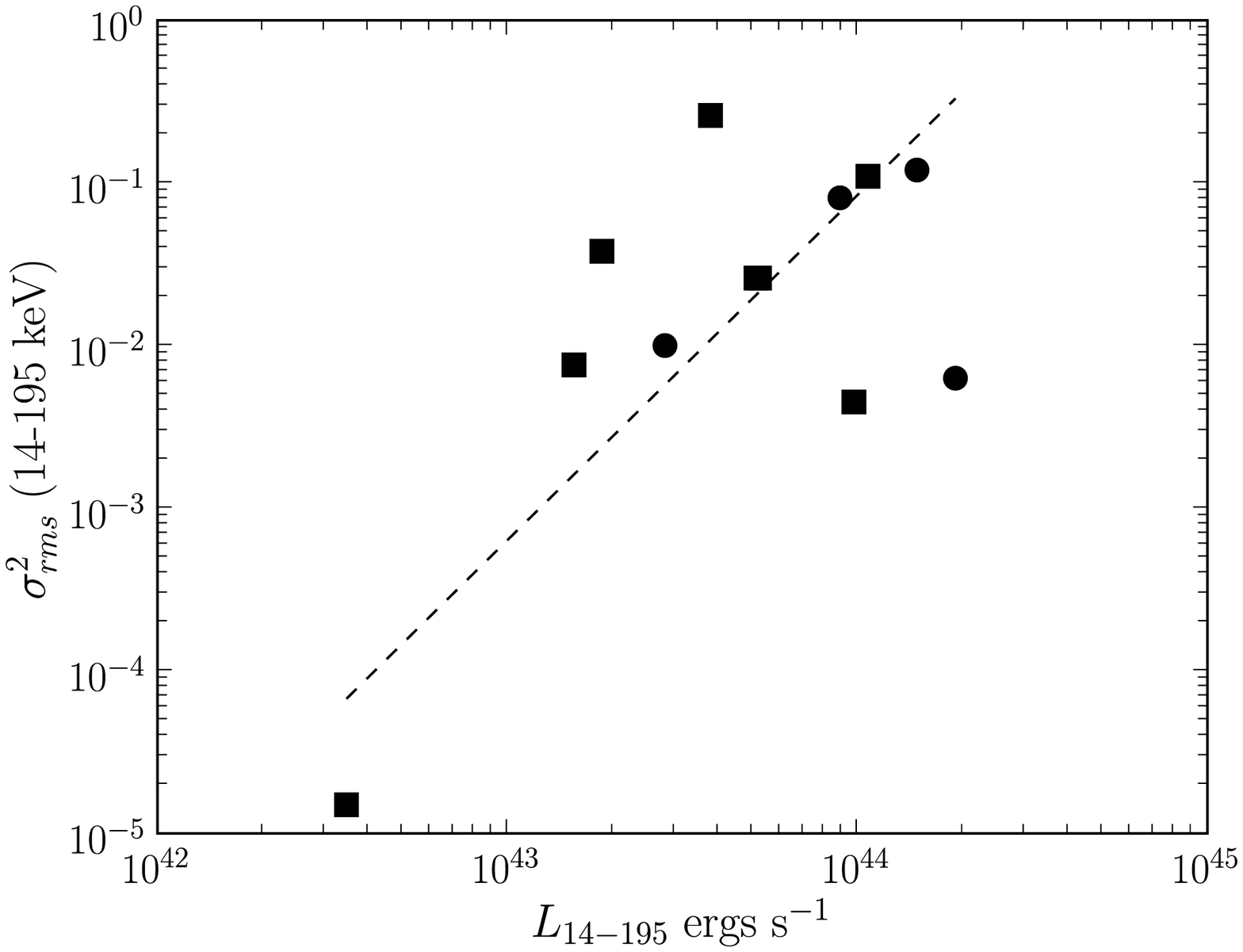}
\caption{Plotted are the excess variability measurements in the 16\,d (left) and 64\,d (right) binned BAT lightcurves versus 14--195\,keV luminosity.  We plot both the target obscured sources (circles) and the comparison unobscured sources from \citet{2008ApJ...674..686W} (squares).  There is no visible difference between the obscured and unobscured sources.  However, our data does show a correlation between excess variability and luminosity in this band of $\sigma^2_{rms} \propto L_{14-195}^{1.70\pm0.48}$ (16\,d) or $\sigma^2_{rms} \propto L_{14-195}^{2.12\pm0.45}$ (64\,d).  A Spearman rank correlation coefficient of 0.76 indicates that the correlation is significant in the 16\,d binned light curves, but not in the 64\,d binned light curves (where $r_s = 0.26$).
\label{fig-batexcess}}
\end{figure}

To illustrate the similarity in $\sigma^2_{rms}$ between the obscured and unobscured sources, we plot excess variance versus 14--195\,keV luminosity in Figure~\ref{fig-batexcess}.  Clearly, the obscured and unobscured sources occupy the same regions in both luminosity and excess variability.  Interestingly, there is a correlation between excess variability and luminosity for these sources (obscured and unobscured).  Fitting the data with an ordinary least squares bi-sector line, we find: \begin{equation} \log\sigma^2_{rms} = (1.70 \pm 0.48) \times \log L_{14-195} + (-75.4 \pm 20.8).
\end{equation}  Computing several statistics to determine the significance, we find a coefficient of determination $R^2 = 0.43$, a Spearman rank correlation coefficient $r_s = 0.76$, and Kendall rank tau correlation coefficient of $\tau = 0.58$.  This shows that the relationship of $\sigma^2_{rms} \propto L^{1.70 \pm 0.48}$ in the 14--195\,keV band is statistically significant.  Further, from an analysis of the BAT light curve of the Compton thick source Circinus, with $\chi^2/dof = 0.75/23$ for a constant source and an excess variance of $\sigma^2_{rms} = 2.8 \times 10^{-3}$, whose BAT luminosity is $\log L = 42.07$, we find that this heavily obscured source also follows this relation.  While the similar relationship between obscured and unobscured sources indicates that the same physical mechanism creates the emission in the 14--195\,keV band, it is unclear why the specific $\sigma^2_{rms} \propto L^{1.70 \pm 0.48}$ relationship exists, but more data points are necessary to increase the statistical significance of this relation.  To test whether this relationship holds with different binning sizes, we also computed the excess variance and $\chi^2$ values for 64\,day bins, the results of which are also included in Table~\ref{tbl-batlc}.  While a similar relationship exists between excess variance and 14--195\,keV luminosity, with $\sigma^2_{rms} \propto L_{14-195}^{2.12 \pm 0.45}$, the significance of this relationship is weak (with similar $R^2$ of 0.42 but $r_s = 0.26$ and Kendall rank $\tau = 0.26$).  Assuming that the 14--195\,keV is composed of both direct and reflected emission, one possible explanation for a possible correlation between luminosity and variability in the 14--195\,keV band is that the more luminous sources have a smaller contribution from reflection, which may be constant on these time scales.  However, since the correlation is weak, particularly in the 64\,day light curves, it is uncertain whether the correlation would be statistically significant in the entire BAT sample.

\subsubsection{Comparison of Suzaku pin and Swift BAT spectra}
One of our main goals in obtaining Suzaku observations of our targets was to obtain simultaneous spectra above and below 10\,keV.  This is especially important since time averaged 13-month BAT spectra have been created for the 9-month BAT AGN sources \citep{2008ApJ...681..113T}.  While these spectra exist, since they are time averaged we have no way of knowing how the spectrum changes with time in the 14--195\,keV band.  As our analysis of the BAT light curves indicates, the sources do vary in flux on the $\approx$ half a month time scales investigated.  To investigate how the shape and flux change for individual sources, it was imperative to obtain multiple Suzaku pin observations for direct comparison with the BAT spectra.  Towards this end, we obtained two Suzaku observations of NGC 1142 (from our AO-1 and AO-2 observations).

\begin{figure*}
\begin{center}
\plotone{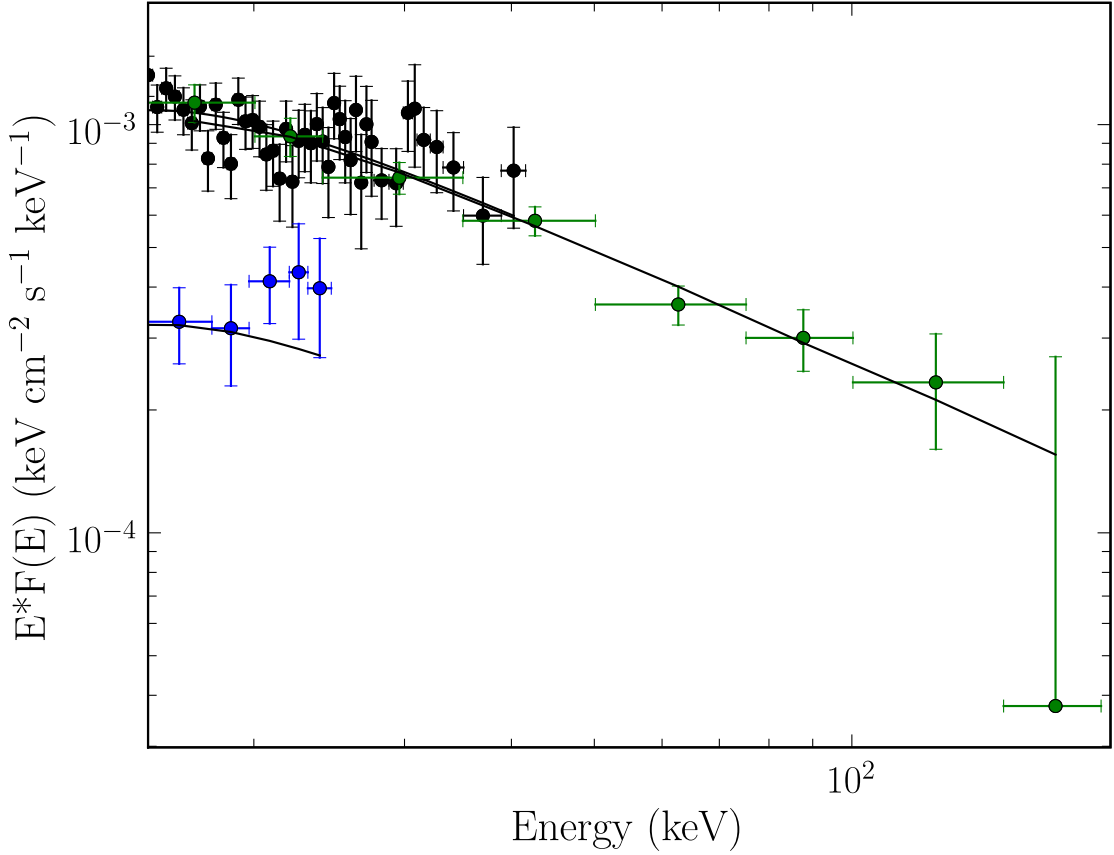} \\
\end{center}
\caption{Comparison of the Swift BAT spectrum (green, time-averaged over 13 months) with the two Suzaku pin observations of NGC 1142 (black = observation 1, blue = observation 2) in the 15--200\,keV band.  The power law component ($\Gamma = 1.53$) and column density (\nh$ = 1.0 \times 10^{23}$\,cm$^{-2}$) are the same for all of the observations, while the flux is allowed to vary by a constant factor.  From these unfolded spectra, it is clear that the flux is much lower in the second Suzaku observation while the first observation has a spectrum consistent with the time-averaged BAT spectrum.\label{fig3}}
\end{figure*}

 To test the variability or lack thereof in the $> 10$\,keV spectra of NGC 1142, we simultaneously fit the pin spectra and BAT spectrum with a simple absorbed power law model ({\tt tbabs}*{\tt pow}).  With no variation between the spectra, this model yields a poor fit with  \nh$ = 1.52 \times 10^{24}$\,cm$^{-2}$, $\Gamma = 1.72$, and $\chi^2$/dof$ = 230$/61.  The residuals from the fit show that while the first Suzaku spectrum and the BAT spectrum are not badly fit, the second Suzaku observation is.  The fit can be improved by allowing the column densities to vary ($\chi^2 = 76$/59 when \nh$ (BAT) = 2.8 \times 10^{23}$, \nh (Suzaku1)$ = 2.1 \times 10^{24}$, and \nh (Suzaku2)$ = 9.17 \times 10^{24}$\,cm$^{-2}$) or allowing the power law index of the second Suzaku observation to flatten ($\chi^2 = 82.7$/59 when \nh$ = 6.3 \times 10^{23}$\,cm$^{-2}$, $\Gamma$ (BAT, Suzaku1)$ = 1.83$, and $\Gamma$ (Suzaku2) $= 0.66$).  The best statistical fit, $\chi^2 = 67$/59, is obtained when the fluxes of the spectra are allowed to vary (shown in Figure~\ref{fig3}), with \nh$ = 1.0^{+12.9}_{-1.0} \times 10^{23}$\,cm$^{-2}$ and $\Gamma = 1.53^{+0.24}_{-0.13}$.     The measured fluxes in the 15--50\,keV band are 4.4 (BAT), 3.3 (Suzaku1), and 1.6 (Suzaku2) $\times 10^{-11}$\,\flux.

Thus, we clearly find flux variation from the Suzaku pin spectra of NGC 1142, the only source for which we have multiple observations.  This variation can be well modeled as flux variability alone, not statistically requiring variations in the spectral shape.  However, with few data points, particularly in the dimmer Suzaku 2 observation, we can not rule out additional variations.  It is important to note that the analysis of the Swift BAT light curve of NGC 1142 found it to be the most variable source ($\chi^2 = 3.87$).  However, from our analysis of the XIS light curves and {\it XMM-Newton} and XIS energy spectra we found NGC 1142 the least variable source in the 0.3--10\,keV band.  Therefore, this source is not necessarily a typical example of variability in the BAT AGN catalog.  It does show, though, that it is important to allow for the flux normalization to vary for the BAT spectra when jointly fitting the BAT spectra with other X-ray spectra (i.e. from {\it XMM-Newton} or Suzaku).

\subsection{Discussion of Variability}
  As already mentioned, one possible indicator of a Compton thick source is a lack of long-term variability.  The AGN source in Circinus is one of the closest and well-studied Compton thick sources and the luminosity and spectral shape has not changed appreciably over the decade it has been monitored in the X-ray band.  This is clearly not the case for our five ``hidden'' AGN.  While there is no evidence of strong short term variability during the Suzaku observations ($\approx $40 -- 80\,ks),  there is clear variability between the {\it XMM-Newton}/Suzaku observations which are separated by $\approx 1$ -- 1.5 years.  This variability is seen in both the spectral parameters (i.e. \nh, $\Gamma$ (though with large error bars on the $\Gamma$ from the partial covering model this is not statistically significant)) and flux.  The flux variability is found both above and below 10\,keV, with a change in the NGC 1142 pin flux by a factor of two and order of magnitude changes in the 0.3--10\,keV fluxes of NGC 6921 and MCG +04-48-002.

From our analysis of the variability in the Swift BAT 14--195\,keV light curves (16 day and 64 day time scales), we found that there is potentially no difference between unobscured and obscured sources, a result that must be quantified with an analysis of the entire Swift BAT AGN sample.  If this holds for the entire sample, it suggests that the same physical mechanism underlies both types of sources.  Additionally, an analysis of the light curve of Circinus shows that this Compton thick source is also consistent with the correlation we found between variability and BAT luminosity in the sample (see Figure~\ref{fig-batexcess}).  As we suggested, the possible correlation between 14--195\,keV luminosity and variability may indicate that the relative contribution from a reflection component is lower at high luminosity and higher at low luminosity.  Since Circinus has a lower 14--195\,keV luminosity and is reflection dominated, it supports our hypothesis.  However, the correlation is weak, particularly in the 64\,day binned light curves.

At softer X-ray flux (below 10\,keV), the unobscured and obscured sources do differ.  In Figure~\ref{fig-comparespec}, we showed that $\Gamma$ tends to be steeper/higher in the lowest luminosity observation of an individual source.  This contradicts our results comparing {\it XMM-Newton} and Swift XRT spectra of unobscured AGN in  \citet{2008ApJ...681..113T}.  We also found that unlike unobscured AGN, our obscured sources have larger scale variability on average in the hard X-ray band (2--10\,keV) than the soft X-ray band (0.5--2\,keV) (Figure~\ref{fig-fmax}).  This illustrates a key difference between the obscured and unobscured sources.  One possible explanation for this difference in variability in the soft band is that the soft emission seen in the obscured sources is not direct emission.  Given the connection between nuclear star forming regions and Seyfert 2 nuclei \citep{2001sgnf.conf...88V}  in addition to the lower soft band luminosity, it is possible that the soft emission in the obscured sources is dominated by galactic emission.  However, the fact that some variation is seen shows that there likely is a component from the AGN as well in the soft band.  

From our analysis of NGC 1142, we found that the variability above 10\,keV is not correlated with the variability below 10\,keV.  In fact, the flux barely changed in the 2--10\,keV band (Figure~\ref{fig-simultaneous}) while there is a factor of 2 change in the Suzaku pin spectra (Figure~\ref{fig3}).  One possible explanation is that there is a time delay between the 14--195\,keV and 2--10\,keV bands, where the softer X-rays are the result of some type of reprocessing (i.e. reflection) of the direct emission seen in the 14--195\,keV band.  If this is the case, observing the source in the accompanying low 2--10\,keV flux state would place constraints on the physical location of the reprocessing material.  However, given the anti-correlation between $\Gamma$ and luminosity in the 2--10\,keV band (for obscured sources), there may be a more complex interplay between direct and reprocessed emission in a clumpy absorbing region.  

\section{Detailed Spectral Analysis}\label{spectra}
In the previous section, we conducted a detailed study of variability in the Suzaku observations, between the Suzaku and {\it XMM-Newton} spectra, the Swift BAT light curves, and between the Suzaku pin and Swift BAT spectra.  This study helped us to conclude that the soft emission ($< 2$\,keV) in all but ESO 506-G027 is not necessarily direct AGN emission (from the lack of variability and good statistical fit with an ionized gas model).  We also found that the emission above 10\,keV is variable and that in order to include the time-averaged BAT spectra in our spectral fits, we must allow the flux normalization to vary.  The presence of variability in all of our sources suggests that these AGN are not heavily Compton thick ($\tau >> 1$ towards Compton scattering).

In this section, we examine the spectral signatures of the sources, in particular looking for the contribution of reflection for the sources with Suzaku HXD pin data.  Since reflection is the main signature of a Compton thick source, this will provide further clues to the nature of these sources.  In addition to looking for reflection signatures, we provide detailed analysis of the spectra for all of our sources, characterizing the high signal-to-noise Suzaku observations of the 5 hidden AGN.  

\subsection{\bf Suzaku XIS Spectral Fits for NGC 6921 and MCG +04-48-002}  
With an angular separation less than the 6\arcmin \,resolution of Swift's BAT and 1\degr \,field of view of Suzaku's HXD, both the BAT and pin spectra are combinations of these two X-ray sources.  Thus, we can not use these spectra to constrain the high energy portion of their spectra.  For NGC 6921, the low number of counts (330 counts in XIS1) for this source, due to the lower flux in the Suzaku observation, can not be used for a more detailed analysis than that presented in \S~\ref{var}.  Here we found that the column density was not well constrained ($\log N_{\mathrm H} \approx 25$, a column density above the limitations of {\tt tbabs}, and with errors encompassing the entire allowed range of \nh).  It is possible that the column density of the source is heavily Compton thick in this observation, i.e. $\log N_{\mathrm H} \ga 1/\sigma_T$ preventing transmission of non-reflected emission.  However, as in the previous {\it XMM-Newton} spectrum, a strong Fe K-$\alpha$ EW, a signature of reflected emission, is not present.  Thus, the spectra of this source continue to be a puzzle.

For MCG +04-48-002, the superior spectral resolution of the Suzaku XIS detectors allows us to better constrain the parameters of the observed Fe K-$\alpha$ feature.  Extracting spectra of the calibration source, Fe-55, we found that the errors on detectable line widths range from 0.013\,keV -- 0.024\,keV.  Therefore, we can detect line widths above this range.  To determine the line width and energy of the Fe K-$\alpha$ line, we fit the combined XIS0+XIS3 and XIS1 spectra with the partial covering model used in Table~\ref{tbl-pcfabs}.  We extended the energy range to 12\,keV to better constrain the continuum.  Within this energy range, we find that the column density and covering fraction are the same as in  Table~\ref{tbl-pcfabs}, within the errors.  The spectral index is slightly steeper, measured as $\Gamma = 2.36^{+0.48}_{-0.46}$.  From the residuals of the spectral fit, we found evidence for an \ion{Fe}{25} K-$\alpha$ emission line in the combined XIS0+XIS3 spectrum (not seen in the XIS1 spectrum).  Adding this feature improved $\chi^2$ by 4.5, showing the line to be marginally significant.  We could not constrain the width, fixing this to 0.01\,keV.  From our spectral fit, we find an energy of $6.68^{+0.06}_{-0.07}$\,keV and EW of $70^{+41}_{-53}$\,eV for \ion{Fe}{25} K-$\alpha$.   We find $E = 6.38^{+0.04}_{-0.04}$\,keV, $\sigma = 0.05^{+0.07}_{-0.05}$\,keV, and EW of $184^{+88}_{-72}$\,eV for Fe I K-$\alpha$.  The energies of these lines are perfectly consistent with the known optical redshift and identification as fluorescent \ion{Fe}{1} K$\alpha$ and He-like Fe.

\subsection{\bf Joint Suzaku XIS, HXD, and Swift BAT Spectral Fits}
For the remaining sources, NGC 1142, Mrk 417, and ESO 506-G027, we performed joint fits of the Suzaku XIS, HXD pin, and Swift BAT spectra.  The inclusion of the BAT spectra provides an extension of the spectrum beyond the HXD pin high energy limit for our sources.  Since the 14--195\,keV flux is not constant over the 13 month period used to construct the BAT spectra, we included a constant value which was allowed to vary for the BAT spectra.

\begin{figure}
\centering
\includegraphics[width=12cm]{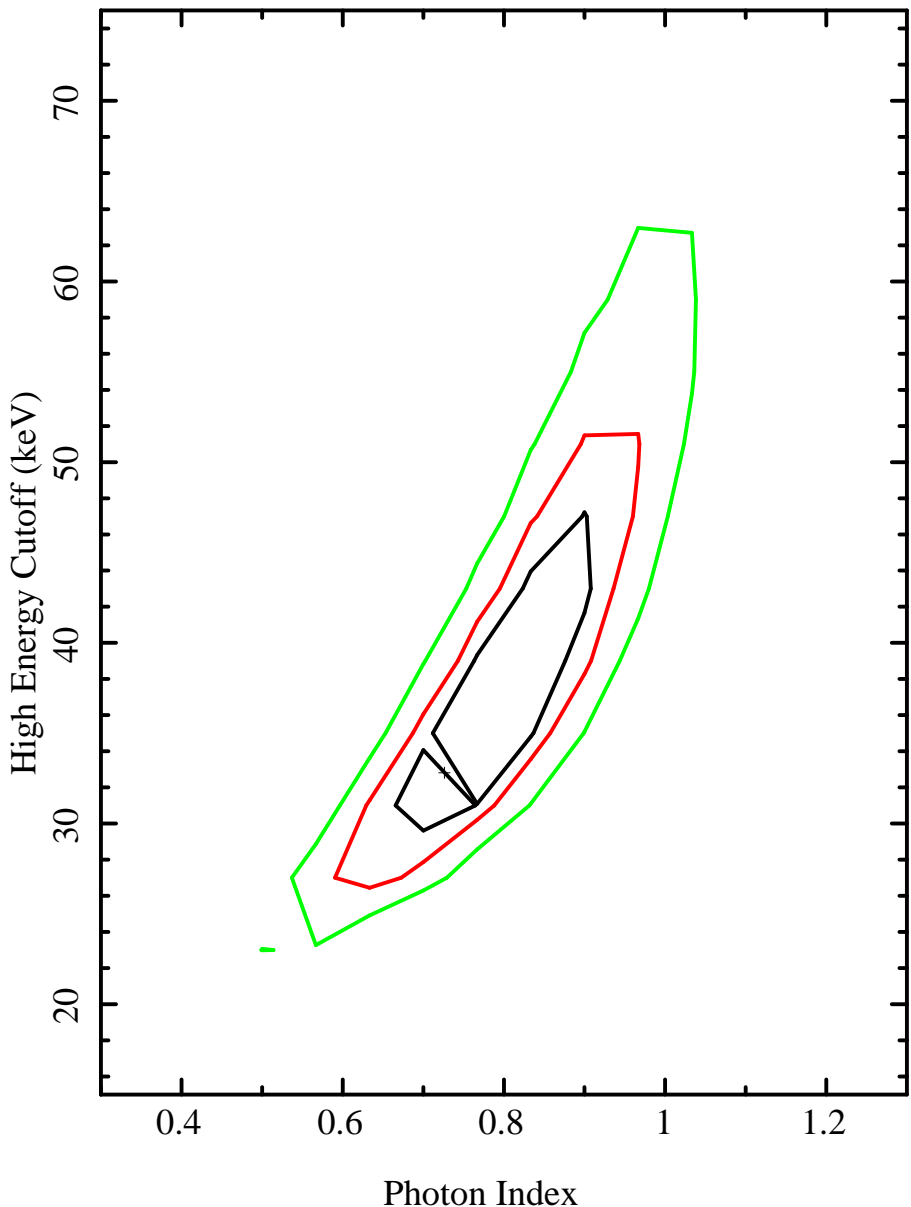}
\caption{Using the partial covering model for the long observation of NGC 1142 (Table~\ref{tbl-cutoff}), we present the 99\%, 90\%, and 68\% error contours for the high energy cuttoff and photon index ($\Gamma$).  This shows that the cutoff energy is dependent of the power law slope, using this partial covering model.  A flatter slope yields a lower cutoff energy.
\label{fig-ecutoff}}
\end{figure}

As a first step to our fitting process, we used the partial covering model described in \S~\ref{var} (a partially covered power law with an Fe K-$\alpha$ line) to fit the combined XIS + pin + BAT spectra.
We fixed the pin flux to be 1.12 times the XIS flux (as described in the Suzaku documentation).  We also allowed the energy and width of the Fe K-$\alpha$ line to vary.  As a next step, we replaced the power law model with a cut-off power law -- a power law model with a high energy cut-off.  As in \S~\ref{var}, we included an {\tt apec} model for the spectra of NGC 1142.  We present the important results from these spectral fits in Table~\ref{tbl-cutoff}.  One result is that adding a high energy cutoff to the power law is statistically significant for all of the spectra ($\Delta\chi^2 > 12$).  The cutoff energies we measure are $\approx 50$--$80$\,keV with error bars of $\approx 20$\,keV, similar to or lower than high energy cutoffs measured in some AGNs with joint XMM-Newton/Integral spectral fits based on a simple absorbed cutoff power law model \citep{2006MNRAS.371..821M,2008AA...483..151P,2008arXiv0809.0255M}.  In Figure~\ref{fig-ecutoff}, we plot the error contours (using the {\tt steppar} command) for the cutoff energy versus power law index for the highest resolution observation (NGC 1142 obs. 1).  As shown, the computed energy cutoff depends on the power law slope (i.e. a flatter slope gives a lower cutoff energy).  Other models examined below give higher cutoff energies or show no strong constraint on cutoff energy (i.e. the reflection model and a double partial covering model).

Another important result is that none of the measured columns are in the Compton-thick regime (\nh$ > 1.4\times 10^{24}$\,cm$^{-2}$).  Adding the high energy data constrains the columns to lower values, though still highly obscured ($\log N_{\mathrm H} > 23$).  Finally, the measured power law slopes for our sources are still flat ($\Gamma \approx 1.0$), much flatter than the average slope of 1.75 measured for the entire BAT sample in the 0.3--10\,keV band \citep{2009ApJ...690.1322W}.  Since the power law model is an approximation to the AGN emission, it is likely that the inclusion of a reflection model or a more accurate Compton scattering model is crucial.  

Before including more complicated models of the AGN emission, we investigate the presence of emission and/or absorption features around the Fe K-\,$\alpha$ emission line.  To quantify the importance of these additional features, we include the $\Delta\chi^2$ values on adding an \ion{Fe}{1} K-$\alpha$ line as well as additional lines often seen in high signal-to-noise spectra (\ion{Fe}{25} K$\alpha$, \ion{Fe}{1} K$\beta$, and \ion{Ni}{1} K$\alpha$) in Table~\ref{tbl-felines}.  Also, the errors on each of the parameters for these features are included.  Since the {\tt pcfabs} model places the 7.11\,keV Fe edge at a redshift of 0, we substituted this model with {\tt zpcfabs}, a model which has an additional parameter to shift this feature to the redshift of the AGN.  

Clearly, the \ion{Fe}{1} K$\alpha$ line is significant ($\Delta\chi^2 > 50$) in all of these sources, with the spectrum of Mrk 417 showing the weakest line.  The width of this line, $<EW> \approx 67$\,eV, is consistent with velocities from about 1800 -- 3000\,km\,s$^{-1}$.  These results agree with those from Chandra grating results, which place the origin of the Fe K-$\alpha$ line in a region between the narrow (500 -- 700\,km\,s$^{-1}$) and broad (3000 -- 10000\,km\,s$^{-1}$) line regions \citep{2004ApJ...604...63Y}.  Of the remaining emission lines we fit the spectra with, Fe I K$\beta$ is detected in the spectra of NGC 1142 (both observations) and ESO 506-G027 and \ion{Fe}{25} K$\alpha$ is only detected in the spectrum of ESO 506-G027.  \ion{Ni}{1} K$\alpha$ is not detected in any of the spectra.  As described in \citet{2007PASJ...59S.283Y} and elsewhere, the ratio of \ion{Fe}{1} K$\beta$/\ion{Fe}{1} K$\alpha$ can be used to constrain the ionization state of the gas.  From the Suzaku observations of NGC 1142 and ESO 506-G027, this ratio is $\approx 10$\%, consistent with neutral iron.  However, the errors are large (ranging from 2--50\%), requiring higher quality data to better constrain these values.

In addition to these features, the addition of an Fe edge at 7.11\,keV (in addition to the edge built into the {\tt zpcfabs} model) was significant ($\Delta\chi^2 = 7.5$) in the long observation of NGC 1142.  The best-fit parameters for the energy and optical depth of this feature (added with a {\tt zedge} model) are $E = 7.16^{+0.04}_{-0.16}$\,keV and $\tau = 0.364^{+0.077}_{-0.079}$.  Since the 7.11\,keV edge is a possible signature of reflection, this could mean that reflection is significant only in the first observation of NGC 1142.  Alternatively, the Fe edge is also an indicator of the Fe abundance.  Using the optical depth ($\tau$) calculated from the edge model and the cross section for H like Fe at 7.11\,keV ($\sigma = 5.304 \times 10^{-22}$\,cm$^2$;  \citet{1993ADNDT..54..181H}), we calculated the column density of iron (\nh$ = \tau/\sigma$) and divided this by the hydrogen column density obtained from the partial covering model.  Assuming an ISM  iron abundance of $2.69 \times 10^{-5}$ the hydrogen abundance \citep{2000ApJ...542..914W}, this yields abundances of approximately 3.5 solar for NGC 1142 (observation 1).  The lack of a detection in the second NGC 1142 observation suggests that reflection may be the best explanation for this source.  However, if the slope of the power law is set to 1.0, as in the longer observation, the upper limit on the iron edge optical depth is 0.363, in line with the value obtained for the first observation.  

\subsubsection*{\bf Alternative Models} 
While the reduced $\chi^2$ values from the partially covered cutoff power law models indicate good fits to the data ($\chi^2 \approx 1.0$), the flat power law slopes ($\Gamma \approx 1.0$) are unphysical.  An alternative model including a transmitted, scattered, and reflected component was applied by \citet{eguchi2008} (Model B) to the Suzaku spectra of ``hidden'' Swift-detected AGN, including the long observation of NGC 1142.  This model is represented as {\tt tbabs}$_{Gal}$*({\tt ztbabs}$_{trans}$*{\tt cutoffpl}$_{trans}$ + {\tt const}*{\tt cutoffpl}$_{scat}$ + {\tt ztbabs}$_{refl}$*{\tt pexrav}$_{refl}$ + Fe lines) in {\tt XSPEC}.  The parameters of the reflection model ({\tt pexrav} \citep{1995MNRAS.273..837M}) include photon index, folding energy, abundance, inclination angle, reflection scaling factor ($R = \Omega/2\pi$, which we allowed to vary between -5 and 0 (negative values in this model account for the reflected spectrum alone)), and the normalization.  In Model B of \citet{eguchi2008}, the normalization of the reflected component is fixed to the same value as the transmitted and scattered component (whose spectral indices are also fixed to the same value, along with the $\Gamma$ value for the reflected component).  We fix the folding energy to 300\,keV (the default value, also used in \citet{eguchi2008}), abundances to solar, and cosine of the inclination angle to the default of 0.45.  In Table~\ref{tbl-modelb}, we include the best fit parameters using this reflection model.  While the first observation of NGC 1142 was already included in \citet{eguchi2008}, we include our own fit, which uses the {\tt wilm} ISM abundances (\citet{eguchi2008} use {\tt angr} abundances \citep{anders89}) and is binned differently than in \citet{eguchi2008}, who bin the spectra by 50\,cts/bin.  For all of the sources, our fits include the \ion{Fe}{1} K$\alpha$ line.   We note that using the reflection model, we can not constrain a cut off energy when we allow the cutoff energy to be a free parameter.  We found that the model prefers no cutoff energy (the value floated to the upper limits imposed), showing that there is a strong trade off of reflection versus cutoff energy.  Further, whether the cutoff energy is set at a lower value of 100\,keV or 300\,keV, we find that the photon index from these fits does not change appreciably and is much steeper than the values of $\Gamma \approx 1.0$ obtained with the partial covering model.


The results of this reflection model (Table~\ref{tbl-modelb}) indicate good fits to the spectra with reduced $\chi^2 \approx 1.0$ and more ``normal'' power law slopes ($<\Gamma> = 1.76$).  These power law index values are more typical of the slopes found for the BAT sample \citep{2009ApJ...690.1322W}.
However, the reflection component measurements, $R$, show the sources to have mild reflection components at most (with the possible exception of ESO 506-G027).  A value for $|R| > 1$, as found for the long observation of NGC 1142,  is unphysical, as described in \citet{eguchi2008}.  Our value is similar to that from the analysis presented in \citet{eguchi2008}, where they classify NGC 1142 as mildly or hardly Compton thick and explain the high value of $R$ as the result of the direct emission being completely blocked non-uniform material in the line of sight \citep{2007ApJ...664L..79U}.  These results suggest that none of these ``hidden'' AGN are Compton thick in the sense that they are not reflection dominated and have column densities below $1.5 \times 10^{24}$\,cm$^{-2}$ (with the possible exception of ESO 506-G027).  The classification as not heavily Compton-thick is supported by the fact that the sources do vary between observations in flux as well as spectral shape.  Particularly, this argument applies to NGC 1142, whose pin spectrum is nearly three times dimmer in the second Suzaku observation. 

\begin{figure}
\centering
\plotone{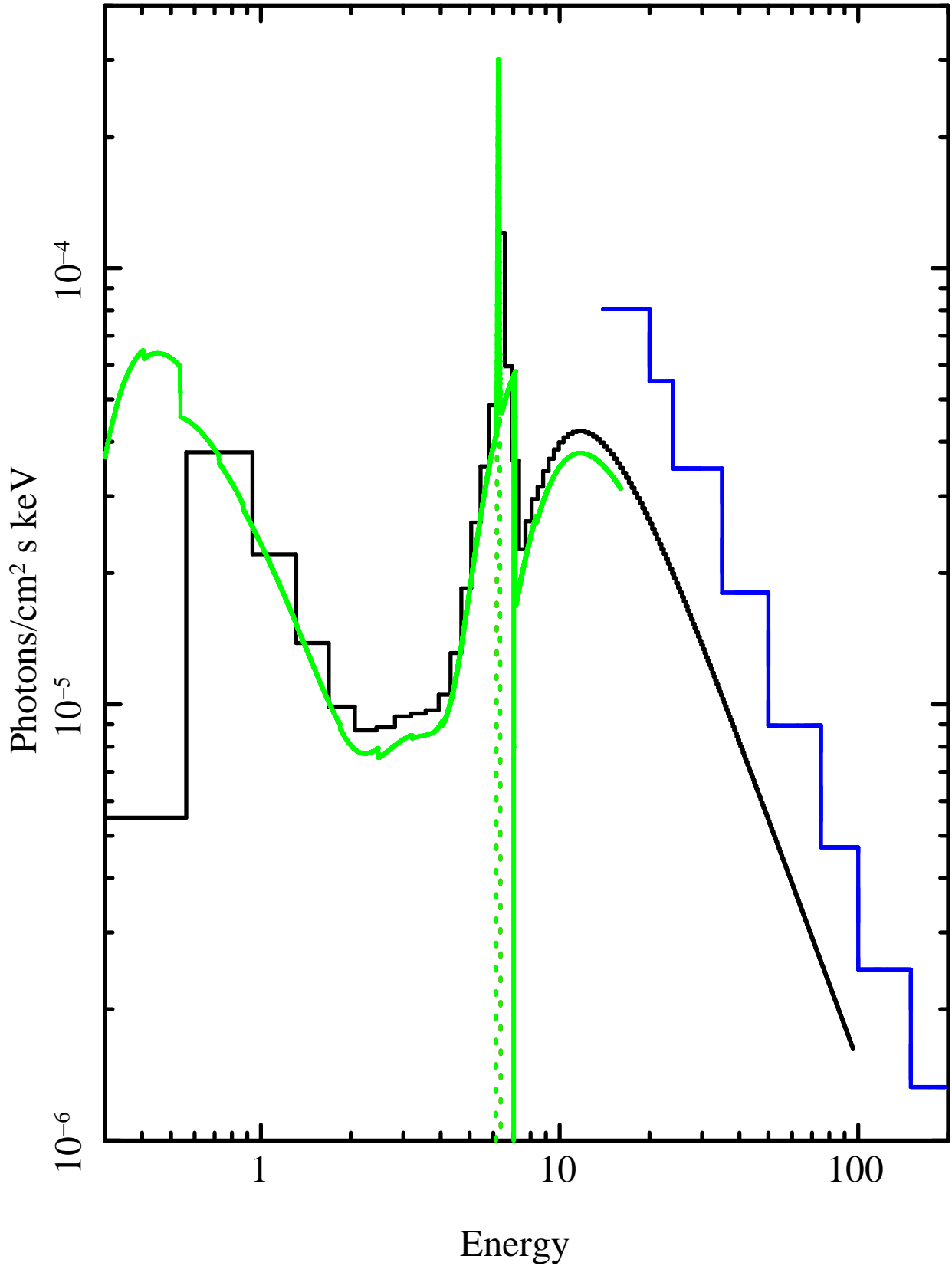}
\caption{Plotted is the double partial covering model used to fit the combined Suzaku XIS, pin, and Swift BAT spectra, implemented as {\tt tbabs}$_{Gal}$*{\tt pcfabs}*{\tt pcfabs}*({\tt pow} + Gaussian lines)*{\tt const} in {\tt XSPEC}.
\label{fig-dblpcfabs}}
\end{figure}

\begin{figure}
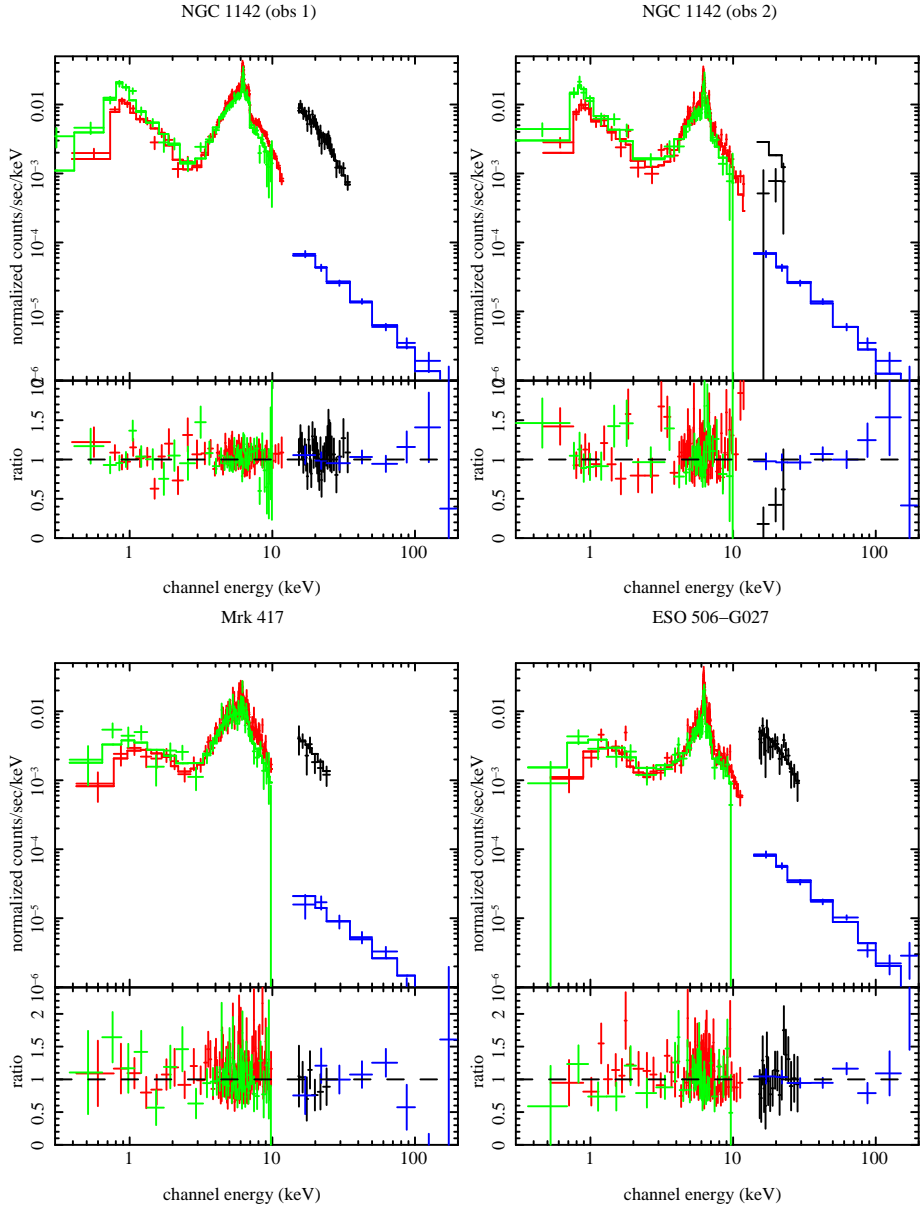

\centering
\includegraphics[width=6cm]{f12a.ps}
\includegraphics[width=6cm]{f12b.ps} \\
\includegraphics[width=6cm]{f12c.ps}
\includegraphics[width=6cm]{f12d.ps}\\
\caption{Plotted is the observed Suzaku + Swift BAT spectrum fit with the double partial covering model, along with the ratio of the data to the model, for each of the observations for NGC 1142, Mrk 417, and ESO 506-G027.  
The double partial covering model (Table~\ref{tbl-dblpcfabs}), shown, yielded similar $\chi^2/dof \approx 1.0$ values as the reflection model fits (Table~\ref{tbl-modelb}).  The Suzaku XIS spectra are shown rebin to a signal-to-noise of 10 (NGC 1142 obs-1) or 5 (for the remaining sources).
\label{fig-dblpcfabsspectra}}
\end{figure}

Another model which is a good fit to the data and does not require reflection is the double partial covering model.   This model assumes that two partially covering absorbers, with different column densities and covering fractions, are within the line of sight to the direct emission.  Thus, this model includes one unabsorbed component plus three absorbed components with different column densities, contrary to Model C in \citet{eguchi2008} which includes two different column densities for the transmitted component.  In Figure~\ref{fig-dblpcfabs}, the double partial covering model is plotted, which is represented in {\tt XSPEC} as {\tt tbabs}$_{Gal}$*{\tt zpcfabs}$_1$*{\tt zpcfabs}$_2$*({\tt cutoffpl} + Fe lines).  Fits to the individual spectra are shown in Figure~\ref{fig-dblpcfabsspectra} and the key parameters from this fit are listed in Table~\ref{tbl-dblpcfabs}.  Since this model constrains the cutoff energy to a lower value in the long observation of NGC 1142, $E_{cutoff} = 143^{+57}_{-73}$\,keV, than in the reflection model where it is unconstrained, we fix $E_{cutoff} = 100$\,keV for the observations where it could not easily be constrained with the double partial covering model.

With this double partial covering model, the power law indices are more consistent with the average value for AGNs in the BAT sample \citep{2009ApJ...690.1322W}.  Further, with this model we can explain the difference between the spectra of NGC 1142 in two ways -- a difference in absorbing columns as well as a steeper spectrum in the second, dimmer observation.  Contrary to the results of \citet{2005AA...432...15P}, \citet{2004AA...422...85P}, and \citet{2006ApJ...646L..29S}, as well as our own results from Seyfert 1 sources \citep{2008ApJ...674..686W}, the accretion rate or flux is anti-correlated with spectral index for the two observations of NGC 1142.  The same is true of the Fe K$\alpha$ equivalent width, contrary to the results of  \citet{1993ApJ...413L..15I}, \citet{2004MNRAS.347..316P}, \citet{2006ApJ...644..725J}, and others.  This highlights another key difference between Sy 1s and the hidden sources, suggesting that there likely is a complex environment of varying absorbers.  This result is consistent with those of \citet{2002ApJ...571..234R}, who concluded that variations in column density are ubiquitous in Sy 2s and the product of clumpy absorbers.

\subsubsection*{The Difference Spectrum for NGC 1142}
With two Suzaku spectra of NGC 1142, we created a difference spectrum -- a spectrum constructed from subtracting the lower flux observation from the higher flux observation.  To do this, we used the {\it Ftool} {\tt mathpha} to subtract the observation 1 combined xis 0+3 source spectrum from the observation 2 source spectrum.  We did the same for the background spectra.  We then used the {\it Ftools} {\tt addrmf} and {\tt addarf} to add the response files from the two observations, weighting them by their respective exposure times (observation 1 was weighted as 0.688 and observation 2 was weighted as 0.312).  In the same way, we created the pin difference spectrum between the two observations.  We binned the spectra in the same manner as in \S~\ref{data}.

\begin{figure}
\centering
\plotone{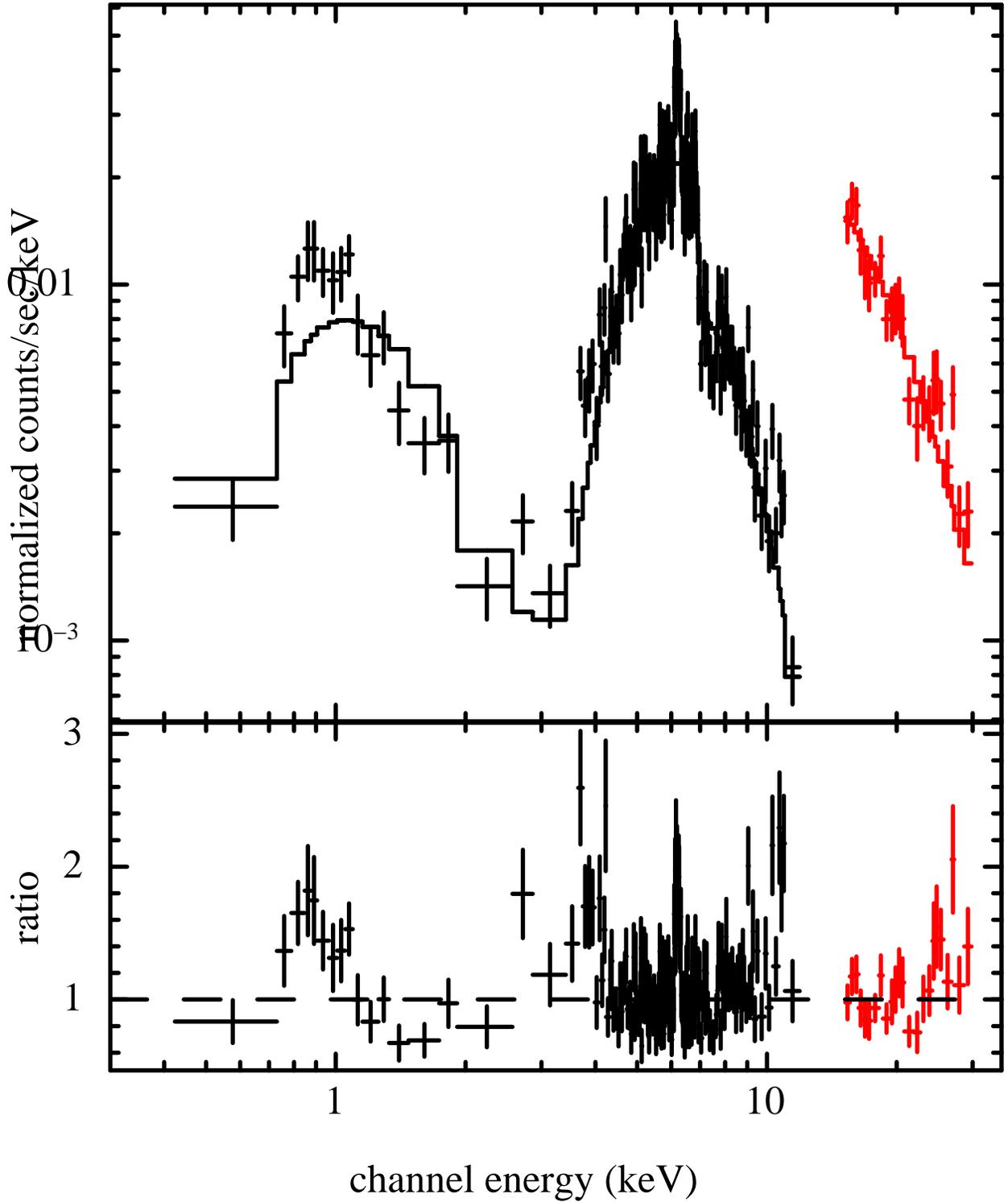}
\caption{Plotted is the difference XIS 0+3 spectrum and difference pin spectrum for NGC 1142, created from subtracting the low flux Suzaku observation from the high flux Suzaku observation.  The model used is a partial covering cutoff power law model ({\tt zpcfabs}*{\tt cutoffpl}), with a constant parameter to allow for flux differences between the XIS and pin spectra.  As seen, this model is not an adequate fit with significant residuals at soft energies and a prominent \ion{Fe}{1} K$\alpha$ feature.  This shows that both the soft emission and the iron line change between observations. 
\label{fig-diffzpcfabs}}
\end{figure}

Unlike the difference spectra of other sources, such as MCG--5--23--16 \citep{2007PASJ...59S.301R}, the difference spectrum of NGC 1142 is not a pure power law.  A partial covered cutoff power law is not a good statistical fit to the data, with $\chi^2/dof = 953/436$.  As shown in Figure~\ref{fig-diffzpcfabs}, there is emission seen below 1\,keV.  This soft emission shows that the soft spectrum did in fact change between the two observations.  Adding an {\tt apec} model to the base partial covered cutoff power law model ({\tt zpcfabs}*{\tt cutoffpl}) improves the fit by $\Delta\chi^2 = 44$.  The temperature of this component ($kT = 0.63^{+0.02}_{-0.04}$\,keV) is consistent with the values we previously found for both observation 1 and 2.  

In addition to a soft component, emission and absorption features are present in the difference spectrum.  The \ion{Fe}{1} K$\alpha$ line is very significant, with $\Delta\chi^2 = 195$.  Both the energy and width of this line are consistent with the previous measurements.  The equivalent width is the same as in observation 1, with $EW = 225^{+30}_{-32}$\,eV.  The \ion{Fe}{1} K$\beta$ line is also present, but much less significant with $\Delta\chi^2 = 4$.  Since these lines are seen as emission features in the difference spectrum, this shows that the flux of these lines scales with the continuum flux (i.e. the line flux is higher when the continuum is higher).

Further, three other features are also present.  At low energy, an emission line is present at $E = 1.09 \pm 0.02$\,keV, with $\Delta\chi^2 = 15$ and $EW = 56$\,eV.  The energy of this line is consistent with \ion{Fe}{22} L ($E = 1.053$\,keV).  Additionally, a 7.2\,keV absorption feature is significant in the difference spectrum ($\Delta\chi^2 = 9$).  This suggests that the edge energy is changing between observations or that there is an absorption feature at this energy which is reacting to changes in the spectrum.  The final feature seen is another ``absorption'' feature, with $\Delta\chi^2 = 15$ and $E = 7.67 \pm 0.01$\,keV ($EW = 98$\,eV).

Upon adding these features, the best fit model to the difference spectrum yields an acceptable fit of $\chi^2/dof = 670/422$ (1.59).  The best-fit parameters for absorption and power law index are: \nh$ = 7.7 \pm 0.8 \times 10^{23}$\,cm$^{-2}$, $Cvr = 0.995 \pm 0.001$, and $\Gamma = 1.69^{+0.17}_{-0.15}$.  Compared to the best-fit parameters for observations 1 and 2 (Table~\ref{tbl-cutoff}), the difference spectrum has a higher column density, slightly higher covering fraction, and steeper power law index (particularly in comparison with observation 1).

\subsection{Discussion of Spectral Properties}
From our detailed spectral analysis, we found that a reflection component is very weak in the spectra of our target sources.  Also, none of the sources have a 1\,keV EW \ion{Fe}{1} K$\alpha$ line.  Since our sources do not meet these two criteria for Compton-thick spectral classification, we classify them as Compton-thin.  All of the spectra are well-fit by a partial covering or double partial covering model, implying that multiple absorbing columns lie along our sight line to the hidden sources.

Through our spectral fits, we find that the derived column densities and power law components change depending upon the model.  For instance, very flat power law components are obtained using a power law model for NGC 1141, Mrk 417, and ESO 506-G027 (Table~\ref{tbl-cutoff}) while steeper values are found using the reflection model or double partial covering model.  This shows that a flat power law is not a good indicator of a Compton thick or reflection dominated spectrum and that other criteria like a 1\,keV EW \ion{Fe}{1} K$\alpha$ emission line or lack of long-term variability (except in the case of `changing-look' Compton-thick sources \citep{2003MNRAS.342..422M}) should be met.  Additionally, we found that the cutoff energy for the power law component is also model dependent.  With a simple partial covering model, we obtained cutoff energies of $\approx 100$\,keV, while the cutoff could not be constrained with the reflection model (and was therefore set to 300\,keV).  Using the double partial covering model, we constrained the cutoff energy to $\approx 150$\,keV in the highest signal-to-noise observation of NGC 1142.  Therefore, caution must be taken in constraining cutoff energies, since these values are highly model dependent.

\begin{figure}
\includegraphics[width=8.5cm]{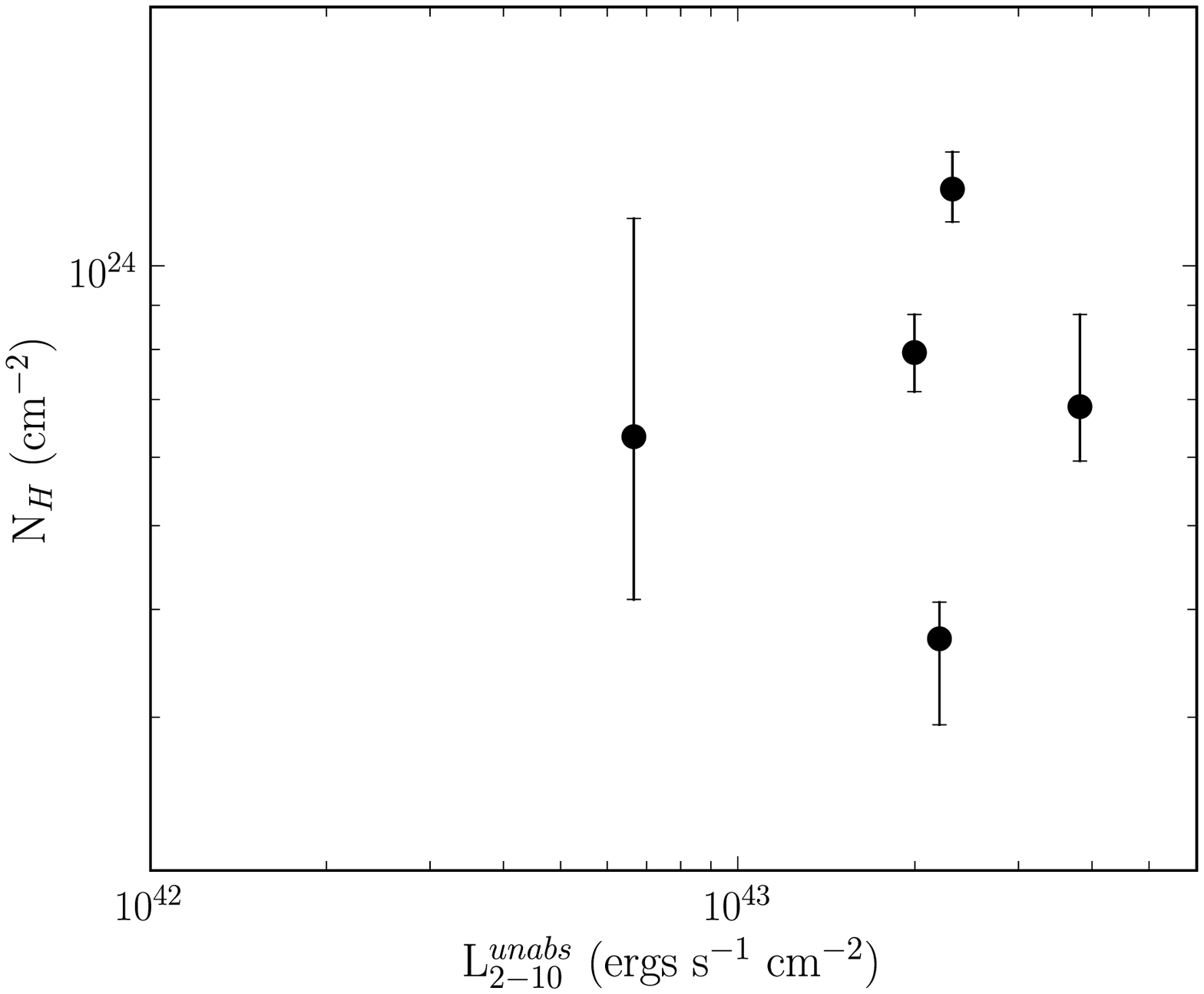}
\hspace{0.1cm}
\includegraphics[width=8.5cm]{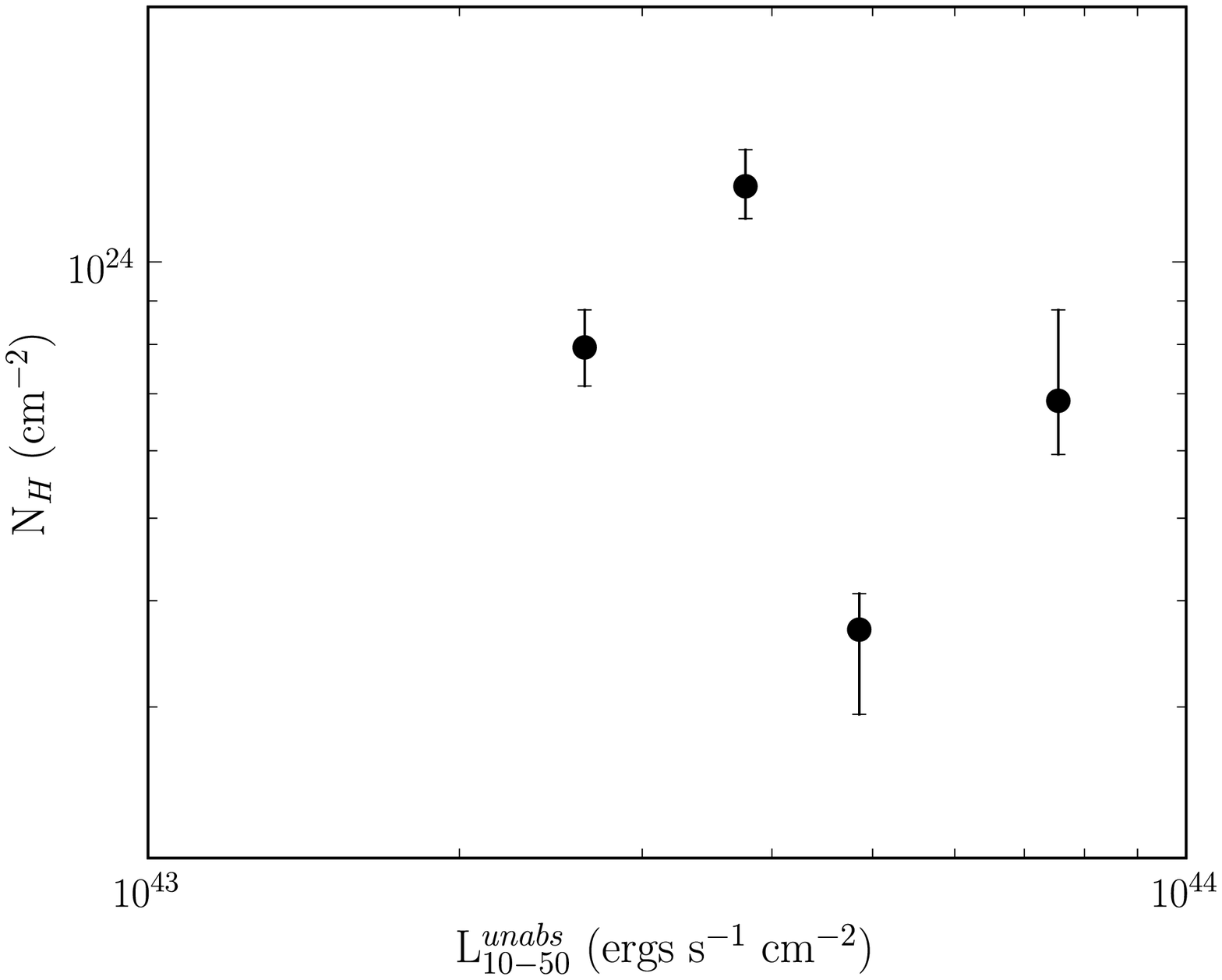}
\vspace{0.1cm}
\includegraphics[width=8.5cm]{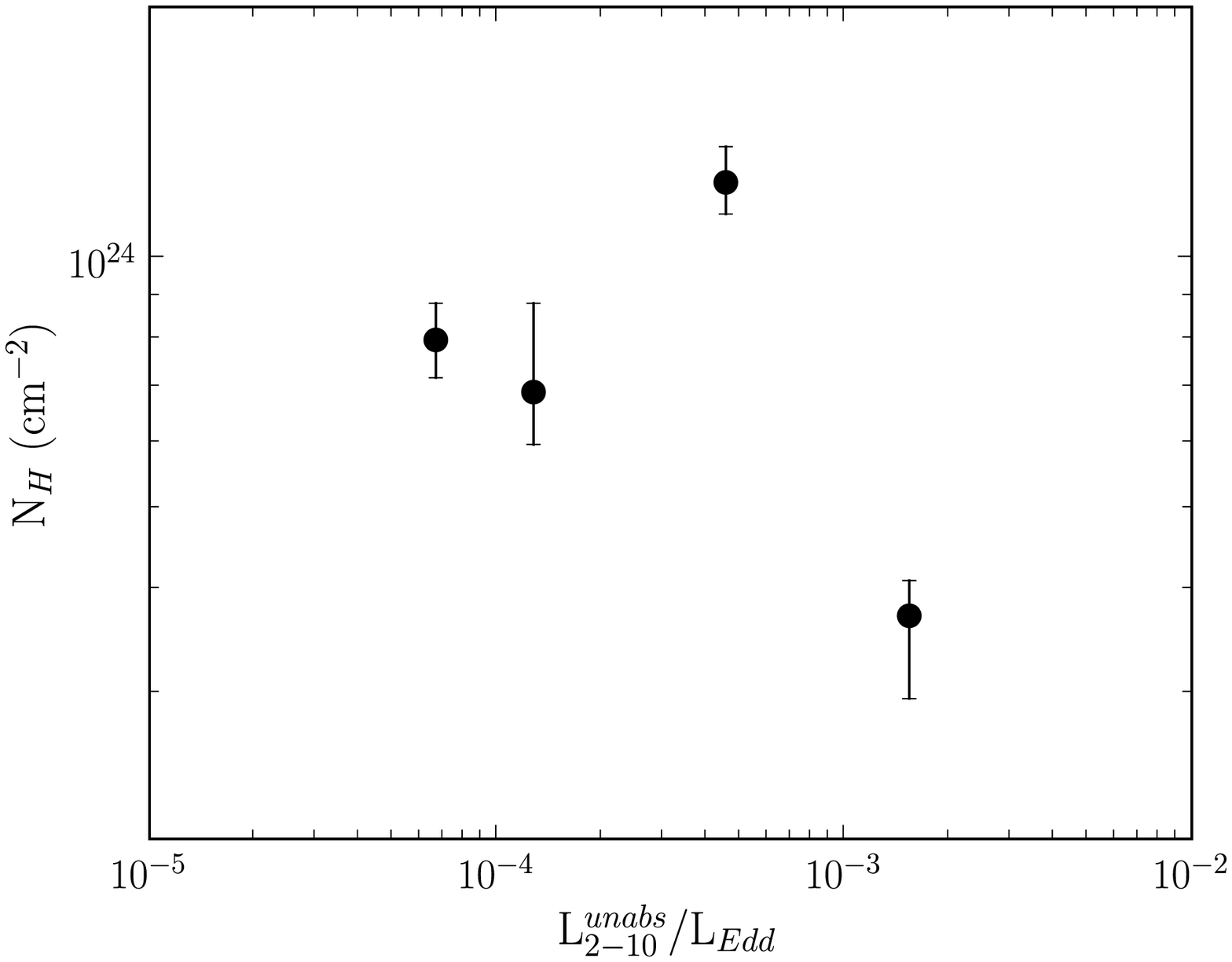}
\hspace{0.1cm}
\includegraphics[width=8.5cm]{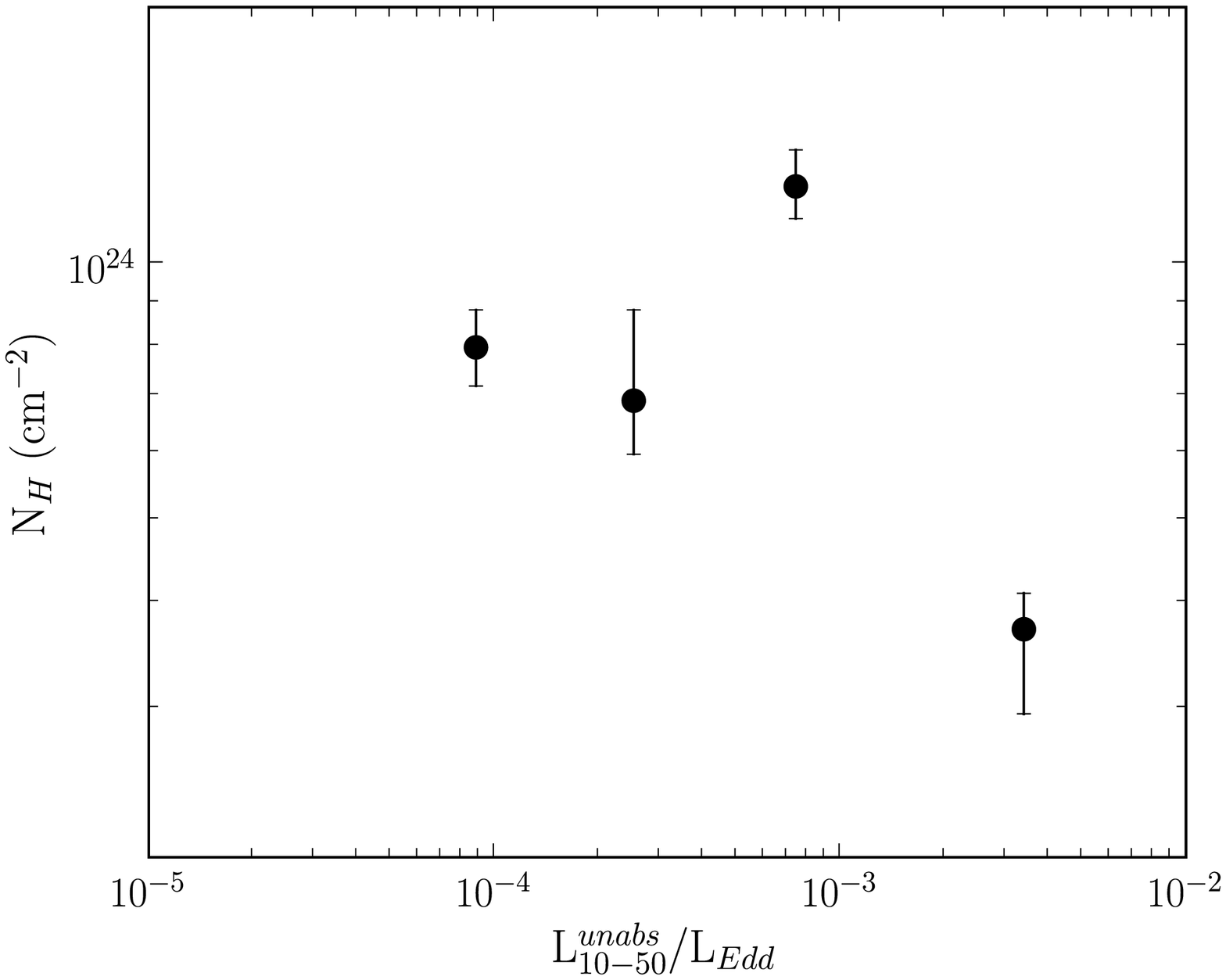}

\caption{Plotted is the line of sight column density (\nh) versus the unabsorbed luminosity in the 2--10\,keV band (top left), the 10--50\,keV band (top right), the ratio of 2--10\,keV luminosity to the Eddington luminosity, and the ratio of 10--50\,keV luminosity to Eddington luminosity.  There is no correlation seen between these values.
\label{fig-nhlum}}
\end{figure}

\begin{figure}
\includegraphics[width=8.5cm]{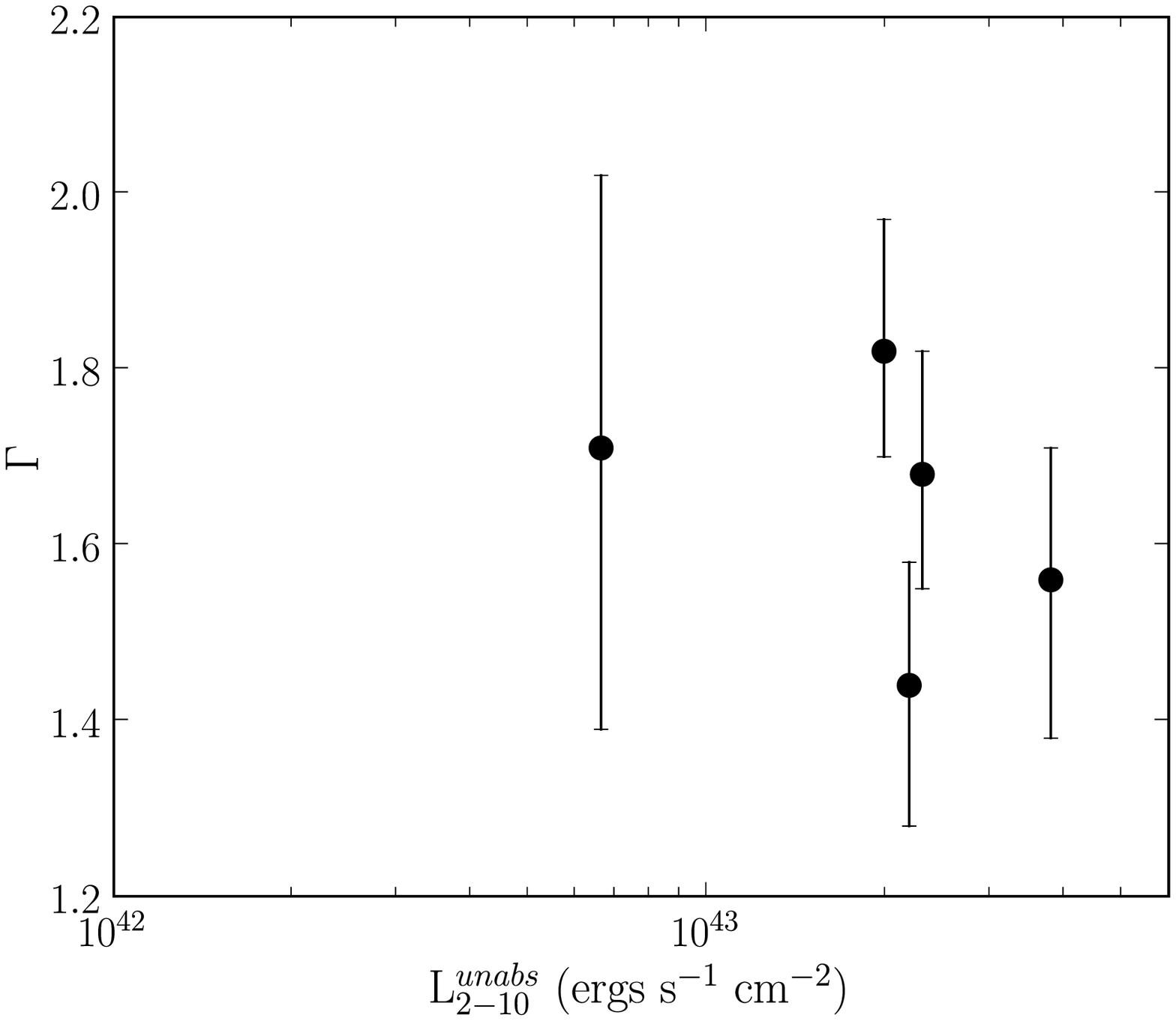}
\hspace{0.1cm}
\includegraphics[width=8.5cm]{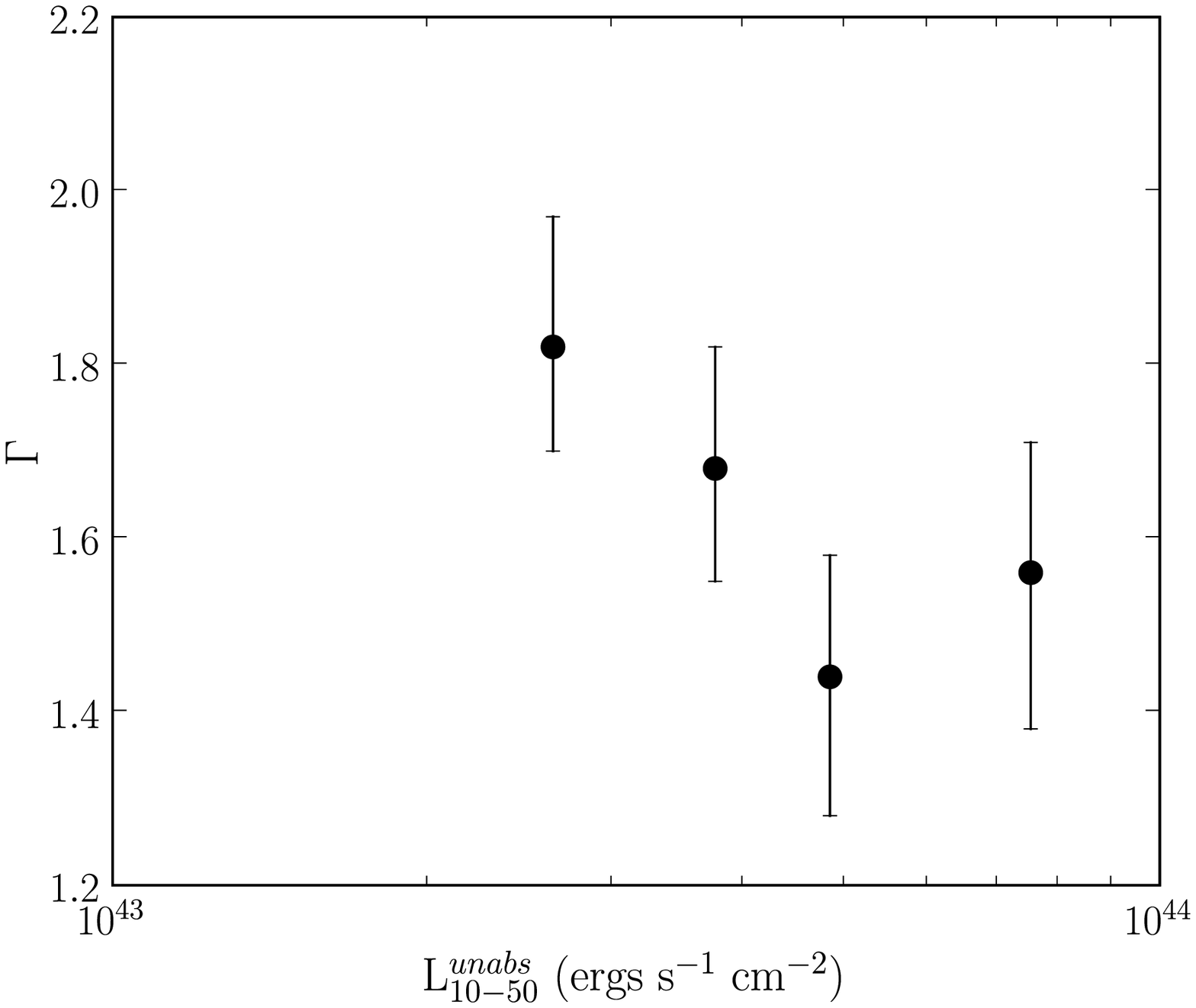}
\vspace{0.1cm}
\includegraphics[width=8.5cm]{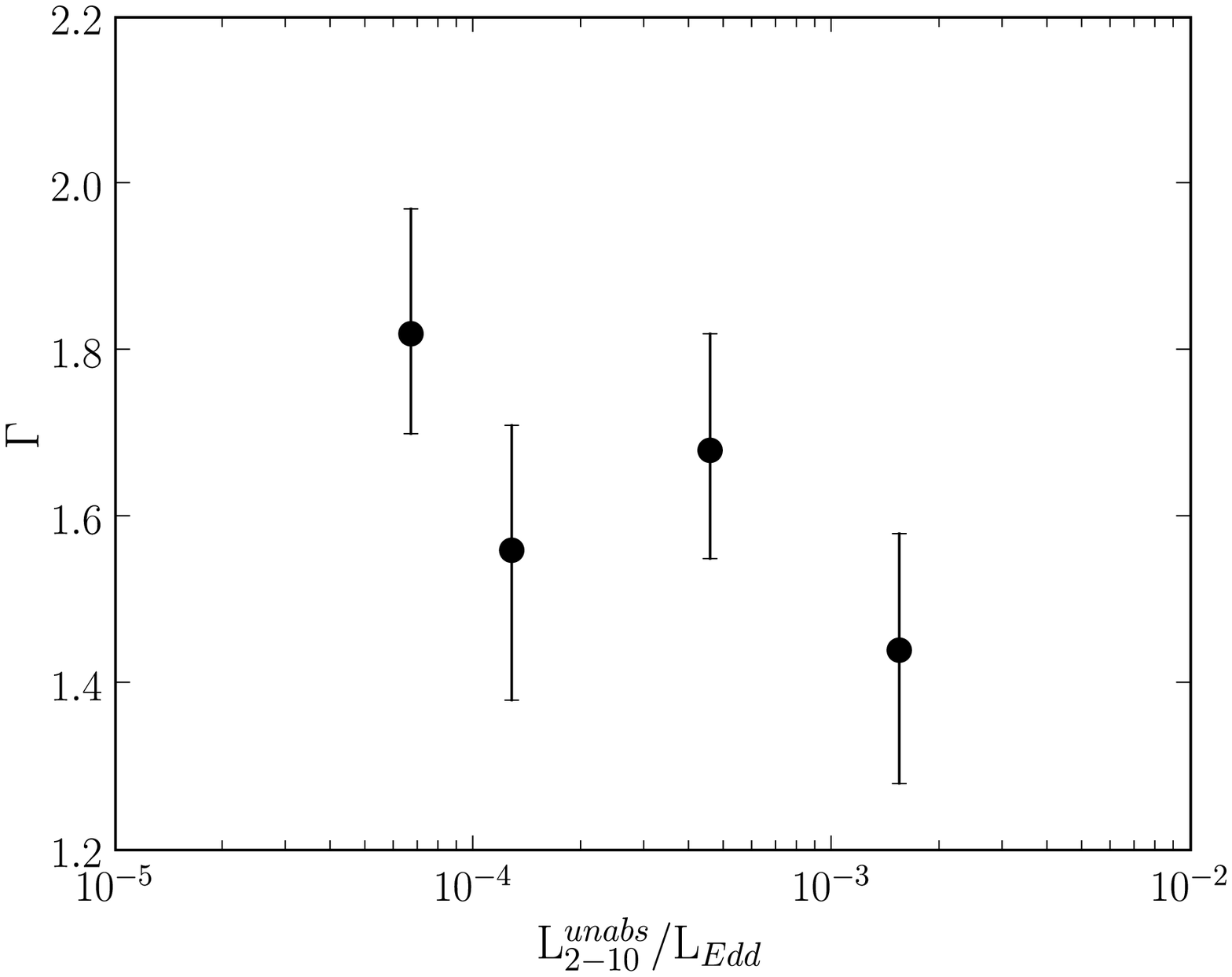}
\hspace{0.1cm}
\includegraphics[width=8.5cm]{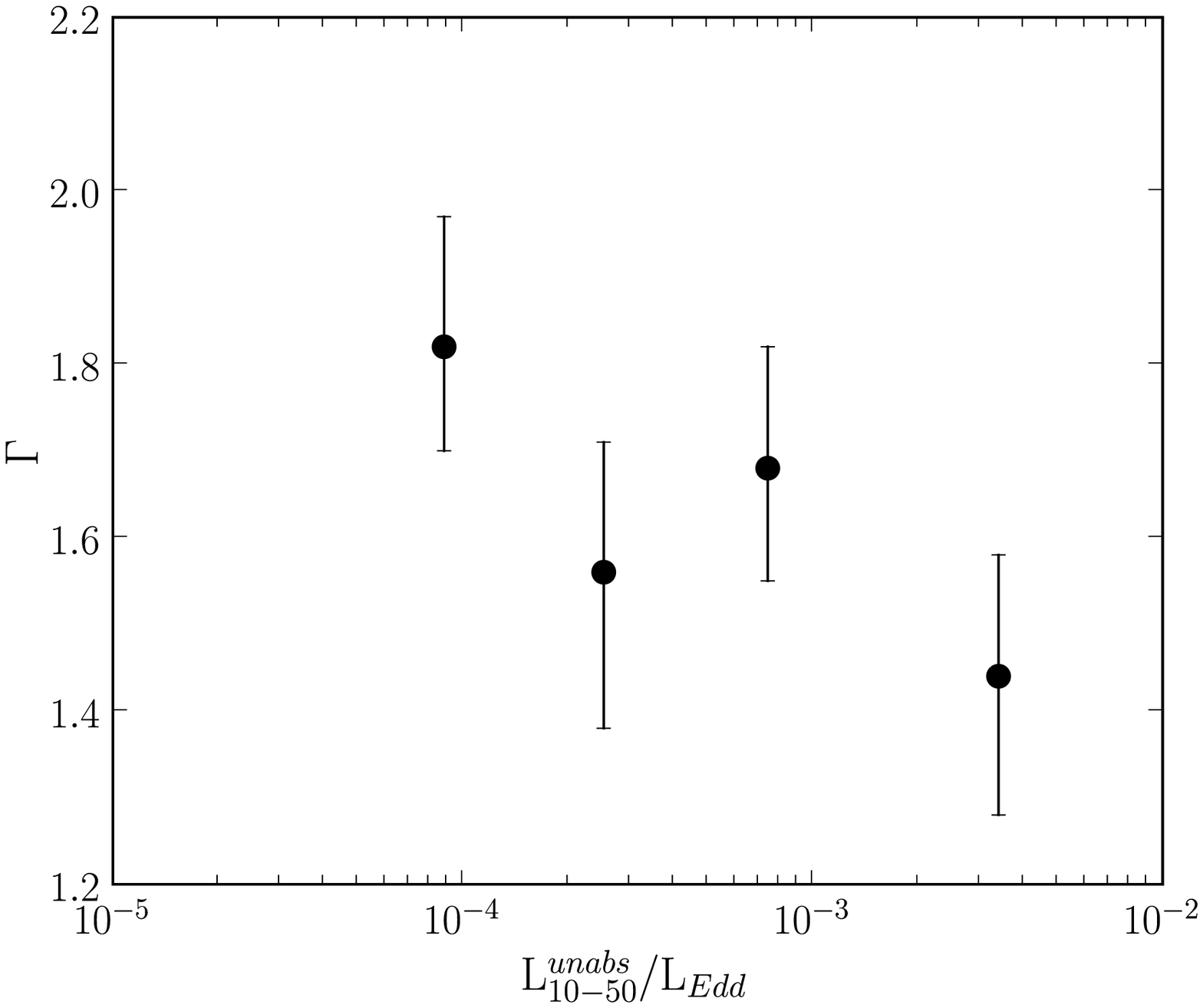}

\caption{Plotted is the spectral index ($\Gamma$) versus the unabsorbed luminosity in the 2--10\,keV band (top left), the 10--50\,keV band (top right), the ratio of 2--10\,keV luminosity to the Eddington luminosity, and the ratio of 10--50\,keV luminosity to Eddington luminosity.  There is evidence of a possible anti-correlation between $\Gamma$ and the computed Eddington ratios, however, the errors on $\Gamma$ are too large for this to be statistically significant.
\label{fig-gammalum}}
\end{figure}

\begin{figure}
\includegraphics[width=8.5cm]{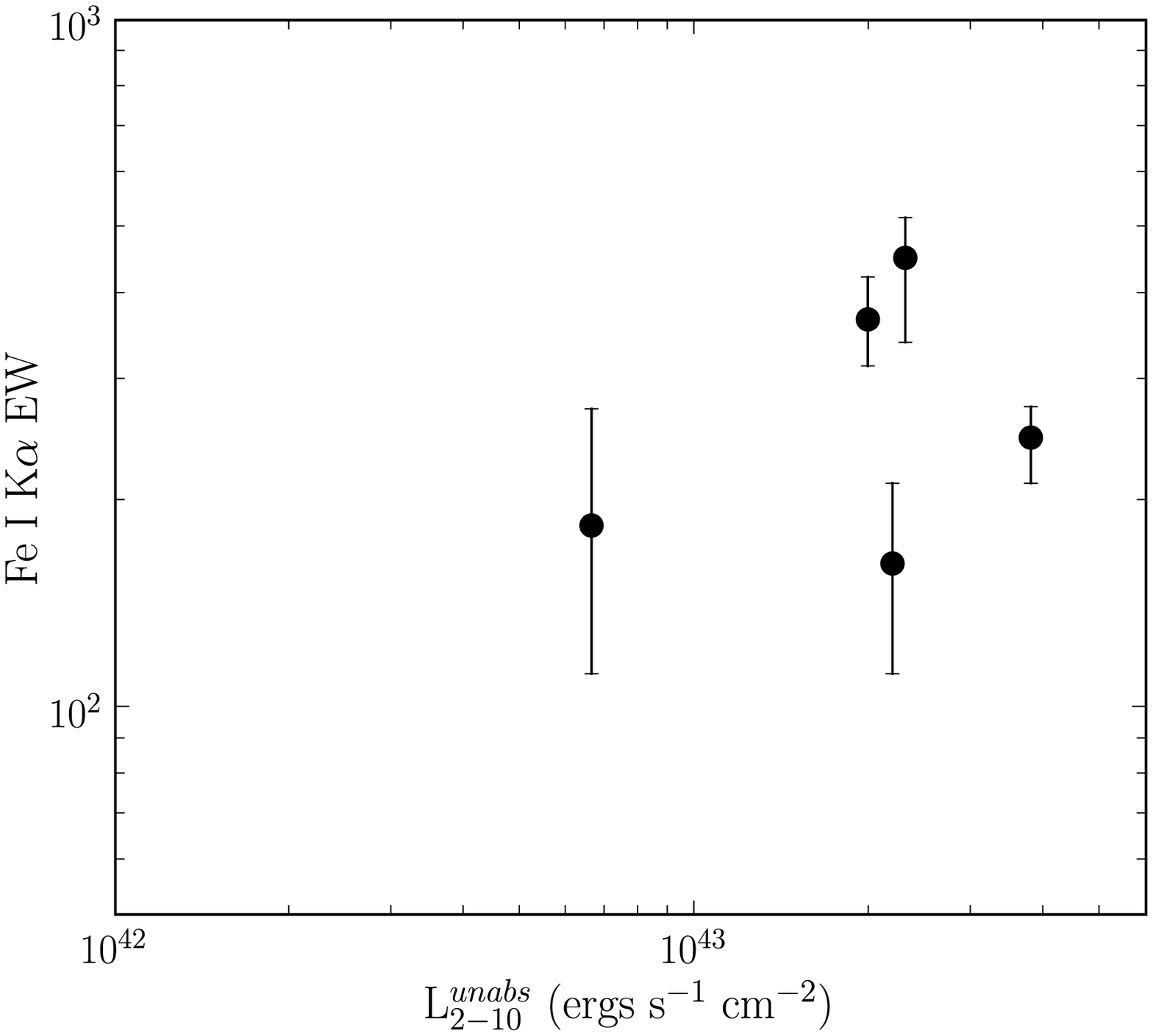}
\hspace{0.1cm}
\includegraphics[width=8.5cm]{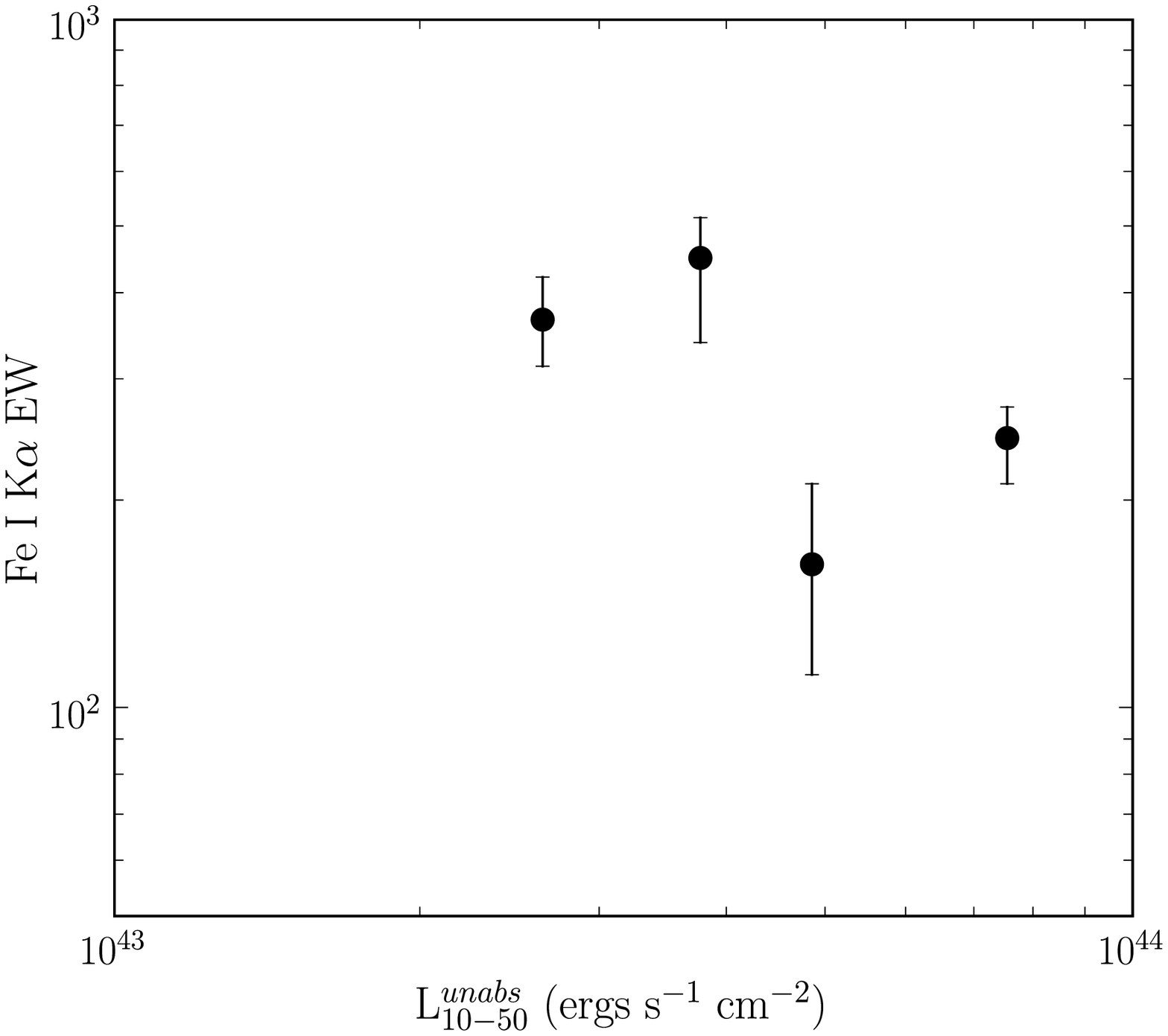}
\vspace{0.1cm}
\includegraphics[width=8.5cm]{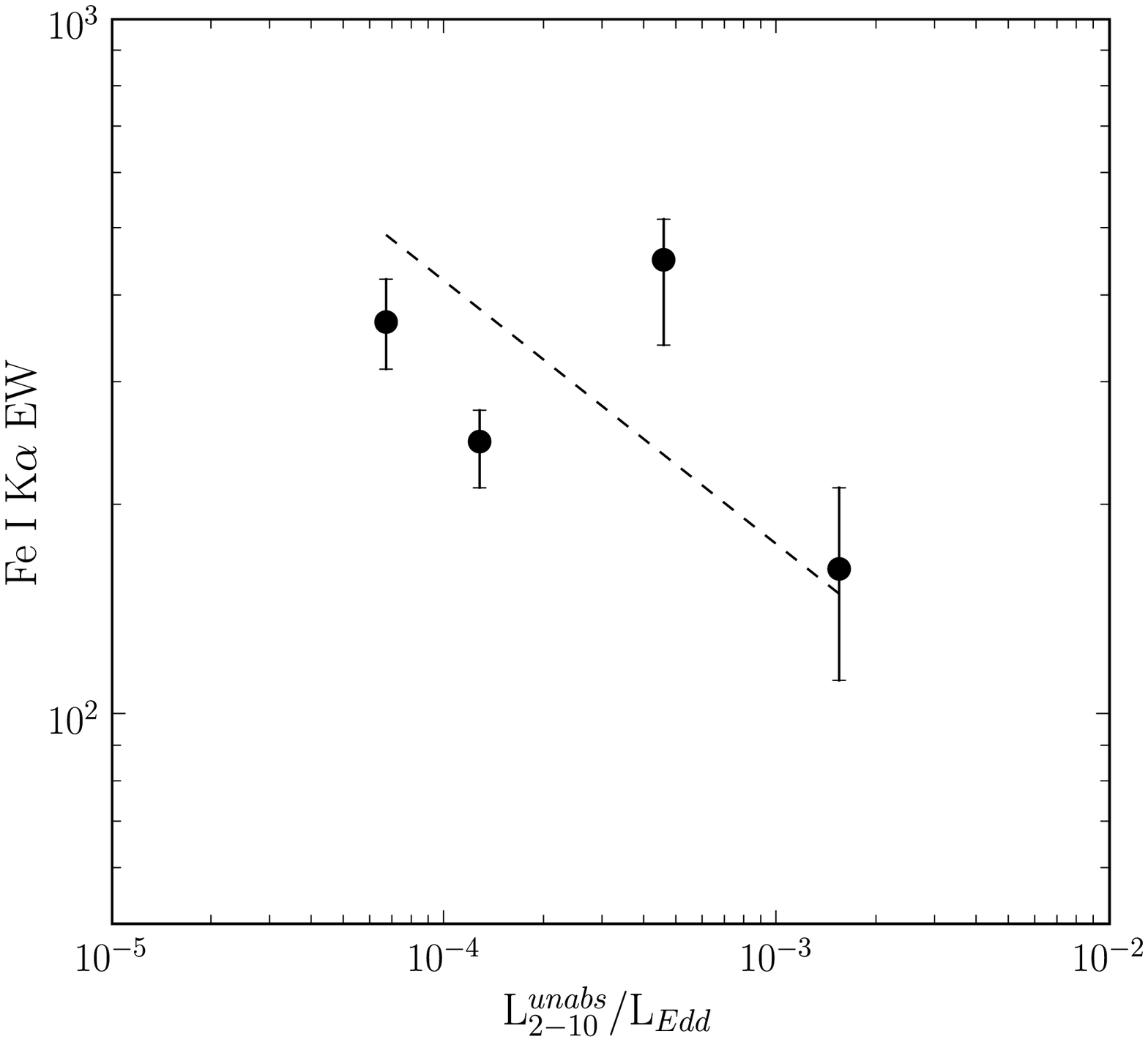}
\hspace{0.1cm}
\includegraphics[width=8.5cm]{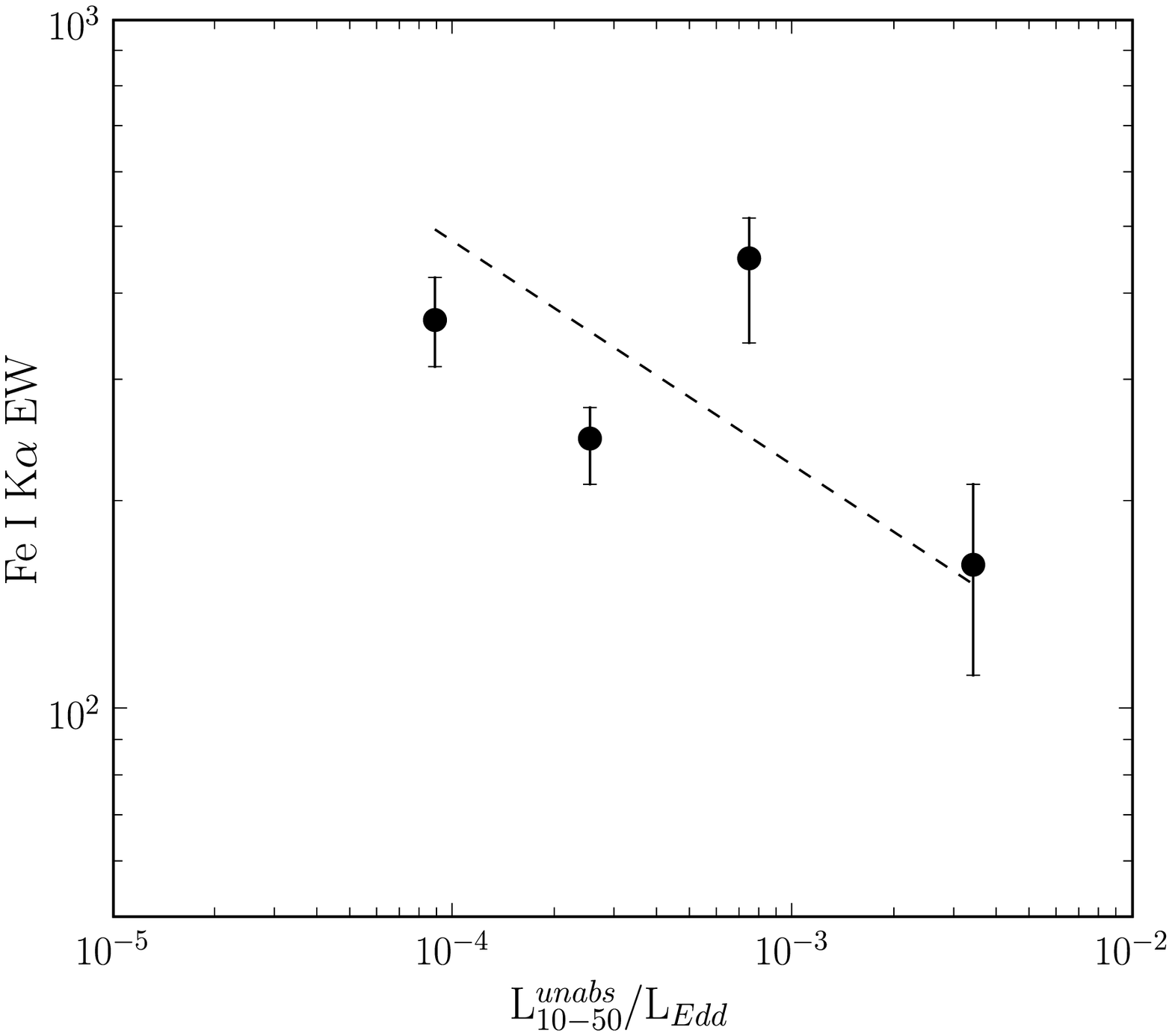}

\caption{Plotted is the \ion{Fe}{1} K$\alpha$ equivalent width versus the unabsorbed luminosity in the 2--10\,keV band (top left), the 10--50\,keV band (top right), the ratio of 2--10\,keV luminosity to the Eddington luminosity, and the ratio of 10--50\,keV luminosity to Eddington luminosity.  There is no obvious correlation in the luminosity plots, but a slight correlation ($R^2 =0.27$ and 0.34, respectively) emerges in the Eddington ratio plots.  The correlations agree with the correlation we found in our binned EW versus corrected $L_{2-10}/L_{Edd}$ plot in \citet{2009ApJ...690.1322W}, with $EW \propto (L^{unabs}_{2-10}/L_{Edd})^{-0.38 \pm 0.16}$ and $EW \propto (L^{unabs}_{10-50}/L_{Edd})^{-0.32 \pm 0.12}$.
\label{fig-felum}}
\end{figure}

In Figures~\ref{fig-nhlum},~\ref{fig-gammalum}, and~\ref{fig-felum}, we plot the relationship between \nh, $\Gamma$, and \ion{Fe}{1} K$\alpha$ EW and luminosity (absorption corrected in the 2--10\,keV and 10--50\,keV bands) and Eddington ratio ($L/L_{Edd}$).  The model parameters are obtained from the double partial covering model (Table~\ref{tbl-dblpcfabs}) and from Table~\ref{tbl-pcfabs} for MCG +04-48-002 (shown only in the top left panel of each figure).  The Eddington luminosity was obtained from the 2MASS derived masses shown in \citet{2009ApJ...690.1322W} ($1.3 \times 10^{38} \times (M_{BH}/M_{\sun})$\,ergs\,s$^{-1}$).  As shown in Figure~\ref{fig-nhlum}, there is no relationship between \nh\, and either luminosity or Eddington ratio.  However, as shown in Figure~\ref{fig-gammalum}, while there is no apparent correlation between $\Gamma$ and luminosity, there is possibly an anti-correlation between $\Gamma$ and $L/L_{Edd}$ (in both the 2--10 and 10--50 keV bands).  In the variability section, we already mentioned this behavior, which is opposite the correlation seen between $\Gamma$ and luminosity in other samples (i.e.  \citet{2004AA...422...85P,2005AA...432...15P,2006ApJ...646L..29S}).  In \citet{2009ApJ...690.1322W}, we had found no correlation of any kind between $\Gamma$ and luminosity or Eddington rate but interpreted this as a result of not having a sample of similar objects (like the PG quasars used in the samples mentioned).  In 
\citet{2008ApJ...674..686W}, we did see the correlation for multiple observations of individual unobscured sources (comparing Swift XRT and {\it XMM-Newton} spectra).  If there is an anti-correlation for the absorbed hidden sources, this suggests either that the model being used is wrong (even the double partial covering model may be too much of a simplification) and/or that there is a different physical mechanism responsible for the spectra of these sources.  However, since the error bars are very large on the $\Gamma$ measurements, the anti-correlation is not statistically significant.

In Figure~\ref{fig-felum}, we plot the \ion{Fe}{1} K$\alpha$ EW versus luminosity and Eddington ratio.  As shown, there is no correlation seen between the EW and luminosity.  However, there is a slight correlation between EW and the Eddington ratios.  Fitting a line to each plot, we find: \begin{equation} \log EW = (-0.38 \pm 0.16) \times \log (L^{unabs}_{2-10}/L_{Edd}) + (1.11 \pm 0.50)\end{equation}
\begin{equation} \log EW = (-0.32 \pm 0.12) \times \log (L^{unabs}_{10-50}/L_{Edd}) + (1.38 \pm 0.34).\end{equation}
While this correlation is not as statistically significant as the correlation we found between EW binned by $L^{corr}_{2-10}/L_{Edd}$ in  Figure 9 of  \citet{2009ApJ...690.1322W}, with $R^2 = 0.27$ and 0.34, the result is consistent.  In  \citet{2009ApJ...690.1322W}, we found $EW \propto (L^{corr}_{2-10}/L_{Edd})^{-0.26 \pm 0.03}$, which is also consistent with the results of \citet{2007AA...467L..19B} for radio quiet AGN ($EW \propto (L_{bol}/L_{Edd})^{-0.19 \pm 0.05}$).  Therefore, the accretion rate and \ion{Fe}{1} K$\alpha$ emission are tied together and the mechanism which creates this line is the same between the unabsorbed and absorbed sources.

\section{Summary}\label{summary}
In this paper, we present an analysis of Suzaku observations of five absorbed sources from the 9-month Swift BAT AGN catalog, NGC 1142, Mrk 417, ESO 506-G027, NGC 6921, and MCG +04-48-002.  These nearby sources ($<z> = 0.023$) were only recently detected in the X-rays by Swift's BAT.  Through our study of these sources, we conducted an analysis of both their variability properties and detailed spectral properties with Suzaku.

From a study of variability during the 40\,ks Suzaku observations (as well as the 80\,ks AO-1 observation of NGC 1142), we found little short term variability (i.e. only small amplitude variability of $< 0.1$\,ct\,s$^{-1}$ in the combined XIS0+XIS1+XIS3 light curves) on 128\,s or 5760\,s time scales.  On longer time scales of 0.5--1.5\,yrs, between the {\it XMM-Newton} and Suzaku observations of these sources, variability is present.  Both the shape (i.e. through changing \nh~\ and/or $\Gamma$) and flux change, most notably for NGC 6921 and MCG +04-48-002 whose fluxes change by an order of magnitude or more.  Further, this result is consistent with the results of \citet{2002ApJ...571..234R}, suggesting that variability is ubiquitous in absorbed sources.  Comparing the change in flux of the absorbed sources with a sample of unobscured sources from \citet{2008ApJ...674..686W}, we found that the obscured sources show significantly higher levels of hard band (2--10\,keV) variability.  This highlights a potentially important difference between absorbed and unabsorbed sources and is in agreement with previous findings which show that the soft component in the absorbed sources is at a larger scale \citep{2006AA...448..499B}.

When we examined variability in the 14--195\,keV band, through 400\,d Swift BAT light curves, we found that both the obscured and unobscured sources showed similar levels of variability.  Interestingly, the 16\,d binned light curves showed a correlation between excess variance and luminosity: $\sigma^2_{rms} \propto L^{(1.70 \pm 0.48)}_{14-195}$, which is also found in the 64\,day binned light curves ($\sigma^2_{rms} \propto L^{(2.12 \pm 0.45)}_{14-195}$), where the correlation is weaker.  This is, however, the first time such a correlation has been noted.  One clear implication of this analysis is that AGN do vary above 10\,keV.  This variability does not appear to be correlated with variability in the softer bands, which is evident from NGC 1142, the least variable source below $10$\,keV and the most variable above $10$\,keV.

From a detailed analysis of the Suzaku spectra of these ``hidden'' sources, we present a few interesting results.  Among these, we constrain the properties of the \ion{Fe}{1} K$\alpha$ feature (central energy, width, intensity, and equivalent width), finding the width of the line consistent with velocities from about 1800 -- 3000\,km\,s$^{-1}$.  Thus, our Suzaku results agree with Chandra grating results, placing the origin of the Fe K-$\alpha$ line in a region between the narrow (500 -- 700\,km\,s$^{-1}$) and broad (3000 -- 10000\,km\,s$^{-1}$) line regions \citep{2004ApJ...604...63Y}.  We also detect additional features in the spectra, including a significant 7.11\,keV Fe edge in the spectra of NGC 1142, \ion{Fe}{25} K$\alpha$ in the spectra of ESO 506-G027 and MCG +04-48-002, and \ion{Fe}{1} K$\beta$ in the spectra of NGC 1142 and ESO 506-G027.  

Finally, based on variations in flux, a small reflection component, and \ion{Fe}{1} K$\alpha$ $EW << 1$\,keV, it appears that none of our sources are true Compton-thick objects.  A more likely explanation for their properties is heavy absorption ($\log N_{\mathrm H} > 23$) in a clumpy environment \citep{2002ApJ...571..234R}.  In support of this, we find good fits to the spectra with a double partial covering model, a model where two absorbers partially cover the central emission, with no reflection component required.  

\acknowledgements
L.W. would like to thank Alex Markowitz (UCSD) for advice on generating light curves and processing the Suzaku data and Yuxuan Yang (University of Illinois) for access to the Suzaku spectrum of Circinus.  We also thank Satoshi Eguchi for discussions on the Suzaku data sets.

{\it Facilities:} \facility{Swift ()}, \facility{Suzaku ()}

\bibliography{ms-astroph.bib}

\newpage

\clearpage

\clearpage

\clearpage

\clearpage
\clearpage

\clearpage

\newpage

\clearpage

\clearpage

\clearpage

\clearpage

\clearpage

\newpage

\begin{deluxetable}{llllllll}
\tabletypesize{\small}
\tablecaption{Details for the Suzaku Observations\label{tbl-1}}
\tablewidth{0pt}
\tablehead{
\colhead{Source} & \colhead{RA (h m s)} & 
\colhead{Dec (\degr\,\arcmin\,\arcsec)} & \colhead{$z$} & \colhead{Obs ID}  & \colhead{Date} & \colhead{Exp (s)\tablenotemark{*}} & \colhead{Ct Rate\tablenotemark{*}} }
\startdata
NGC 1142 -- 1 & 02 55 12.2 & 00 11 01.0 & 0.028847 & 701013010 & 2007-01-23 & 101630, 80580 & 0.060, 0.081 \\
NGC 1142 -- 2 & \nodata & \nodata & \nodata & 702079010 & 2007-07-21 & 40570, 36540 & 0.041, 0.020 \\
Mrk 417 & 10 49 30.9 & 22 57 51.8 & 0.032756 & 702078010 & 2007-05-18 & 41507, 13865 & 0.04, 0.022 \\
ESO 506-G027 & 12 38 54.6 & -27 18 28.1 & 0.025024 & 702080010 & 2007-08-02 & 35721, 41761 &  0.035, 0.027\\
NGC 6921 & 20 28 28.9 & 25 43 24.3 & 0.014287 & 702081010 & 2007-04-18 & 41299 & 0.008 \\
MCG +04-48-002 & 20 28 35.0 & 25 44 00.0 & 0.013900 & 702081010 & 2007-04-18 &  41299 & 0.026 \\
\enddata
\tablenotetext{*}{Exposure time and count rate for XIS1 and PIN.  NGC 6921 and MCG +04-48-002
are in the same observation, however, NGC 6921 is very dim.}
\end{deluxetable}

\begin{deluxetable}{l | cll | cll}
\tabletypesize{\small}
\tablecaption{Variability in Suzaku Observations in 128s and 5760s Bins\label{tbl-2}}
\tablewidth{0pt}
\tablehead{
\colhead{Source} & \colhead{$<$Ct Rate$>$\tablenotemark{a}}  &
\colhead{$\sigma^2_{rms}$\tablenotemark{b}} & \colhead{$\chi^2$/dof\tablenotemark{c}}
& \colhead{$<$Ct Rate$>$\tablenotemark{a}}  &
\colhead{$\sigma^2_{rms}$\tablenotemark{b}} & \colhead{$\chi^2$/dof\tablenotemark{c}}
}
\startdata
& \multicolumn{3}{c}{\bf 128 s Bins} & \multicolumn{3}{c}{\bf 5760 s Bins} \\
\hline
NGC 1142 -- 1.1 &   6.93 & $27\pm3.5\times 10^{-6}$ & 1.63/398  & 6.94 & $6.1 \pm 1.4 \times 10^{-7}$ & 2.77/10 \\
NGC 1142 -- 1.2 & 6.99 & $10\pm1.9\times 10^{-6}$ & 1.35/398 & 6.64 & $0.02 \pm 1.1 \times 10^{-8}$ & 0.99/10\\
NGC 1142 --2  &  4.99 & $38\pm1.7\times 10^{-5}$ & 1.77/313 & 5.06 & $0.5 \pm 4.2\times 10^{-8}$ & 1.12/8 \\
Mrk 417 &  5.08 & $28\pm1.1\times 10^{-5}$ & 2.07/324 & 4.89 & $5.6 \pm 2.7\times 10^{-7}$ & 2.15/10 \\
ESO 506-G027 & 3.58 & $8.5\pm1.0 \times 10^{-5}$ & 1.44/327 & 3.32 & $5.0 \pm 4.9\times 10^{-8}$ & 0.31/6\\
MCG+04-48-002 & 4.61 & $140\pm 3 \times 10^{-4}$ & 2.45/326 & 4.24 & $13 \pm 9.1 \times 10^{-6}$ & 2.11/9\\
NGC 6921 & 1.35 & $220\pm 0.008$ & 2.79/326 & 0.89 & $23 \pm 4.3 \times 10^{-2}$ & 11.18/9 \\
\enddata
\tablenotetext{a}{Average background subtracted count rate ($10^{-2} \times$ cts\,s$^{-1}$) for the 0.1 -- 12\,keV band (averaged between the XIS0, XIS1, and XIS3 observations).\\}
\tablenotetext{b}{Corresponding excess variability measurements ($\times 10^{-3}$), as defined in \citet{1997ApJ...476...70N}.\\}
\tablenotetext{c}{Reduced $\chi^2$ value divided by the number of bins (dof).}
\end{deluxetable}

\begin{deluxetable}{l | cl | cl | cl}
\tabletypesize{\small}
\tablecaption{Variability in Suzaku Observations in 5760s Bins at Specified Energies\label{tbl-varenergies}}
\tablewidth{0pt}
\tablehead{
\colhead{Source} & \colhead{$<$Ct Rate$>_L$\tablenotemark{a}}  &
 \colhead{$\chi^2$/dof$_L$\tablenotemark{b}}
& \colhead{$<$Ct Rate$>_M$\tablenotemark{a}}  &
 \colhead{$\chi^2$/dof$_M$\tablenotemark{b}} & \colhead{$<$Ct Rate$>_H$\tablenotemark{a}}  &
 \colhead{$\chi^2$/dof$_H$\tablenotemark{b}}
}
\startdata

NGC 1142 -- 1 &   1.14 & 1.05/20 & 4.20  & 0.63/20 & 1.34 & 2.29/20 \\
NGC 1142 --2  &  1.29 & 1.14/8 & 2.68 & 1.17/8 & 0.86 & 1.60/8 \\
Mrk 417 &  0.73 & 2.74/10 & 3.04 & 2.18/10 & 1.08 & 1.12/10 \\
ESO 506-G027 & 0.56 & 0.52/6 & 1.99 & 1.06/6 & 0.66 & 0.93/6\\
MCG+04-48-002 & 1.21 & 3.86/9 & 2.04 & 1.88/9 & 0.87 & 1.00/9\\
NGC 6921 & 1.05 & 9.18/9 & 0.05 & 1.24/9 & 0.11 & 3.37/9 \\
\enddata
\tablenotetext{a}{Average background subtracted count rate ($10^{-2} \times$ cts\,s$^{-1}$) for the 0.1 -- 3\,keV band (L), 3 -- 7\,keV band (M), and 7 -- 12\,keV band (H), all averaged between the XIS0, XIS1, and XIS3 observations.\\}
\tablenotetext{b}{Reduced $\chi^2$ value divided by the number of bins (dof).}
\end{deluxetable}

\begin{deluxetable}{llllllll}
\tabletypesize{\small}
\tablecaption{Spectral Fits (0.3--10\,keV) with the Partial Covering Model\label{tbl-pcfabs}}
\tablewidth{0pt}
\tablehead{
\colhead{\bf Source} & \colhead{\bf N$_H$\tablenotemark{1}} & \colhead{\bf Cvr.\tablenotemark{1}} & \colhead{\bf $\Gamma$} & \colhead{Fe EW\tablenotemark{2}} & \colhead{$\chi^2$/dof} & \colhead{\bf F$_{S}$, F$_{H}$\tablenotemark{3}} & \colhead{\bf L$_S$, L$_H$\tablenotemark{4}}
}
\startdata
NGC 1142 - 1 & $8.52^{+0.24}_{-0.20}$ & $0.997^{+0.001}_{-0.002}$ & $2.32^{+0.14}_{-0.06}$ & $233^{+19}_{-18}$ & 1212.2/797 & 0.12, 3.60 &  43.78, 43.61\\
NGC 1142 - 2 & $8.59^{+0.88}_{-0.41}$ & $0.995^{+0.002}_{-0.003}$ & $2.40^{+0.23}_{-0.26}$ & $305^{+78}_{-47}$ & 257.3/180 & 0.10, 2.19 &  43.60, 43.42\\
Mrk 417 & $4.85^{+5.28}_{-3.41}$ & $0.974^{+0.010}_{-0.007}$ & $1.10^{+0.16}_{-0.02}$ & $126^{+30}_{-31}$ & 331.7/314 & 0.04, 2.99 &  42.55, 43.21\\
ESO 506-G027 & $8.90^{+8.39}_{-6.64}$ & $0.976^{+0.009}_{-0.008}$ & $1.20^{+0.21}_{-0.12}$ & $510^{+72}_{-68}$ & 304.3/236 & 0.05, 2.22 &  42.44, 43.06\\
NGC 6921 & $307$$^{*}$ & $0.554^{*}$ & 1.75$^{\dagger}$ & -- & 110.4/83 & 0.02, 0.05 &  41.26, 39.69\\
MCG +04-48-002 & $6.36^{+5.02}_{-2.24}$ & $0.990^{+0.005}_{-0.008}$ & $1.71^{+0.31}_{-0.32}$ & $157^{+70}_{-55}$ & 244.7/408 & 0.08, 2.29 & 42.48, 42.80 \\

\enddata
\tablenotetext{1}{Absorption (in addition to the Galactic value) is modeled with the partial covering model.  N$_H$ is in units of $\times 10^{23}$\,cm$^{-2}$ while Cvr. is the partial covering fraction.\\}
\tablenotetext{2}{Fe K EW in \,eV, assuming an Fe K-$\alpha$ line with width 0.01\,keV at an energy of 6.4\,keV.}
\tablenotetext{3}{The units for quoted fluxes are $\times 10^{-12}$\,erg\,s$^{-1}$\,cm$^{-2}$.  $F_S$ is the observed 0.5--2.0\,keV flux and $F_H$ is the observed 2.0 -- 10.0\,keV flux.}
\tablenotetext{4}{\,Logarithm of the absorption corrected (unabsorbed) luminosity in the 0.5--2.0\,keV ($L_S$) and 2.0--10.0\,keV ($L_H$) bands.}
\tablenotetext{*}{These parameters are upper limits.}
\tablenotetext{\dagger}{This parameter was fixed to the indicated value.}
\end{deluxetable}

\begin{deluxetable}{lllllll}
\tabletypesize{\small}
\tablecaption{Results of Simultaneous {\it XMM-Newton} and Suzaku Spectral Fits (0.3--10\,keV) with the Partial Covering Model\label{tbl-simultaneous}}
\tablewidth{0pt}
\tablehead{
\colhead{\bf Source} & \colhead{Soft Var.\tablenotemark{1}} & \colhead{Hard Var.\tablenotemark{1}} & \colhead{\bf N$_H$\tablenotemark{2}} & \colhead{\bf $\Gamma$\tablenotemark{2}} & \colhead{Fe EW\tablenotemark{2}} 
}
\startdata
NGC 1142 & 0.37 & 0.52 & 47.1 & 38.4 & 2.9 \\ 
Mrk 417 & 0.55 & 0.80 & 23 & 249.4 & 0.3 \\
ESO 506-G027 & 0.65 & 0.53 & 77.6 & -12.5 & 22.5 \\
NGC 6921 & 0.88 & 1.98 & 98.84 & 46.7 & 0.1 \\
MCG +04-48-002 & 1.10 & 1.60 & 3 & 12.2 & 0 \\
\enddata
\tablenotetext{1}{Observed $(F_{max} - F_{min})/F_{avg}$ in the soft band (0.5--2\,keV) and hard band (2--10\,keV).\\}
\tablenotetext{2}{$\Delta \chi^2$ when absorption (\nh and covering fraction for the {\tt pcfabs} model), $\Gamma$ and its normalization, or the Fe K normalization ({\tt zgauss}) are allowed to vary between the observations.\\}

\end{deluxetable}

\begin{deluxetable}{l | cll | cll | l}
\tabletypesize{\footnotesize}
\tablecaption{Variability in the BAT lightcurves (16 d, 64 d)\label{tbl-batlc}}
\tablewidth{0pt}
\tablehead{
\colhead{Source} & \colhead{$<$Ct Rate$>$\tablenotemark{a}}  &
\colhead{$\sigma^2_{rms}$\tablenotemark{b}} & \colhead{$\chi^2$/dof\tablenotemark{c}} &
\colhead{$<$Ct Rate$>$\tablenotemark{a}}  &
\colhead{$\sigma^2_{rms}$\tablenotemark{b}} & \colhead{$\chi^2$/dof\tablenotemark{c}}
& \colhead{$L_{14-195}$\tablenotemark{d}}}
\startdata
\multicolumn{1}{c}{} & \multicolumn{3}{c}{\bf 16 d} & \multicolumn{3}{c}{\bf 64 d} & \multicolumn{1}{c}{}\\
\hline
NGC 1142 & 1.46 & $67\pm 7.3 \times 10^{-4}$ & 3.87/23 &  1.42 & $120\pm1.3 \times 10^{-3}$ & 11.51/5 & 44.17 \\
Mrk 417 &  0.51 & $210\pm 0.83$ & 0.84/23 & 0.41 & $81\pm4.5 \times 10^{-3}$ & 0.39/5 & 43.95 \\
ESO 506-G027 & 1.28 & $100\pm0.01$ & 0.99/23 & 1.15 & $6.3\pm3.6 \times 10^{-4}$ & 1.49/10 & 44.28 \\
MCG +04-48-002/NGC 6921 & 1.14 & $24\pm2.0\times 10^{-3}$ & 1.12/23 & 1.10 & $10\pm1.0\times 10^{-5}$ & 1.02/5 & 43.45\\

\hline

Mrk 352 & 0.50 & $38\pm0.30$ & 1.49/23 & 0.45 & $38\pm5.8 \times 10^{-3}$ & 1.74/15 & 43.27\\
ESO 548-G081 & 0.68 & $1.6\pm7.7\times 10^{-3}$ & 1.14/23 & 0.76 & $7.6\pm8.2 \times 10^{-5}$ & 0.82/5 & 43.19\\
ESO 490-G026 & 0.65 & $300\pm0.09$ & 1.38/23 & 0.69 & $26\pm3.5 \times 10^{-3}$ & 1.45/5 & 43.71 \\
2MASX J09043699+5536025 & 0.29 & $420\pm0.43$ & 0.87/23 & 0.30 & $110\pm4.3 \times 10^{-3}$ & 0.28/10 & 44.03\\
MCG +04-22-042 &  0.52 & $4800\pm4000$ & 1.18/23 & 0.57 & $4.5\pm1.1 \times 10^{-3}$ & 1.12/10 & 43.99 \\
UGC 06728 & 0.43 & $9.3\pm0.19$ & 1.17/23 & 0.53 & $0.02\pm3.8\times 10^{-5}$ & 1.09/15 & 42.54 \\
WKK 1263 & 0.41 & $60\pm59$ & 1.22/23 & 0.40 & $260\pm0.58$ & 0.32/5 & 43.58 \\
MCG +09-21-096 &  0.58 & $58\pm9.8\times 10^{-3}$ & 1.37/23 & 0.58 & $26\pm1.0 \times 10^{-3}$ & 1.91/5 & 43.72\\
\enddata
\tablenotetext{a}{Average count rate ($10^{-4} \times$ cts\,s$^{-1}$) for the 14 -- 195\,keV band BAT lightcurves.\\}
\tablenotetext{b}{Corresponding excess variability measurements ($\times 10^{-3}$), as defined in \citet{1997ApJ...476...70N}.\\}
\tablenotetext{c}{Reduced $\chi^2$ value divided by the number of bins (dof).}
\tablenotetext{d}{\,Logarithm of the 14--195\,keV luminosity, from \citet{2008ApJ...681..113T}}
\end{deluxetable}

\begin{deluxetable}{llllllll}
\tabletypesize{\small}
\tablecaption{Spectral Fits (0.3--195\,keV) with the Partial Covered Cut-off Power Law Model\label{tbl-cutoff}}
\tablewidth{0pt}
\tablehead{
\colhead{\bf Source} & \colhead{\bf N$_H$\tablenotemark{a}} & \colhead{\bf Cvr.\tablenotemark{a}} & \colhead{\bf $\Gamma$} & \colhead{$E_{cutoff}$ (keV)} & \colhead{$\Delta\chi^2$\tablenotemark{b}} & \colhead{$\chi^2$/dof}
}
\startdata
NGC 1142 - 1 & $4.30^{+0.16}_{-0.10}$ & $0.970^{+0.004}_{-0.003}$ & $0.73^{+0.20}_{-0.11}$ & $32.8^{+30.0}_{-5.3}$ & 35.4 & 914.0/830 \\
NGC 1142 - 2 & $4.61^{+0.29}_{-0.29}$ & $0.975^{+0.006}_{-0.007}$ & $1.46^{+0.18}_{-0.12}$ & $80.6^{+62.4}_{-27.3}$ & 12.5 & 470.1/313 \\
Mrk 417 & $3.30^{+0.34}_{-0.21}$ & $0.981^{+0.003}_{-0.011}$ & $1.25^{+0.14}_{-0.17}$ & $49.3^{+32.4}_{-23.8}$ & 23.3 & 361.5/336 \\
ESO 506-G027 & $6.15^{+0.34}_{-0.61}$ & $0.979^{+0.005}_{-0.010}$ & $1.22^{+0.14}_{-0.22}$ & $67.4^{+30.6}_{-21.6}$ & 28.0 & 324.0/268 \\

\enddata
\tablenotetext{a}{Absorption (in addition to the Galactic value) is modeled with the partial covering model.  N$_H$ is in units of $\times 10^{23}$\,cm$^{-2}$ while Cvr. is the partial covering fraction.\\}
\tablenotetext{b}{$\Delta\chi^2$ between the partially covered power law and partially covered cut-off power law model.}
\end{deluxetable}

\begin{deluxetable}{llcccc}
\tabletypesize{\footnotesize}
\tablecaption{Emission Features with the Partial Covered Cut-off Power Law Model\label{tbl-felines}}
\tablewidth{0pt}
\tablehead{
\colhead{\bf Component} & \colhead{\bf Parameter\tablenotemark{1}} & \colhead{\bf NGC 1142 -- 1} & \colhead{\bf NGC 1142 -- 2} & \colhead{\bf Mrk 417} &
\colhead{\bf ESO 506-G027}  
}
\startdata
Fe I K$\alpha$ & $\Delta\chi^2$ & 531.9 & 173.7 & 54.3 & 231.2 \\
 & E$_c$ & $6.394^{+0.008}_{-0.014}$ & $6.394^{+0.007}_{-0.007}$ & $6.356^{+0.035}_{-0.036}$ & $6.393^{+0.023}_{-0.010}$ \\
 & $\sigma$ & $0.054^{+0.011}_{-0.018}$ & $0.064^{+0.019}_{-0.019}$ & $0.092^{+0.051}_{-0.046}$ & 0.057\tablenotemark{\dagger} \\
 & $I$ & $40^{+4}_{-6}$ & $51^{+8}_{-7}$ & $20^{+6}_{-6}$ & $72^{+10}_{-18}$ \\
 & $EW$ & $247^{+27}_{-35}$ & $367^{+56}_{-53}$ & $162^{+50}_{-50}$ & $451^{+65}_{-111}$ \\
\hline
Fe XXV K$\alpha$ & $\Delta\chi^2$ & 0.7 & 0 & 0.5 & 6.8 \\
 & E$_c$ & 6.722 & 6.67\tablenotemark{*} & 6.67\tablenotemark{*} & $6.657^{+0.107}_{-0.162}$ \\
 & $I$ & 0.5\tablenotemark{\dagger} & 6\tablenotemark{\dagger} & 3\tablenotemark{\dagger} & $8.2^{+9.9}_{-5.5}$ \\
 & $EW$ & 17\tablenotemark{\dagger} & 29\tablenotemark{\dagger} & 22\tablenotemark{\dagger} & $34^{+41}_{-23}$ \\
\hline
Fe I K$\beta$ & $\Delta\chi^2$ & 5.0 & 5.3 & 0.1 & 4.4 \\
 & E$_c$ & $7.029^{+0.096}_{-0.069}$ & $6.999^{+0.055}_{-0.061}$ & 7.06\tablenotemark{*} & $7.048^{+0.112}_{-0.076}$ \\
 & $I$ & $3^{+14}_{-1}$ & $4.9^{+3.9}_{-3.7}$ & 13\tablenotemark{\dagger} & $5.7^{+4.3}_{-4.1}$ \\
 & $EW$ & 118\tablenotemark{\dagger} & $45^{+37}_{-11}$ & 146\tablenotemark{\dagger} & $39^{+32.9}_{-31.2}$ \\ 
\hline
Ni I K$\alpha$ & $\Delta\chi^2$ & 1.7 & 0.6 & 1.3 & 0.4 \\
 & E$_c$ & 7.451 & 7.45\tablenotemark{*} & 7.45\tablenotemark{*} & 7.45\tablenotemark{*} \\
 & $I$ & 8\tablenotemark{\dagger} & 6\tablenotemark{\dagger} & 7\tablenotemark{\dagger} &11\tablenotemark{\dagger} \\
 & $EW$ & 63\tablenotemark{\dagger} & 66\tablenotemark{\dagger} & 124\tablenotemark{\dagger} & 99\tablenotemark{\dagger} \\ 
\enddata
\tablenotetext{1}{The parameters from fitting the feature with a gaussian model.  These parameters include the $\Delta\chi^2$ value from adding this component, the central energy of the component in keV, the intensity in units of $10^{-6}$\,photons\,cm$^{-2}$\,s$^{-1}$, the width of the line ($\sigma$, where not indicated this is fixed to 0.01\,keV), and the equivalent width (eV) with respect to the absorbed cutoff power law model.\\}
\tablenotetext{*}{This value is fixed.\\}
\tablenotetext{\dagger}{This value is an upper limit.\\}
\end{deluxetable}

\begin{deluxetable}{lcccc}
\tabletypesize{\small}
\tablecaption{Spectral Fits (0.3--195\,keV) with the Reflection Model\label{tbl-modelb}}
\tablewidth{0pt}
\tablehead{
 \colhead{\bf Parameter\tablenotemark{1}} & \colhead{\bf NGC 1142 -- 1} & \colhead{\bf NGC 1142 -- 2} & \colhead{\bf Mrk 417} &
\colhead{\bf ESO 506-G027}  
}
\startdata
\nh$^{trans}$ ($\times 10^{23}$\,cm$^{-2}$)& $11.15^{+1.33}_{-1.43}$ & $7.98^{+0.44}_{-0.81}$ & $5.55^{+0.42}_{-0.37}$ & $8.41^{+0.73}_{-1.17}$ \\
$\Gamma$ & $1.92^{+0.09}_{-0.11}$ & $1.91^{+0.06}_{-0.17}$ & $1.45^{+0.13}_{-0.11}$ & $1.52^{+0.10}_{-0.05}$ \\
$f_{scat}$ ($\times 10^{-2}$) & $0.65^{+0.16}_{-0.17}$\tablenotemark{*} & $0.33^{+0.17}_{-0.05}$ & $0.17^{+0.06}_{-0.04}$ & $0.19^{+0.03}_{-0.02}$ \\
 \nh$^{refl}$ ($\times 10^{23}$\,cm$^{-2}$) & $2.55^{+0.66}_{-0.55}$ & 4.55\tablenotemark{\dagger} & $5.24^{+2.23}_{-1.70}$ & 1.41\tablenotemark{\dagger} \\
 $R$ & $-2.20^{+1.83}_{-0.85}$ & $-0.29^{+0.15}_{-0.15}$ & $-0.19$\tablenotemark{\dagger} & $-0.46^{+0.33}_{-0.18}$ \\
$kT_{apec} (keV)$ & $0.71^{+0.04}_{-0.06}$ & $0.61^{+0.06}_{-0.06}$ & \nodata & \nodata \\
 $F_{2-10}^{unabs}$ ($\times 10^{-11}$\,ergs\,s$^{-1}$\,cm$^{-2}$) & 1.99 & 1.24 & 1.06 & 1.70 \\
 $F_{10-50}^{unabs}$ ($\times 10^{-11}$\,ergs\,s$^{-1}$\,cm$^{-2}$)& 4.48 & 1.54 & 2.31 & 3.00 \\
 $\chi^2/dof$ & 898.5/882 & 349.7/321 & 322.0/342 & 263.8/242 \\
\enddata
\tablenotetext{1}{The parameters from a reflection model represented as {\tt tbabs}$_{Gal}$*({\tt ztbabs}$_{trans}$*{\tt cutoffpl}$_{trans}$ + {\tt const}*{\tt cutoffpl}$_{scat}$ + {\tt ztbabs}$_{refl}$*{\tt pexrav}$_{refl}$ + Fe lines) in {\tt XSPEC}.  The transmitted, scattered, and reflected power law components are fixed to have the same spectral index ($\Gamma$) and cutoff energy (300\,keV).\\}
\tablenotetext{\dagger}{Upper limit on the indicated parameter.}
\tablenotetext{*}{Note that in \citet{eguchi2008} $f_{scat}$ is the constant factor recorded here divided by the reflection factor, $R$, for values of $R > 1.0$.}
\end{deluxetable}

\begin{deluxetable}{lcccc}
\tabletypesize{\small}
\tablecaption{Spectral Fits (0.3--195\,keV) with the Double Partial Covering Model\label{tbl-dblpcfabs}}
\tablewidth{0pt}
\tablehead{
 \colhead{\bf Parameter\tablenotemark{a}} & \colhead{\bf NGC 1142 -- 1} & \colhead{\bf NGC 1142 -- 2} & \colhead{\bf Mrk 417} &
\colhead{\bf ESO 506-G027}  
}
\startdata
\nh$^1$ ($\times 10^{23}$\,cm$^{-2}$)& $6.89^{+1.92}_{-0.93}$ & $7.96^{+0.85}_{-0.79}$ & $3.71^{+0.38}_{-0.76}$& $12.31^{+1.28}_{-1.03}$\\
$Cvr^1$ & $0.896^{+0.051}_{-0.175}$ & $0.978^{+0.007}_{-0.014}$ & $0.969^{+0.013}_{-0.067}$ & $0.969^{+0.011}_{-0.015}$\\
\nh$^2$ ($\times 10^{23}$\,cm$^{-2}$)& $3.77^{+1.59}_{-1.36}$ & $0.88^{+1.04}_{-0.56}$ & $0.51^{+0.99}_{-0.33}$ & $1.44^{+0.46}_{-0.65}$\\
$Cvr^2$ & $0.957^{+0.024}_{-0.054}$ & $0.681^{+0.080}_{-0.137}$ & $0.642^{+0.221}_{-0.389}$ & $0.816^{+0.067}_{-0.106}$\\
$\Gamma$ & $1.56^{+0.15}_{-0.18}$ & $1.82^{+0.15}_{-0.12}$ & $1.44^{+0.14}_{-0.16}$ & $1.68^{+0.14}_{-0.13}$\\
$E_{cutoff}$ (keV) & $143^{+57}_{-73}$ & 100\tablenotemark{*} & 100\tablenotemark{*}& 100\tablenotemark{*} \\
$kT_{apec}$ (keV) & $0.66^{+0.03}_{-0.03}$ & $0.62^{+0.03}_{-0.09}$ & \nodata & \nodata\\
$\tau_{7.11 keV}$ & $0.26^{+0.08}_{-0.06}$ &  0.12\tablenotemark{\dagger}& \nodata & \nodata \\
 $F_{2-10}^{unabs}$ ($\times 10^{-11}$\,ergs\,s$^{-1}$\,cm$^{-2}$) & 2.35 & 1.23 & 1.05 & 1.90 \\
 $F_{10-50}^{unabs}$ ($\times 10^{-11}$\,ergs\,s$^{-1}$\,cm$^{-2}$)& 4.66 & 1.63 & 2.32 & 3.10 \\
$\chi^2/dof$ & 898.5/881 & 338.9/319 & 316.1/338 & 242.7/240\\

\enddata
\tablenotetext{a}{Absorption (in addition to the Galactic value) is modeled with the partial covering model.  N$_H$ is in units of $\times 10^{23}$\,cm$^{-2}$ while Cvr. is the partial covering fraction.  Our double partial covering model is represented in {\tt XSPEC} as: {\tt tbabs}*({\tt apec} + {\tt zpcfabs}*{\tt zpcfabs}*{\tt zedge}*({\tt cutoffpl} + Fe lines)). The photo-ionized model ({\tt apec}) and Fe edge at 7.11\,keV are only used where they are statistically significant.\\}
\tablenotetext{\dagger}{Upper limit on the indicated parameter.}
\tablenotetext{*}{The value could not be constrained and is fixed to 100\,keV.}
\end{deluxetable}

\end{document}